\documentclass[notitlepage,floatfix,superscriptaddress,nofootinbib,tightenlines,preprintnumbers,twocolumn]{revtex4-2}
\usepackage{amsmath,verbatim,latexsym,amssymb,indentfirst,mathrsfs,mathtools,amsthm,bbm,bm,hyperref,url,cancel,subcaption,enumitem,float}
\usepackage[font=small,labelfont=bf,format=plain,justification=raggedright,singlelinecheck=false]{caption}
\usepackage{verbatim,indentfirst}
\usepackage{siunitx}
\usepackage[title,titletoc]{appendix}
\hypersetup{nolinks=true}

\usepackage{graphicx}
\usepackage{epstopdf}

\renewcommand{\thesection}{\Roman{section}}
\renewcommand{\thesubsection}{\Roman{section} \Alph{subsection}}
\renewcommand{\thesubsubsection}{\Roman{section} \Alph{subsection} \arabic{subsubsection}}
\makeatletter
\def\p@subsection{}
\makeatother
\makeatletter
\def\p@subsubsection{}
\makeatother

\makeatletter
\newcommand\footnoteref[1]{\protected@xdef\@thefnmark{\ref{#1}}\@footnotemark}
\makeatother

\usepackage{soul}

\usepackage{color}
\usepackage[dvipsnames]{xcolor}
\usepackage[normalem]{ulem}

\newcommand{\NATS}{{\rm NATS}}

\newcommand{\dyn}{{\rm dyn}}  

\newcommand{\stat}{{\rm stat}}

\newcommand{\hc}{ {\rm h.c.} }

\newcommand{\tot}{ {\rm tot} }
\newcommand{\Tr}{{\rm Tr}}   
\def\id{\mathbbm{1}}   

\newcommand{\Sys}{\mathcal{S}}  
\newcommand{\0}{ {(0)} }
\newcommand{\1}{ {(1)} }
\newcommand{\2}{ {(2)} }

\newcommand{\JParen}{ {(j)} }
\newcommand{\KParen}{ {(k)} }
\newcommand{\LParen}{ \bm{(} }
\newcommand{\RParen}{ \bm{)} }

	\definecolor{blue(pigment)}{rgb}{0.2, 0.2, 0.6}
 
\renewcommand\th{ {\rm th} }

\newcommand*{\bra}[1]{\langle #1\rvert}
\newcommand*{\ket}[1]{\lvert #1 \rangle}
\newcommand*{\braket}[2]{\langle #1 \lvert #2 \rangle}
\newcommand*{\ketbra}[2]{\lvert #1 \rangle\!\langle #2 \rvert}
\newcommand*{\expval}[1]{\left\langle  #1  \right\rangle}

\hypersetup{final}

\begin{document}

\title{Kubo–Martin–Schwinger relation for energy eigenstates of SU(2)-symmetric quantum many-body systems}

\author{Jae~Dong~Noh}
\email{jdnoh@uos.ac.kr}
\affiliation{Department of Physics, University of Seoul, Seoul 02504, Korea}

\author{Aleksander~Lasek}
\affiliation{Joint Center for Quantum Information and Computer Science, NIST and University of Maryland, College Park, Maryland 20742, USA}

\author{Jade~LeSchack}
\affiliation{Joint Center for Quantum Information and Computer Science, NIST and University of Maryland, College Park, Maryland 20742, USA}

\author{Nicole~Yunger~Halpern}
\email{nicoleyh@umd.edu}
\affiliation{Joint Center for Quantum Information and Computer Science, NIST and University of Maryland, College Park, Maryland 20742, USA}
\affiliation{Institute for Physical Science and Technology, University of Maryland, College Park, MD 20742, USA}

\date{\today}

%
%
\begin{abstract}
The fluctuation–dissipation theorem (FDT) is a fundamental result in statistical mechanics. It stipulates that, if perturbed out of equilibrium, a system responds at a rate proportional to a thermal-equilibrium property. Applications range from particle diffusion to electrical-circuit noise. To prove the FDT, one must prove that common thermal states obey a symmetry property, the Kubo–Martin–Schwinger (KMS) relation. Energy eigenstates of certain quantum many-body systems were recently proved to obey a KMS relation. The proof relies on the eigenstate thermalization hypothesis (ETH), which explains how such systems thermalize internally. This KMS relation contains a finite-size correction that scales as the inverse system size. Non-Abelian symmetries conflict with the ETH, so a non-Abelian ETH was proposed recently. Using it, we derive a KMS relation for SU(2)-symmetric quantum many-body systems' energy eigenstates. The finite-size correction scales as usual under certain circumstances but can be polynomially larger in others, we argue. We support the ordinary-scaling result numerically, simulating a Heisenberg chain of 16–24 qubits. The numerics, limited by computational capacity, indirectly support the larger correction. This work helps extend into nonequilibrium physics the effort, recently of interest across quantum physics, to identify how non-Abelian symmetries may alter conventional thermodynamics.
\end{abstract}

{\let\newpage\relax\maketitle}

%
%
%
The fluctuation–dissipation theorem (FDT) forms a pillar of nonequilibrium statistical mechanics~\cite{85_Kubo_Book,06_Mazenko_Book}. The theorem arises in linear-response theory, describing how quickly a system responds to a perturbation. One quantifies the responsiveness with a second derivative of a free energy. This derivative is proportional to a two-time thermal correlation function, according to the FDT. Knowing an equilibrium property, therefore, one can infer about a nonequilibrium response. Famous instances of the FDT include the Einstein–Smoluchowski relation, which relates a particle's mobility to its diffusion constant and temperature. Classical and quantum systems obey the FDT. Yet the quantum FDT governs a system that can be in a thermal quantum state. A symmetry property of this state, called the Kubo–Martin–Schwinger (KMS) relation, underlies the FDT.

Thermal states are mixed; in contrast, closed, isolated quantum systems are in pure states. How to extend the FDT to such systems is therefore unclear. Yet generic such many-body systems thermalize internally: a small subsystem will likely approach a thermal state, the rest of the system serving as an effective environment. This thermalization has drawn attention across theory and experiment recently~\cite{DAlessio_16_From,Gogolin_2016,21_Simulator_review}. The eigenstate thermalization hypothesis (ETH) explains why the thermalization occurs~\cite{DeutschThermalization1991,SrednickiThermalization1994,RigolThermalization2008}. Noh \emph{et al.} recently proved that, if a system obeys the ETH, its energy eigenstates approximately obey the KMS relation~\cite{Noh_20_Numerical}. If $N$ denotes the system size, the finite-size correction is $O(N^{-1})$. 

Non-Abelian symmetries have recently been found to partially preserve, and partially alter, conventional thermodynamic results. A lack of non-Abelian symmetries implicitly underlies derivations of the thermal state's form~\cite{Halpern_2018_HeatBaths2,Halpern_2016_microcanonical}, the Onsager relations (another pillar of linear-response theory)~\cite{Manzano_22_Non}, and more. If a Hamiltonian respects a non-Abelian symmetry, it conserves quantities (\emph{charges}) that fail to commute with each other. Such noncommutation typifies quantum physics, as exemplified by uncertainty relations, measurement disturbance, and quantum error correction. The \emph{quantum thermodynamics of noncommuting charges} has therefore emerged recently~\cite{guryanova_2016_thermodynamics,Lostaglio_2017_MaxEnt,Halpern_2016_microcanonical} and spread across many-body physics; quantum computation; atomic, molecular, and optical (AMO) physics; and high-energy physics~\cite{Majidy_23_Noncommuting,25_Campbell_Roadmap}. Non-Abelian symmetries alter thermodynamic-entropy production~\cite{Manzano_22_Non,Upadhyaya_24_Non}, entanglement entropy~\cite{22_Xie_Entanglement,Majidy_23_Non,Majidy_23_Critical,24_Bianchi_Non,24_Moharramipour_Symmetry,24_Zhang_Operator,25_Li_Highly,25_Kumar_Thesis}, quantum-computational universality~\cite{Marvian_24_Theory}, thermalization-resistant regimes of matter~\cite{Potter_16_Symmetry,Majidy_24_Noncommuting}, and more. The testing of these theoretical results recently began with a trapped-ion experiment~\cite{Kranzl_23_Experimental}. 

Non-Abelian symmetries conflict with the conventional ETH~\cite{MurthyNAETH}. To reconcile the two, Murthy \emph{et al.} recently proposed a \emph{non-Abelian ETH}~\cite{MurthyNAETH}. Numerics indicate that certain spin systems obey this ansatz~\cite{JaeDongXXZ,24_Lasek_Numerical,25_Patil_Eigenstate}. Do they obey a KMS relation?

We introduce a \emph{fine-grained KMS relation} obeyed by quantum many-body systems subject to SU(2) symmetry and the non-Abelian ETH. The relation depends not only on a temperature, as usual, but also on analogous parameters associated with spin (or, more generally, angular-momentum) quantum numbers. 
The proof requires a certain parameter regime, as well as approximations of Clebsch--Gordan coefficients introduced by the non-Abelian symmetry. Under certain conditions, we argue, the FDT's finite-size correction is $O( N^{-1} )$, as in the absence of any non-Abelian symmetry. Under other conditions, we contend, the finite-size correction can be polynomially larger. Hence non-Abelian symmetry appears able to alter a standard thermodynamic result by an amount polynomial in the system size.

To support our analytical arguments, we numerically simulate a one-dimensional (1D) chain of 16--24 qubits. We directly observe evidence for the $O(N^{-1})$ scaling. Limited by computational capacity, we observe indirect evidence for the larger correction. This work shows how non-Abelian symmetry may sometimes preserve, and sometimes alter, a result crucial for nonequilibrium statistical mechanics.

The rest of the paper is organized as follows. Section~\ref{sec_Review_FDT} reviews the conventional KMS relation; and Sec.~\ref{sec_Review_NAETH}, the non-Abelian ETH. In Sec.~\ref{sec_FDT_Thermo}, we prove the fine-grained KMS relation for an SU(2)-symmetric system in a mixed state; and, in Sec.~\ref{sec_FDT_Analytics}, the fine-grained KMS relation for a closed SU(2)-symmetric system. Numerics in Sec.~\ref{sec_Num} support the analytical results. Section~\ref{sec_Outlook} highlights research opportunities established by this work.

\section{Review of conventional KMS relation}
\label{sec_Review_FDT}

We derive the conventional quantum KMS relation and FDT~\cite{85_Kubo_Book,06_Mazenko_Book,23_Cugliandolo_Sorbonne} pedagogically in App.~\ref{app_Review_FDT}. Here, we review three highlights: the setup, KMS relation, and FDT. We set $\hbar = 1$. 

Consider a quantum system $\Sys$ that has observables $A$ and $B$~\cite{23_Cugliandolo_Sorbonne}. Upon beginning in a state $\rho$, $\Sys$ evolves under a Hamiltonian $H$ during the times 0 to $t'$ and $t' + \epsilon$ to $t$. During the infinitesimal interval $\epsilon$, $\Sys$ evolves under a perturbed Hamiltonian $H' \coloneqq  H - hB$ of perturbation strength $h \in \mathbb{R}$. 

How sharply does the perturbation change the time-$t$ expectation value 
$\expval{ A(t) }_\rho  \equiv  \Tr \LParen A(t) \rho \RParen$? In the zero-field limit, the \emph{response function} answers this question:
$R_{AB} (t, t')  
\coloneqq  \frac{\partial \expval{A(t)}_\rho }{\partial (h \epsilon)}
\big\lvert_{h = 0} \,  \Theta (t - t') .$
The Heaviside function encodes causality. The response function is proportional to a two-time correlator, according to the \emph{Kubo formula}:
$R_{AB}(t, t')
=  i \langle [ A(t), B(t') ] \rangle_\rho  \,  \Theta (t - t')$.

The correlator exhibits a symmetry---the KMS relation---under certain conditions. Assume that $\rho$ is the canonical state
$\rho_\th \coloneqq e^{-\beta H} / \Tr (e^{-\beta H})$. This state is stationary, or invariant under time evolution. Correlators in $\rho_\th$, therefore, depend on time only through differences $t - t'$. The Kubo formula's correlator has a Fourier transform 
$\bar{C}_{AB}(\Omega)$ that obeys the KMS relation 
$\bar{C}_{AB}(\Omega)
=  e^{\beta \Omega} \, \bar{C}_{AB}(-\Omega)$.

Applying the KMS relation to the Kubo formula, we can derive the FDT. The FDT depends on the anticommutator correlator
$\langle \{ A(t), B(t') \} \rangle_{\rho_\th} \, ,$ which has the Fourier transform
$\bar{C}_{ \{ AB \} } (\omega)$. This transform is proportional to the response function's transform:
${\rm Im} \LParen  \tilde{R}_{AB}(\omega)  \RParen
= \tanh (\beta \omega / 2)  \bar{C}_{ \{ AB \} } (\omega)$.
According to this FDT, a perturbation $-hB$ generates a nonequilibrium response determined by equilibrium correlations. In the derivation, only the KMS relation depends on the thermal state---and so, in generalizations to systems with more symmetries, on charges. Hence we focus on the KMS relation.

\section{Review of non-Abelian ETH}
\label{sec_Review_NAETH}

The rest of the main text features the following setup. Consider a system (which is closed here and in Sections~\ref{sec_FDT_Analytics}--\ref{sec_Num} but not in Sec.~\ref{sec_FDT_Thermo}) of $N \gg 1$ qubits. The Hamiltonian $H$ has eigenvalues $E_\alpha$, an SU(2) symmetry,\footnote{
The results in~\cite{MurthyNAETH} are expected to extend to other non-Abelian symmetries.}
and no other symmetry [apart from the U(1) symmetry equivalent to energy conservation]. Let $S_{a=x,y,z}$ denote the global spin's $a^\th$ component; $s_\alpha$, the total spin quantum number; and $m$, the total magnetic spin quantum number. $H$ shares an eigenbasis $\{ \ket{\alpha, m} \}$ with $\vec{S}^2$ and $S_z$:
$H \ket{\alpha, m}  =  E_\alpha \ket{\alpha, m}$,
$\vec{S}^2  \ket{\alpha, m}  =  s_\alpha (s_\alpha + 1) \ket{\alpha, m}$, and 
$S_z  \ket{\alpha, m}  =  m  \ket{\alpha, m}$.

The non-Abelian ETH governs $H$ and a local operator. Such operators form a space spanned by a basis of \emph{spherical tensor operators} $T^\KParen_q$~\cite{shankar2008principles}. $T^\KParen_q$ transforms irreducibly under SU(2) elements, so we analyze $T^\KParen_q$s for convenience, without loss of generality. The rank $k$ resembles a spin quantum number; and $q$, a magnetic spin quantum number. For example, some $T^\1_1$ operators are proportional to single-site raising operators; some $T^\1_{-1}$s, to lowering operators; and some $T^\1_0$s, to single-site spins' $z$-components.

The spherical tensor operators obey the \emph{Wigner--Eckart theorem}, a fundamental result in AMO physics~\cite{shankar2008principles}: consider representing $T^\KParen_q$ as a matrix relative to $\{ \ket{\alpha, m} \}$. Each matrix element equals a product of two factors, according to the theorem:
\begin{align}
   \label{eq_WE}
   \bra{\alpha, m}  T^\KParen_q  \ket{\alpha' , m'}
   =  \braket{s_\alpha \, , m}{s_{\alpha'} \, , m' \, ; k, q} \,
   \langle \alpha || T^\KParen || \alpha' \rangle .
\end{align}
The \emph{Clebsch--Gordan} coefficient
$\braket{s_\alpha \, , m}{s_{\alpha'} \, , m' \, ; k, q}$ interrelates the tensor product 
$\ket{ s_{\alpha'} \, , m' \, ; k, q}  \coloneqq  \ket{ s_{\alpha'} \, , m' } \ket{k, q}$ and the eigenstate $\ket{s_\alpha \, , m}$ of total-spin operators.
The \emph{reduced matrix element} $\langle \alpha || T^\KParen || \alpha' \rangle$ depends on no magnetic-type quantum numbers.

The non-Abelian ETH posits a form for the reduced matrix elements~\cite{MurthyNAETH}. 
Let $\omega_{\alpha' \alpha} \coloneqq E_{\alpha'} - E_\alpha$ denote a difference between the energies associated with the matrix column and row. Define the difference
$\nu_{\alpha' \alpha} \coloneqq s_{\alpha'} - s_\alpha$ analogously.\footnote{
$\omega_{\alpha' \alpha}$ differs from the $\omega$ in~\cite{MurthyNAETH} by a minus sign, and an analogous statement concerns $\nu_{\alpha' \alpha}$. Appendix~\ref{app_Corr} motivates our definitions, below Eq.~\eqref{eq_S_w_dis_help1}.}
$S_\th (E, s)$ denotes the thermodynamic entropy at $E$, $s$, and any fixed $m$ value~\cite{24_Lasek_Numerical}. If $D_\tot (E, s)$ denotes the corresponding density of states, $S_\th (E, s) = \log \LParen D_\tot (E, s) \RParen$.
This paper's logarithms are base-$e$.
$\mathcal{T}^\KParen (\mathcal{E}, \mathcal{S})$ and $f^{(T)}_\nu (\mathcal{E}, \mathcal{S}, \omega)$ denote smooth, real functions; and $R^{(T)}$ denotes a matrix of erratically varying numbers~\cite{Foini_19_ETH, Pappalardi_22_FreeETH, Wang_22_ETH,24_Lasek_Numerical}. According to the non-Abelian ETH~\cite{MurthyNAETH}, 
\begin{align}
   \label{eq_NAETH} &
   \langle \alpha || T^\KParen || \alpha' \rangle
   = \mathcal{T}^\KParen 
   \left( \frac{E_\alpha + E_{\alpha'} }{2} \, , 
           \frac{s_\alpha + s_{\alpha'} }{2}  \right) \,
   \delta_{\alpha, \alpha'}
   \\ \nonumber & \qquad \qquad \qquad \;
   + e^{- S_\th \big( \frac{E_\alpha + E_{\alpha'} }{2} \, ,  
                         \frac{s_\alpha + s_{\alpha'} }{2} \big) / 2 }  \,
   \\ \nonumber & \qquad \qquad \times
   f_{ \nu_{\alpha' \alpha} }^{(T)} \left( \frac{E_\alpha + E_{\alpha'} }{2} \, , 
           \frac{s_\alpha + s_{\alpha'} }{2} \, , \omega_{\alpha' \alpha}  \right) 
   R^{(T)}_{\alpha \alpha'} \, .
\end{align}
The first, \emph{diagonal} term is nonzero only on the matrix's block-diagonal. This term determines the time-averaged expectation value of $T^\KParen_q$~\cite{MurthyNAETH}. The second, \emph{off-diagonal} term controls the operator's time dependence. This term influences our KMS relation.

\section{Fine-grained KMS relation for SU(2)-symmetric system in a mixed state}
\label{sec_FDT_Thermo}

This section sketches arguments detailed in Apps.~\ref{sec_Thermo_props}--\ref{app_Thermo_KMS}. We review a thermal state suited to SU(2) symmetry, then introduce a variation on that state. Afterward, we introduce correlators and prove a fine-grained KMS relation for them.

Consider a quantum many-body system whose Hamiltonian $H$ conserves the total-spin components $S_{a = x,y,z}$ and no other nontrivial operators. The appropriate whole-system thermal state has been argued to be the \emph{non-Abelian thermal state} (NATS)~\cite{Jaynes_57_Info_II,guryanova_2016_thermodynamics,Lostaglio_2017_MaxEnt,Halpern_2016_microcanonical,NYH_20_Noncommuting,Kranzl_23_Experimental}:
$\tilde{\rho}_\NATS
\coloneqq  \exp ( - \beta [ H - \sum_{a=x,y,z} \mu_a S_a ] ) 
/ \tilde{Z}_\NATS
=  \exp ( - \beta [ H - \mu S_z ] ) / \tilde{Z}_\NATS$.
The $\mu_a$s serve as effective chemical potentials, and
$\tilde{Z}_\NATS$ normalizes the state. The final equality follows from rotating the $z$-axis onto $\mu_x \hat{x} + \mu_y \hat{y} + \mu_z \hat{z}$. Each exponential contains only extensive (additive) observables, consistently with a conventional derivation of the thermal state's form~\cite{landau_1980_statistical,Halpern_2016_microcanonical}. The observable
\begin{align}
   S  \coloneqq  \sum_{\alpha, m}  s_\alpha
   \ketbra{\alpha, m}{\alpha, m} 
\end{align}
is not extensive. Yet one might know the value of $\expval{S}$. Information-theoretic arguments~\cite{Jaynes_57_Info_II} therefore suggest a \emph{modified NATS},
\begin{align}
   \label{eq_Mod_NATS}
   \rho_\NATS
   \coloneqq  \exp \left( - \beta \left[ H - \mu S_z - \gamma S \right] \right)
   / Z_\NATS \, .
\end{align}
$\gamma$ serves as an effective chemical potential; and $Z_\NATS$, as the partition function. We posit that $\ket{\alpha, m}$ most closely locally resembles a $\rho_\NATS$ whose 
\begin{align}
   \label{eq_Correspond}
   \expval{H} = E_\alpha , \quad
   \expval{S_z} = m , \quad \text{and} \quad
   \expval{S} = s_\alpha .
\end{align} 
One might debate about whether $\rho_\NATS$ deserves the label \emph{thermal state}, however. We define $\expval{ . }  \coloneqq  \Tr ( . \rho_\NATS )$.

Let us derive a KMS relation satisfied by $\rho_\NATS$. For convenience, we focus on operators $A \coloneqq A^{(k')}_{-q}$ and $B \coloneqq B^\KParen_q$. Their $k, k' = O(1)$, such that the operators are local; in each operator, each term acts nontrivially on just an $O(1)$ number of qubits. $A$ and $B$ participate in the time-domain correlator
\begin{align}
   C_{AB}^\NATS (t)
   \coloneqq  \expval{ A(t) B } . 
\end{align}
Fourier-transforming yields the frequency-domain correlator
\begin{align}
   \label{eq_Corr_freq}
   & \bar{C}_{AB}^\NATS (\Omega)
   \coloneqq \int_{-\infty}^\infty  dt  \, C_{AB} (t)  \,
   e^{i \Omega t} \\
   & =  \frac{2 \pi}{Z}  \sum_{\alpha, m, \alpha', m'}
   e^{- \beta \left( E_\alpha - \mu m - \gamma s_\alpha \right) }
   \\ \nonumber & \times
   \bra{\alpha, m} A \ketbra{\alpha', m'}{\alpha', m'}  B  \ket{\alpha, m}  \,  
   \delta \left( \Omega - \left[ E_{\alpha'} - E_\alpha \right] \right) .
\end{align}

An obstacle hinders the proof of a KMS relation for $\bar{C}_{AB}^\NATS (\Omega)$:
$B$ and $e^{\beta \gamma S}$ appear not to participate in any simple commutation relation. Hence we define a \emph{fine-grained correlator}. It follows from introducing Kronecker delta functions for $m$ and $s_\alpha$. These functions are discrete-variable analogs of the Dirac delta function for energy, which is continuous in the thermodynamic limit:
\begin{align}
   & \hat{\bar{C}}^\NATS_{AB} (\Omega, \Delta m, \Delta s)
   \coloneqq  \frac{2 \pi}{Z}  \sum_{\alpha, m, \alpha', m'}
   e^{- \beta \left( E_\alpha - \mu m - \gamma s_\alpha \right) }
   \nonumber \\ \nonumber & \times
   \bra{\alpha, m} A \ketbra{\alpha', m'}{\alpha', m'}  B  \ket{\alpha, m}  \,  
   \delta \left( \Omega - \left[ E_{\alpha'} - E_\alpha \right] \right)
   \nonumber \\ & \times
   \delta_{m' (m + \Delta m)}  \,
   \delta_{ s_{\alpha'}  (s_\alpha + \Delta s) } \, .
\end{align}
Summing over $\Delta m$ and $\Delta s$ yields the conventional correlator~\eqref{eq_Corr_freq}:
\begin{align}
   \label{eq_Corr_Sum}
   \bar{C}_{AB}^\NATS (\Omega)
   = \sum_{\Delta m, \Delta s}
   \hat{\bar{C}}^\NATS_{AB} (\Omega, \Delta m, \Delta s) .
\end{align}
The fine-grained correlator obeys the \emph{fine-grained KMS relation}, one of our main results:
\begin{align}
   \hat{\bar{C}}^\NATS_{AB} (\Omega, \Delta m, \Delta s)
   & = e^{\beta (\Omega - \mu \Delta m - \gamma \Delta s) }  
   \\ \nonumber & \quad \; \times
   \hat{\bar{C}}^\NATS_{BA} (-\Omega, -\Delta m, -\Delta s) .
\end{align}

%
%
%
\section{Fine-grained KMS relation for energy eigenstates of SU(2)-symmetric systems}
\label{sec_FDT_Analytics}

Let us sketch the derivation, detailed in Apps.~\ref{app_Corr}--\ref{app_KMS_Anom}, of the fine-grained KMS relation for energy eigenstates of SU(2)-symmetric systems. Consider the setup in Sec.~\ref{sec_Review_NAETH}, as well as the operators $A$ and $B$ in Sec.~\ref{sec_FDT_Thermo}. The operators participate in the time-domain correlator
\begin{align}
   C_{AB}(t)
   & \coloneqq  \bra{\alpha, m} A(t) B \ket{\alpha, m} .
\end{align}
Fourier-transforming yields the frequency-domain correlator
\begin{align}
   \bar{C}_{AB} (\Omega)
   & \coloneqq \int_{-\infty}^\infty  dt  \,
   C_{AB} (t)  \,
   e^{i \Omega t} \, .
\end{align}
$\bar{C}_{AB}(\Omega)$ can contain a nonzero \emph{static} contribution, which multiplies a $\delta(\Omega)$, under certain conditions~\cite{Static_corr}. 
The static correlator can signal long-term memory (Sec.~\ref{sec_Outlook} and~\cite{Static_corr}). We focus on the dynamical contribution, which does not contain or multiply any $\delta (\Omega)$, to $\bar{C}_{AB}(\Omega)$. Like in Sec.~\ref{sec_FDT_Thermo}, we define a \emph{fine-grained correlator},
\begin{align}
    \label{eq_Corr_help1}
    & \hat{\bar{C}}_{AB}^\dyn (\Omega, \Delta m, \Delta s; \alpha, m)
    \coloneqq 2 \pi \sum_{\alpha' \neq \alpha}
    \bra{\alpha, m} A \ket{\alpha', m+ q}
    \\ \nonumber & \times
    \bra{\alpha', m + q} B \ket{\alpha, m}  \,
    \delta ( \Omega - [E_{\alpha'} - E_\alpha ] )  \,
    \delta_{(\Delta m) q}  \,
    \delta_{s_{\alpha'}  (s_\alpha + \Delta s) } \, .
\end{align}
Often, we elide the function's final two arguments for conciseness. We have applied a selection rule encoded in the Wigner–Eckart theorem~\eqref{eq_WE}. 
$\hat{\bar{C}}_{AB}^\dyn$ obeys a sum rule analogous to Eq.~\eqref{eq_Corr_Sum}.

Next, we evaluate and simplify Eq.~\eqref{eq_Corr_help1}. We apply the Wigner–Eckart theorem~\eqref{eq_WE} and the non-Abelian ETH~\eqref{eq_NAETH}. To simplify notation, we define the Clebsch–Gordan product
\begin{align}
   \label{eq_CG_prod}
   \mathcal{C} ( \nu | s_{\alpha} \, , m , k , {k'}, {q} )
   & \coloneqq \braket{ s_{{\alpha}} \, , {m} }{ s_{{\alpha}} + {\nu} , {m} + {q} ; {k'} , - {q} }
   \nonumber \\ & \times
   \braket{ s_{{\alpha}} + {\nu} , {m} + {q} }{ s_{{\alpha}} \, , {m} ; {k} , {q} } .
\end{align}
Also, we define a composite function $\mathcal{G}_{AB}(E, s; \Omega, \Delta s)$ [Eq.~\eqref{eq_G_Def_app}] in terms of four factors from the non-Abelian ETH: two $f$ functions and two $R$ factors. The composite function obeys the symmetry relation
\begin{align}
   \label{eq_G_Symm_Main}
   \mathcal{G}_{AB} (E, s; \Omega, \Delta s)
   = \mathcal{G}_{BA} (E, s; -\Omega, -\Delta s).
\end{align}
The correlator~\eqref{eq_Corr_help1} becomes
\begin{align}
   \label{eq_Corr_help2}
   & \hat{\bar{C}}_{AB}^\dyn (\Omega, \Delta m, \Delta s)
   \approx 2 \pi    \,
   \mathcal{C} ( \Delta s | s_\alpha \, , m , k , k' , q ) 
   \nonumber \\ & \qquad \qquad \qquad \qquad \; \: \times
   \frac{ D_\tot \left( E_\alpha + \Omega , 
                               s_\alpha + \Delta s  \right) }{ 
              D_\tot \left( E_\alpha + \frac{\Omega}{2} \, , 
                                 s_\alpha + \frac{\Delta s}{2}  \right)  }
     \\ \nonumber & \qquad \qquad \; \times
     \mathcal{G}_{AB}  \left(  E_\alpha + \frac{\Omega}{2}  \, ,
     s_\alpha + \frac{\Delta s}{2} \, ;  
     \Omega  \, ,  \Delta s  \right)
   \delta_{(\Delta m) \, q} \, .
\end{align}
The correction is exponentially small in $N$. The densities of states, $D_\tot$, come from the $\sum_{\alpha' \neq \alpha}$ in Eq.~\eqref{eq_Corr_help1}.

We can approximate the correlator as follows. We assume that $s_\alpha, E_\alpha \gg 1$, inspired by ETHs' tendency to hold when additive charges lie far from their extreme values~\cite{DAlessio_16_From}. Since $k, k' = O(1)$, and by the rules of addition of angular momentum, $\Delta s = O(1)$. Also, the correlator has a significant magnitude only when the non-Abelian ETH's $f$ function does, as happens only when $\Omega = O(1)$~\cite{24_Lasek_Numerical,DAlessio_16_From,Srednicki_99_Approach,Khatami_13_Fluctuation}. Hence we can Taylor-approximate $\mathcal{G}_{AB}$ about $(E_\alpha, s_\alpha ; \Omega, \Delta s)$. 

We can also Taylor-approximate the ratio of $D_\tot$s. Denote by $\tilde{S}_\th (E, s)$ the thermodynamic entropy at energy $E$ and spin quantum number $s$. By the definitions above Eq.~\eqref{eq_NAETH}, 
$D_\tot(E, s) = \exp \LParen \tilde{S}_\th (E, s) \RParen / (2s + 1)$.
We can calculate properties of $\tilde{S}_\th (E, s)$ by applying standard statistical-physics techniques to $Z_\NATS$ [Eq.~\eqref{eq_Mod_NATS} and App.~\ref{sec_Thermo_props_2}]. To apply those properties here, we invoke the correspondence between $\rho_\NATS$ and $\ket{\alpha, s_\alpha}$ [see the text around Eq.~\eqref{eq_Correspond}]. For example,
$\partial_E \tilde{S}_\th (E, s_\alpha) |_{E_\alpha}  \approx  \beta$.
Equation~\eqref{eq_Corr_help2} approximates to
\begin{align}
   \label{eq_Corr_help3} &
   \hat{\bar{C}}_{AB}^\dyn (\Omega, \Delta m, \Delta s)
   = 2 \pi   \,
   \mathcal{C} ( \Delta s | s_\alpha \, , m , k , k' , q ) 
   \nonumber \\ & \times
   \exp \left( \frac{1}{2} \left[ \beta \Omega 
   + \partial_s \tilde{S}_\th (E, s) |_{E_\alpha, s_\alpha}  \Delta s 
   \right] \right)
   \nonumber \\ & \times
   \mathcal{G}_{AB} \left( E_\alpha \, , s_\alpha \, ; \Omega , \Delta s \right) \,
   \delta_{(\Delta m) \, q}  \,
   \left[ 1 + (\text{correction})  \right] .
\end{align}

To progress further, we assume that $s_\alpha = O (N^\zeta)$, wherein $\zeta \in (0, 1]$. We prove the fine-grained KMS relation under each of two conditions on $m$ in App.~\ref{app_KMS_0th}. Here, we sketch the simpler case. Suppose that the magnetization vanishes: $m = 0$. The derivative in Eq.~\eqref{eq_Corr_help3} approximates to $- \beta \gamma$: 
\begin{align}
   & \hat{\bar{C}}_{AB}^\dyn (\Omega, \Delta m , \Delta s)
   = e^{\frac{\beta}{2} ( \Omega - \gamma \Delta s) }  \,
   2 \pi   \,
   \delta_{(\Delta m) q}  
   \\ \nonumber & \times
   \mathcal{C} ( \Delta s | s_\alpha, 0, k, k' , q )  \,
   \mathcal{G}_{AB} (E_\alpha, s_\alpha; \Omega, \Delta s )  \,
   [1 + (\text{correction}) ] .
\end{align}
We must interrelate this correlator with
$\hat{\bar{C}}_{BA}^\dyn (- \Omega, - \Delta m , - \Delta s)$.
To do so, we apply the symmetry~\eqref{eq_G_Symm_Main}.
Similarly, the Clebsch–Gordan product obeys the symmetry
$\mathcal{C}(\Delta s | s_\alpha, 0 , k, k', q)
= \mathcal{C} (-\Delta s | s_\alpha, 0, k', k, -q) + O (s_\alpha^{-1})$~\cite{24_Lasek_Numerical}.
(In the second case mentioned above, we apply a Clebsch–Gordan property proved in the present paper's App.~\ref{app_CG_property}.) Hence
\begin{align}
   \label{eq_KMS_eigen}
   & \hat{\bar{C}}_{AB}^\dyn (\Omega, \Delta m , \Delta s)
   = e^{\beta (\Omega - \mu \, \Delta m - \gamma \, \Delta s) }  
   \\ \nonumber & \quad \times
   \hat{\bar{C}}_{BA}^\dyn (- \Omega, - \Delta m , - \Delta s)  \,
   [1 + (\text{correction})] .
\end{align}

The correction depends on $s_\alpha^{-1}$, $E_\alpha^{-1}$, and $N$. Its precise size does not matter in App.~\ref{app_KMS_0th}; only its vanishing in the thermodynamic limit does. We expect the correction often to behave approximately as usual (as in the absence of non-Abelian symmetry), as $O(N^{-1})$~\cite{Noh_20_Numerical}.

However, we calculate the finite-size correction at $q=\beta=0$ in App.~\ref{app_KMS_Anom}. We argue that the correction scales as follows: 
(i) When $s_\alpha = O(N)$, the correction is $O(N^{-1})$.
(ii) When $s_\alpha = O(N^{\zeta \in (0, 1)} )$, the correction is 
$O(N^{- \min \{ \zeta, 1 - \zeta \} } )  >  O(N^{-1})$. Furthermore, we argue that $s_\alpha = O(N^{1/2})$ is possible. The argument relies on an exact formula for $\tilde{S}_\th$, the similarity between $\ket{\alpha, m}$ and $\rho_\NATS$, and the equivalence of thermodynamic ensembles. In fact, $s_\alpha = O(N^{1/2})$ is typical, due to subspaces' scalings. In conclusion, we argue that SU(2) symmetry can render the (fine-grained) KMS relation's finite-size correction as small as usual under certain conditions and polynomially larger under others.

\section{Numerics}
\label{sec_Num}

This section numerically supports the fine-grained KMS relation~\eqref{eq_KMS_eigen}. We simulated a Heisenberg XXX chain subject to periodic boundary conditions. The number of qubits, $N$, ranged from 16 to 24. Denote by $\sigma^\JParen_{a=x,y,z} = \sigma^{(N+j)}_a$ the Pauli-$a$ operator of qubit $j$. The Hamiltonian has the form
\begin{align}
    H = -\frac{J}{2} \sum_{j=1}^N \left[ \lambda \vec{\sigma}^\JParen \cdot
        \vec{\sigma}^{(j+1)} + (1-\lambda) \vec\sigma^\JParen \cdot
    \vec\sigma^{(j+2)} \right] .
    \label{eq:Hxxx}
\end{align}
The overall coupling strength $J = 1$. $\lambda$ interrelates the nearest- and next-nearest-neighbor couplings' strengths. We set $\lambda=0.25$, such that the Hamiltonian is nonintegrable~\cite{24_Lasek_Numerical,25_Patil_Eigenstate}. The model has translational invariance, which we leverage by working in the maximally symmetric subspace defined in~\cite{JaeDongXXZ}. We numerically diagonalized $H$, calculating the forms of the eigenstates $\ket{\alpha, m}$. 

We evaluated the fine-grained correlator~\eqref{eq_Corr_help1} on $|\alpha, m\rangle$s. For convenience, we set $A = B = T^\KParen_q$. To introduce the operator, we denote by $\sigma_\pm^\JParen$ the qubit-$j$ raising and lowering operators; and, by $\sigma_\pm^\tot  \coloneqq  \sum_{j=1}^N  \sigma_\pm^\JParen$, the whole-system versions.\footnote{
Identity operators are tensored on implicitly such that each term operates on the total Hilbert space.}
We analyzed the operators~\cite[App.~D]{24_Lasek_Numerical}
\begin{align}
   \label{eq_T00_T20}
   & T^\0_0
   = \frac{-1}{12 N}
   \sum_{j = 1}^{ N }  
   \left[  2  \left( \sigma_+^\JParen  \sigma_-^{(j+1)}  +  \hc  \right)
            +  \sigma_z^\JParen  \sigma_z^{(j+1)}  \right] ,
   \nonumber \\ &
   T^\2_0
   =  \frac{1}{ \sqrt{24} \, N}  
   \sum_{j=1}^N  \left[  \sigma_z^\JParen  \sigma_z^{(j+1)}
   -  \left(  \sigma_+^\JParen  \sigma_-^{(j+1)}
               +  \hc  \right)  \right] ,
\end{align}
and $T^{(4)}_0$. The latter operator appears to lack a clean formula but follows from lowering $T^{(4)}_4$ four times: 
$T^{(4)}_0 
= [ \sigma_-^\tot, [ \sigma_-^\tot, [ \sigma_-^\tot, [ \sigma_-^\tot, T^{(4)}_4 ]]]]$, wherein
\begin{align}
   T^{(4)}_4 = \frac{1}{N}  \sum_{j=1}^N
   \sigma_+^\JParen  \sigma_+^{(j+1)}  \sigma_+^{(j+2)}  \sigma_+^{(j+3)} \, .
\end{align}

A logarithmic ratio facilitates the testing of the fine-grained KMS relation (App.~\ref{app_Sym_for_num}). Under the conditions above, the log-ratio simplifies to
\begin{align}
    \label{eq:LogRatioFunc0}
    & \mathcal{L}_{k,q}(\Omega, \Delta s; \alpha, m)
    \\ \nonumber &
    \coloneqq \ln \left(
    \frac{ \hat{\bar{C}}^{\rm dyn}_{T^{(k)}_{-q} T^{(k)}_q}
    (\Omega, \Delta m{=}q, \Delta s; \alpha, m)}
    { \hat{\bar{C}}^{\rm dyn}_{T^{(k)}_q T^{(k)}_{-q}}
    (-\Omega, \Delta m{=}{-q}, -\Delta s; \alpha,m)}  \right) .
\end{align}
The log-ratio, evaluated at arbitrary $q$ and $m$ values, is related simply to an $\mathcal{L}$ evaluated at $(q, m) = (0, 0)$ [Eq.~\eqref{eq:Ltransform}].
According to the fine-grained KMS relation, 
$\mathcal{L}_{k,q} = \beta(\Omega- \mu q - \gamma \, \Delta s)$,
to within finite-size corrections. 
The parameters ($\beta$, $\beta\mu$, and $\beta\gamma$) are determined by the resemblance between $\rho_{\rm NATS}$ and $|\alpha,m\rangle$. Our numerics focus on (i) eigenstates with $m=0$, such that $\beta\mu=0$ and (ii) the spherical-tensor index $q=0$. Equation~\eqref{eq:Ltransform} implies the extension to general $m$ and $q$.

The eigenstate $|\alpha,m\rangle$ should resemble $\rho_{\rm NATS}$ under the conditions~\eqref{eq_Correspond}. Consider fixing those equations' left-hand sides. No finite-$N$ system's $\ket{\alpha, m}$ necessarily satisfies the equations, because $s_\alpha$ and $m$ are discrete. To mitigate this numerical challenge, we fix $\beta$. Simultaneously, we set the $S$ quantum number to $s$ and the $S_z$ eigenvalue to $0$. Then, we identify the eigenstates $\ket{\alpha, 0}$ whose $s_\alpha = s$. We select the eigenstate whose $E_\alpha$ lies closest to the Hamiltonian's thermal average
with respect to the Boltzmann distribution (over the $s_\alpha$ subspace) at $\beta$. Having introduced our techniques, we test the fine-grained KMS relation at $\Delta s = 0$ then at $\Delta s \neq 0$.

Suppose that $\Delta s=0$. The fine-grained KMS relation assumes the form
$\mathcal{L}_{k,0}(\Omega, \Delta s{=}0 ; \alpha, 0) = \beta\Omega$
in the thermodynamic limit. Finite-size corrections stem from two sources. First, random-matrix theory models the matrix elements~\eqref{eq_WE}~\cite{MurthyNAETH,24_Lasek_Numerical,25_Patil_Eigenstate}.
Therefore, $\mathcal{L}_{k,0}(\Omega, 0; \alpha,0)$ fluctuates from eigenstate to eigenstate. By the non-Abelian ETH, the fluctuation's amplitude decays exponentially as $N$ grows. To suppress the fluctuation, we assemble the set
$\{\mathcal{L}_{k,0}(\Omega, 0;\alpha,0) \, : \,
s_\alpha{=}s, \,
|   E_\alpha-E(\beta)|\leq \Delta E/2\}$ 
of log-ratios from nearby energy eigenstates (associated with energies within a window of size $\Delta E = 0.4$). 
We average $\mathcal{L}$ over the eigenstates, producing 
$\overline{\mathcal{L}}_{k,0}(\Omega,0;s)$. Then, we calculate the standard
deviation. Second, $\overline{\mathcal{L}}_{k,0}(\Omega,0;s)$ deviates from the predicted value due to the finite-size correction discussed in Sec.~\ref{sec_FDT_Analytics} and App.~\ref{sec_Derive_FDT_SU2}.

\begin{figure}
    \includegraphics[width=\columnwidth]{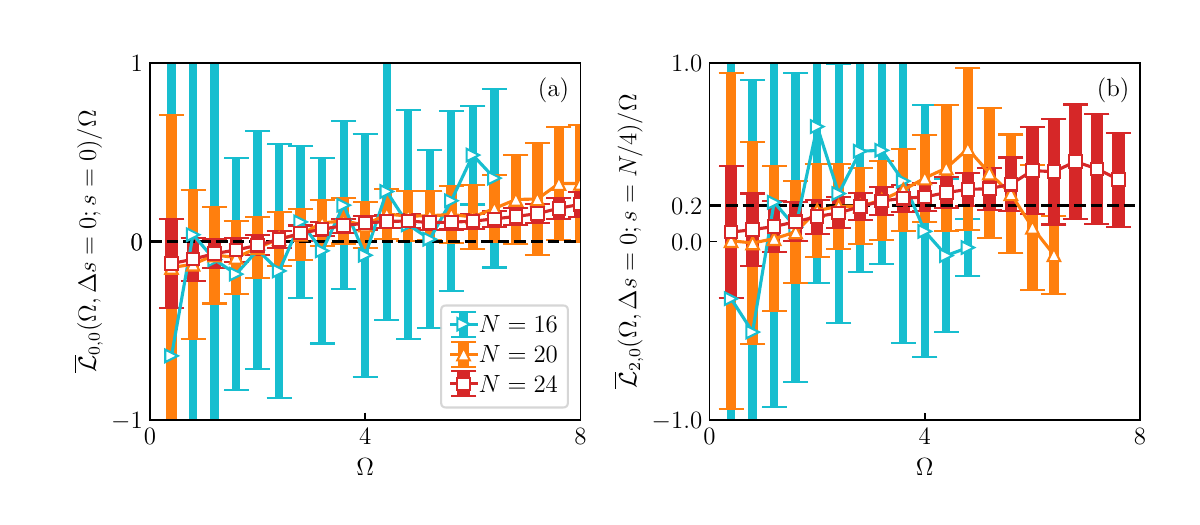}
    \caption{Averages and standard deviations of normalized logarithmic ratios:
    An error bar represents the standard deviation of log-ratios across eigenstates $\ket{\alpha, 0}$ associated with close-together energies.
    (a) Average log-ratio $\overline{\mathcal{L}}_{0,0}/\Omega$, normalized by $\Omega$, of $T^\0_0$ operator. The energy eigenstates used correspond to $s=0$ and $\beta=0$~(dashed black line). Different marker shapes and colors represent different system sizes ($N = 16, 20, 24$).
    (b) Average normalized log-ratio $\overline{\mathcal{L}}_{2,0}/\Omega$ of $T^\2_0$ operator at $s=N/4$ and $\beta=0.2$~(dashed black line). }
    \label{fig:LogRatioNu0}
\end{figure}

Figure~\ref{fig:LogRatioNu0} shows average logarithmic ratios $\overline{\mathcal{L}}$, normalized by $\Omega$, versus $\Omega$. The error bars in represent the eigenstate-to-eigenstate fluctuations (standard deviations). The fluctuations shrink exponentially
as $N$ grows, consistently with the non-Abelian ETH.\footnote{
In Fig.~\ref{fig:LogRatioNu0}(a), $s_\alpha = 0$, unlike in our derivation of the fine-grained KMS relation. However, the fine-grained correlator is simpler in Fig.~\ref{fig:LogRatioNu0}(a)'s context (when $s_\alpha = m = \Delta s = \Delta m = 0$) than when $s_\alpha > 1$: the relevant Clebsch–Gordan products equal each other. The fine-grained KMS relation contains a correction of $O(E_\alpha^{-1})$.
}

\begin{figure}
    \includegraphics[width=\columnwidth]{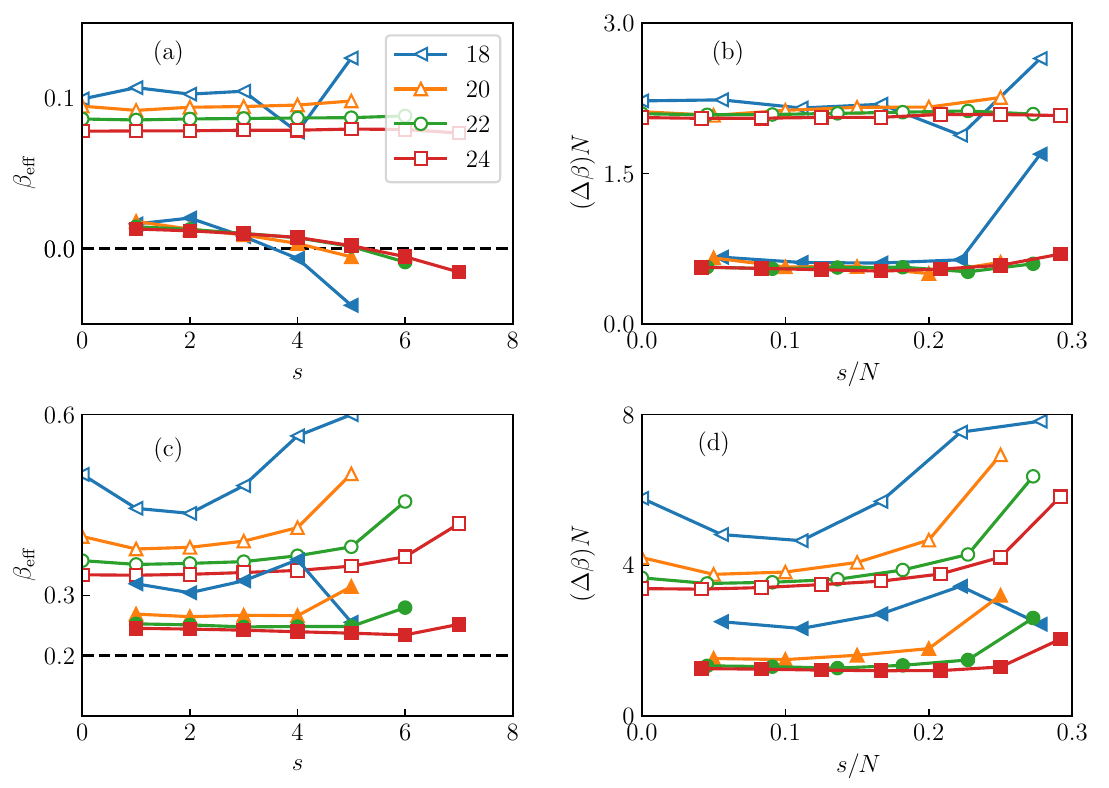}
    \caption{Quantification of finite-size effects in the average normalized logarithmic ratio at $\Delta s=0$: 
    Dashed black curves mark $\beta$, which equals 0.0 in the top row and $0.2$ in the bottom. Open markers were calculated from $T^{(0)}_0$; and filled markers, from $T^{(2)}_0$.
        (a) and (c) Effective inverse temperature $\beta_{\rm eff}$, versus spin quantum number $s$. $\beta_{\rm eff}$ follows from averaging $\overline{\mathcal{L}}(\Omega)/\Omega$ across $\Omega \in [2, 5]$.
        (b) and (d) Finite-size deviation $(\Delta \beta_{\rm eff}) N$ from the fine-grained KMS relation versus spin-quantum-number density $s/N$. $\Delta \beta_{\rm eff}$ equals the root-mean-square of
        $\overline{\mathcal{L}}/\Omega-\beta$ across $\Omega \in [2, 5]$.
        In all four plots, the blue curves jag sharply on the right-hand sides. The reason is that the system is small enough that few energy eigenstates correspond to such large $s$ values.
        }
    \label{fig:BetaEffNu0}
\end{figure}

Figure~\ref{fig:BetaEffNu0} quantifies the finite-size deviation of $\overline{\mathcal{L}}/\Omega$ from the expected value $\beta$. 
The top row corresponds to $\beta = 0.0$; and the bottom row, to $\beta = 0.2$. Define the \emph{effective inverse temperature} $\beta_{\rm eff}$ as the average of 
$\overline{\mathcal{L}}/\Omega$ across $\Omega \in [2, 5]$.
Furthermore, define the finite-size correction $\Delta \beta$ as the root-mean-square of 
$\overline{\mathcal{L}}/\Omega-\beta$ within the same interval. Figures~\ref{fig:BetaEffNu0}(a) and~\ref{fig:BetaEffNu0}(c) depict $\beta_{\rm eff}$. Despite deviating from the prediction $\beta$, $\beta_{\rm eff}$ approaches $\beta$ as $N$ increases, at most $s$ values.

Figure~\ref{fig:BetaEffNu0}(b) shows the root-mean-square deviation, $\Delta \beta$. It decreases as $1/N$ at all $s/N$ values. $T^\0_0$ and $T^{(2)}_0$ exhibit this behavior, but $T^{(2)}_0$ exhibits a weaker finite-size effect. The reason merits further investigation.

In addition to comparing operators, we compare $\beta = 0.0$ [Fig.~\ref{fig:BetaEffNu0}(b)] with $\beta = 0.2$ [Fig.~\ref{fig:BetaEffNu0}(d)]. $\Delta \beta$ adopts its asymptotic $1/N$ scaling at larger system sizes when $\beta = 0.2$. We speculate that the reason is, the density of states is smaller at $\beta = 0.2$ than at $\beta = 0.0$.

\begin{figure}
    \includegraphics[width=\columnwidth]{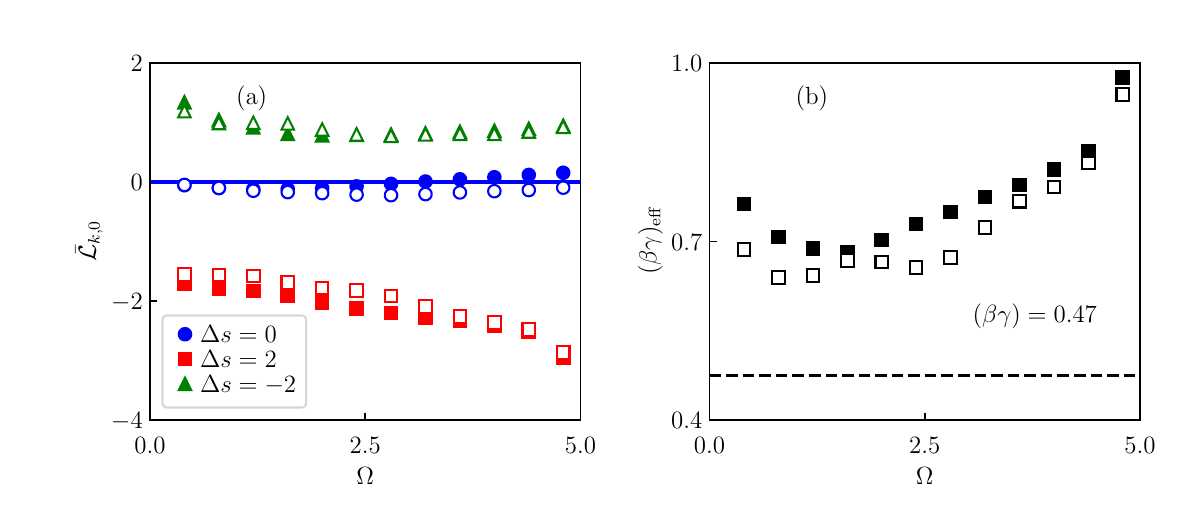}
    \caption{Comparisons of finite-size corrections at various $\Delta s$ values:
    Filled symbols describe the $k{=}2$ tensor; and empty symbols, the rank-4 tensor. The parameters $N = 24$, $s=4$, and $\beta=0.0$.
    (a) Average log-ratios 
        $\overline{\mathcal{L}}_{k,0}(\Omega, \Delta s; s)$. 
        Different marker shapes and colors correspond to different $\Delta s$ values.
        The blue, solid curve represents $\beta\Omega$, the ideal $\mathcal{L}$ at $\Delta s= 0$.
        (b) $(\beta\gamma)_{\rm eff}$ for the $k=2$~(filled symbols) and $k=4$~(empty symbols) operators. The black, dashed curve represents $\beta\gamma$. 
}
    \label{fig:LforNonzeroNu}
\end{figure}

Having tested the fine-grained KMS relation at $\Delta s = 0$, we test it at
$\Delta s \neq 0$. Figure~\ref{fig:LforNonzeroNu}(a) shows two
$T^{(k)}_0$ operators' average log-ratios,
$\overline{\mathcal{L}}_{k,0}(\Omega,\Delta s; s)$. The rank $k=2$ corresponds to filled markers; and $k = 4$, to empty markers. 
Different marker shapes and colors represent different values $\Delta s=0, \pm 2$.
All data come from the largest-system simulations, $N=24$, and $s=4$. 
According to the fine-grained KMS relation, 
\begin{align}
   \label{eq_L_Thermo_limit}
   \mathcal{L}_{k,0}(\Omega, \Delta s; \alpha, 0) 
   = \beta(\Omega - \gamma \, \Delta s)
\end{align}
in the thermodynamic limit. The solid blue curve represents this ideal at $\Delta s = 0$ in Fig.~\ref{fig:LforNonzeroNu}(a). This curve closely tracks the blue circles and disks. Hence the average correction $\overline{\mathcal{L}}_{k,0}(\Omega, \Delta s; s)$ approximates the ideal value well at $\Delta s = 0$.

Figure~\ref{fig:LforNonzeroNu}(b) sharpens the $\Delta s {\neq} 0$ analysis. By Eq.~\eqref{eq_L_Thermo_limit}, the $\mathcal{L}$ at 
$\Delta s=\pm 2$ should be vertically shifted, through a displacement 
$-(\Delta s) \beta\gamma$, from the $\mathcal{L}$ at $\Delta s =
0$. To test this prediction, we define the effective thermodynamic parameter
\begin{align}
   (\beta\gamma)_{\rm eff} 
   \coloneqq \left[ \overline{\mathcal{L}}_{k,0}(\Omega, -2;s) 
   -  \overline{\mathcal{L}}_{k,0}(\Omega, 2;s)\right ]/4 .
\end{align} 
It equals the $\beta \gamma$ associated with $\ket{\alpha, 0}$, plus a finite-size correction, according to the fine-grained KMS relation. Figure~\ref{fig:LforNonzeroNu}(b) shows $(\beta\gamma)_{\rm eff}$ (markers) and $\beta\gamma$ (dashed curve) versus $\Omega$.\footnote{
To approximate the $\beta\gamma$ associated with a finite-$N$ eigenstate $|\alpha,0\rangle$, we used a finite difference:
$\beta\gamma  
\approx - [S_{\rm tot}(E_\alpha,s+1)-S_{\rm tot}(E_\alpha,s-1) ]/2 
= \frac{\partial S_{\rm tot} (E_\alpha, \tilde{s}) }{\partial \tilde{s}} \big|_{ \tilde{s} = s} 
+ \frac{1}{6} \frac{\partial^3 S_{\rm tot} (E_\alpha, \tilde{s}) }{\partial \tilde{s}^3} \big|_{ \tilde{s} = s} + \ldots$ }
The markers deviate from the curve substantially. The finite-size correction is larger at $\Delta s \neq 0$ than at $\Delta s = 0$ [Fig.~\ref{fig:LforNonzeroNu}(a)].

\begin{figure}
    \includegraphics[width=\columnwidth]{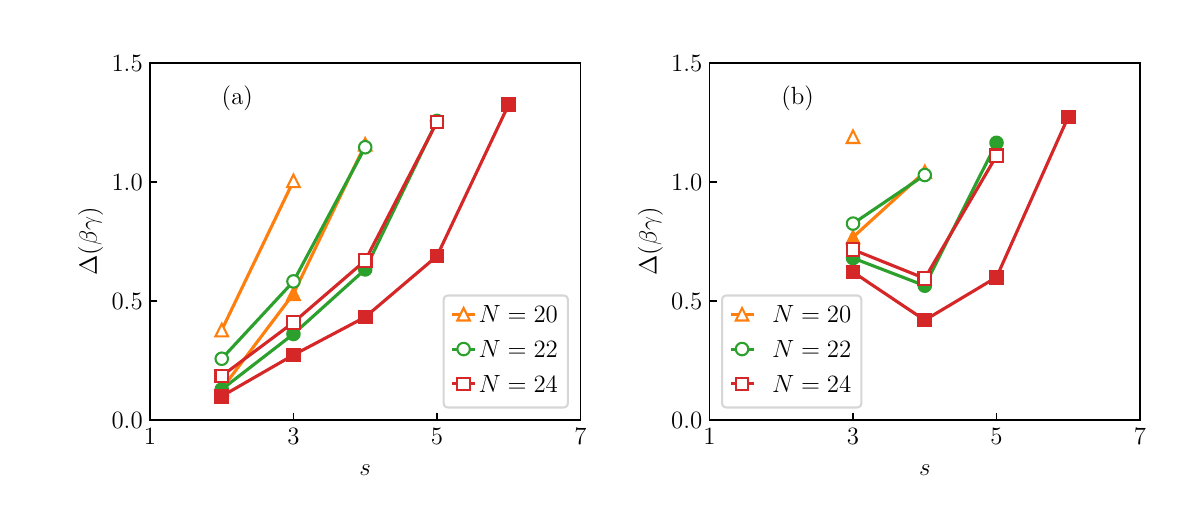}
    \caption{Finite-size correction to the fine-grained KMS relation, as measured by $\Delta(\beta\gamma)$: Figure~(a) describes the $k{=}2$ tensor; and Fig.~(b), the $k=4$ tensor. Empty markers correspond to $\beta = 0$; and filled markers, to $\beta = 0.2$. 
    }
    \label{fig:FSSnonzeroNu}
\end{figure}

We quantify the finite-size correction with $\Delta (\beta\gamma)$, the root-mean-square value of $(\beta\gamma)_{\rm eff} - (\beta\gamma)$, within the interval $\Omega \in [2,5]$. Figure~\ref{fig:FSSnonzeroNu} shows $\Delta(\beta\gamma)$ versus $s$. Different curves correspond to different system sizes $N$. Figure~\ref{fig:FSSnonzeroNu}(a) describes a $k{=}2$ operator; and Fig.~\ref{fig:FSSnonzeroNu}(b), a rank-4 operator. The $\Delta s {\neq} 0$ corrections are a few times larger than the $\Delta s {=} 0$ corrections displayed in Figs.~\ref{fig:BetaEffNu0}(a) and~\ref{fig:BetaEffNu0}(b). As $N$ grows, the correction $\Delta (\beta\gamma)$ decreases, as expected. We cannot perform a finite-size-scaling analysis on the correction, because computational limitations restrict $N$.

According to Fig.~\ref{fig:FSSnonzeroNu}(b), the rank-4 tensor's $\Delta (\beta\gamma)$ correction changes nonmonotonically as $s$ changes. The correction is smallest at a value $s = s_{\rm min}$. At each $s>s_{\rm min}$ value, $\Delta (\beta\gamma)$ decreases rapidly as $N$ increases. At each $s< s_{\rm min}$ value, $\Delta (\beta\gamma)$ decreases relatively slowly as $N$ increases.
To understand this phenomenon's origin, we revisit a formula derived for the log-ratio in the appendices [Eqs.~\eqref{eq_Logratio_app_help2} and~\eqref{eq_L_case2_app}]. First, we define two derivatives of the symmetric function $\mathcal{G}_{AB}$:
$\Gamma_s  
\coloneqq  \partial_{ \tilde{s} }  
\ln  \LParen \mathcal{G}_{AB} \left( E, \tilde{s}; \Omega, \Delta s \right)
   \RParen  |_{E_\alpha \, , \, s}$
and the analogous $\Gamma_E$.
In terms of them, the log-ratio is
$\mathcal{L}_{k,0}(\Omega, \Delta s;\alpha,0) 
= \beta (\Omega - \gamma \Delta s) 
   + (\Gamma_E \Omega + \Gamma_s  \,  \Delta s) + \ldots$
The final parenthesized term is the leading-order finite-size correction in the fine-grained KMS relation. This correction, at nonzero $\Delta s$, is
\begin{equation}
    \Delta (\beta\gamma) = -\Gamma_s \, .
    \label{eq:DeltaGammaCorrection}
\end{equation}
This logarithm is not analytically computable.

To bypass this obstacle, we consider the modified NATS at $\beta=\beta\mu=0$: $\rho_{\rm NATS} = e^{\beta\gamma S} / \tilde{Z}(\beta \gamma)$.
This partition function is exactly calculable. In terms of the scaling variable
$\tilde{\gamma} \coloneqq  \sqrt{\frac{N}{8}} \, \beta \gamma$~(App.~\ref{app_KMS_Anom}), 
\begin{align}
    & \tilde{Z}(\beta \gamma)
    = 2^{N+1}  \tilde{\mathcal{Z}} \left( \tilde{\gamma} \right) ,
    \quad \text{wherein} \\
    & \tilde{\mathcal{Z}} \left( \tilde{\gamma} \right) 
    = \frac{ \tilde{\gamma} }{\sqrt{\pi}} 
    +  \frac{1+2 \tilde{\gamma}^2}{2} \,
    \left[ 1+{\rm erf} \left( \tilde{\gamma} \right) \right] \, 
    e^{\tilde{\gamma}^2} \, .
\end{align}
The expectation value
$\langle S \rangle 
= \partial_{\beta \gamma} \ln ( \tilde{Z} ) 
= \sqrt{\frac{N}{8}}   \,
\frac{d\mathcal{Z}}{d\tilde{\gamma}}$  evaluates to
\begin{equation}
    \langle S \rangle 
    = \sqrt{N / 8}  \;  \mathcal{J} \left( \tilde{\gamma} \right) .
    \label{eq:Sscaling}
\end{equation}
The scaling function $\mathcal{J}(\tilde{\gamma}) 
\coloneqq d \tilde{\mathcal{Z}} / d\tilde{\gamma}$ behaves as 
\begin{equation}
    \mathcal{J} \left( \tilde{\gamma} \right) \simeq \begin{cases}
        2\tilde{\gamma},  &  \tilde{\gamma} \gg 1 \\
        O(1), &  |\tilde{\gamma}| \ll 1 \\
        - 3 / \tilde{\gamma} , &  \tilde{\gamma} \ll -1.
    \end{cases}
\end{equation}
If $|\beta\gamma| = O( N^{-1/2} )$, then $\langle S\rangle = O (N^{1/2}$); and, if
$\beta\gamma = O(1)$, then $\langle S\rangle \sim N$.
The scaling~\eqref{eq:Sscaling} stems from the Hilbert space's decomposition into subspaces labeled by $s_\alpha$. We therefore posit that thermodynamic quantities depend on $s$ through the scaling variable $s/\sqrt{N}$ when 
$s= O( N^{1/2} )$. Applying this ansatz to Eq.~\eqref{eq:DeltaGammaCorrection} implies that, (i) when $s = O( N^{1/2} )$, the finite-size correction is anomalous,
$\Delta (\beta\gamma) = O (N^{-1/2})$, and (ii) when $s = O(N)$, the correction is of $O(N^{-1})$ . The nonmonotonic behavior in Fig.~\ref{fig:FSSnonzeroNu}(b), we believe, reflects the anomalous finite-size correction to the fine-grained KMS relation at $s = O( N^{1/2} )$.

Appendix~\ref{app_Add_num} supports the fine-grained KMS relation~\eqref{eq_KMS_eigen} with further numerics. There, we simulate a Heisenberg model that (i) obeys closed boundary conditions and (ii) lacks symmetries other than SU(2) [and the U(1) symmetry equivalent to energy conservation]. The results echo those in this section.

\section{Outlook}
\label{sec_Outlook}

We have derived and numerically supported a fine-grained KMS relation obeyed by energy eigenstates of quantum many-body systems subject to non-Abelian symmetries. The proof relies on the recently introduced non-Abelian ETH. Under certain conditions, we have argued, the fine-grained KMS relation's finite-size correction scales as $O(N^{-1})$, as usual (as in the absence of any non-Abelian symmetry). Under other conditions, the correction may be polynomially larger. Hence charges' noncommutation may enable an anomalously large deviation from a result applied in nonequilibrium statistical physics. This result exemplifies quantum thermodynamics, given noncommutation's role in quantum phenomena such as uncertainty relations. Our result helps extend into nonequilibrium thermodynamics the rapidly growing program of identifying how charges' noncommutation influences dynamics~\cite{23_Li_SUd,Majidy_23_Critical,Marvian_24_Theory,24_Li_Designs,24_Varikuti_Unraveling,24_Dabholkar_Ergodic,24_Saraidaris_Finite,24_Zadnik_Quantum,24_Bai_Nucleon,25_Cataldi_Disorder}.

Our results differ from two related results. First, Manzano \emph{et al.} proved that charges' noncommutation lowers linear–response entropy production~\cite{Manzano_22_Non}. Yet their result concerns the Onsager relations, not the FDT. Their setup differs, featuring two thermal systems at different effective temperatures. Nor does many-body physics feature in their discovery, which involves no finite-size correction. 

Second, Murthy \emph{et al.} proved that, if a Hamiltonian $H$ and a $T^\KParen_q$ obey the non-Abelian ETH, the operator's time-averaged expectation value can differ from its thermal expectation value by an anomalously large correction~\cite{MurthyNAETH}. The result derives from the non-Abelian ETH’s first term. In contrast, our result derives from the second term, which encodes how $\expval{ T^\KParen_q }$ fluctuates over time. Also,~\cite{MurthyNAETH} contains no numerical evidence and invokes an assumption that we do not. 

Our work establishes six opportunities for future research. First, we argued analytically that the fine-grained KMS relation contains an anomalously large correction under certain conditions, including $s_\alpha = O( N^{1/2} )$.
Other parameter regimes might admit of anomalously large corrections, too. So might non-Abelian symmetry groups other than SU(2). 

Second, one might observe our KMS relation's anomalous correction by simulating larger systems. The simulations may be classical. However, third, experiments can test our KMS relation. The correlator equals an average over a Kirkwood–Dirac quasiprobability distribution, as all correlators do~\cite{18_NYH_Quasi}. Quasiprobabilities resemble probabilities but can violate certain axioms of probability theory. Kirkwood–Dirac quasiprobabilities can assume negative and nonreal values. Yet one can measure Kirkwood–Dirac quasiprobabilities in many ways~\cite{23_Lostaglio_KD,24_DRMAS_Properties,24_Wagner_Quantum}. Furthermore, noncommuting-charge thermodynamics was recently tested in a trapped-ion experiment~\cite{Kranzl_23_Experimental}. Other feasible platforms include superconducting qubits and ultracold atoms~\cite{NYH_20_Noncommuting,NYH_22_How}. 


Third, our work leverages the non-Abelian ETH’s off-diagonal term, expected to illuminate thermalization dynamics~\cite{MurthyNAETH}. By extending this work, one may elucidate dynamical questions---for example, does charges' noncommutation slow thermalization? Does it enlarge the fluctuations, undergone over time, by a local operator’s expectation value? 

Fourth, one can derive an FDT from our KMS condition. Manzano \emph{et al.} derived what might be called an FDT, Eq.~(13) of~\cite{Manzano_22_Non}, in the context of noncommuting charges. The equation's similarity to FDTs was not emphasized, however. Also, the derivation differed substantially from the conventional FDT derivation in App.~\ref{app_Review_FDT}. Starting from our KMS condition could illuminate more directly how noncommuting charges affect the conventional FDT.

Fifth, Appendix~\ref{sec_Derive_FDT_SU2} shows that SU(2) symmetry supports a nonzero static correlator across a relatively wide parameter regime. A static correlator never decays and so evidences memory. We will quantify this memory's capabilities in future work~\cite{Static_corr}.

Sixth, many realistic settings feature perturbations that break symmetries. Do such perturbations alter predictions of the non-Abelian ETH, such as the fine-grained KMS relation? The answer can help bridge the present work to experiments.

\section{Data Availability}
The data that supports the findings of this study are available from the corresponding author upon reasonable request.

\begin{acknowledgments}
The authors thank Sarang Gopalakrishnan, Alexey Gorshkov, and Joonhyun Yeo for helpful conversations.
This work received support from the National Science Foundation 
(QLCI grant OMA-2120757 
and NSF award DMS-2231533),  
the John Templeton Foundation (award no. 62422), and
a National Research Foundation of Korea~(NRF) grant funded by the Korea government (MSIT) (grant No. RS-2024-00348526).
The opinions expressed in this publication are those of the authors and do not necessarily reflect the views of the John Templeton Foundation or UMD.
\end{acknowledgments}

\begin{appendices}

\onecolumngrid

\renewcommand{\thesection}{\Alph{section}}
\renewcommand{\thesubsection}{\Alph{section} \arabic{subsection}}
\renewcommand{\thesubsubsection}{\Alph{section} \arabic{subsection} \roman{subsubsection}}

\makeatletter\@addtoreset{equation}{section}
\def\theequation{\thesection\arabic{equation}}

\section{Review of the conventional KMS relation and FDT for quantum systems}
\label{app_Review_FDT}

This appendix contains a pedagogical three-step review of the quantum KMS relation and FDT. In App.~\ref{app_Kubo}, we derive the Kubo formula for a quantum system's response function. In App.~\ref{app_Conventional_KMS}, we derive the conventional KMS relation. Combining these two ingredients, we prove the conventional FDT in App.~\ref{app_Conventional_FDT}. The overall derivation appears in many references; we follow Sections~3.3.2--3.3.6 of~\cite{23_Cugliandolo_Sorbonne}.

Before embarking upon the derivation, we review the reason for the name \emph{fluctuation--dissipation theorem.} Kubo explains eloquently in his review~\cite{66_Kubo_FDT}. For simplicity, consider a classical statistical mechanical system, such as an electrically charged particle in a fluid. In equilibrium, the particle undergoes Brownian motion: fluid particles bump into the charged particle, which random-walks around the fluid. That is, the charged particle's position fluctuates. Now, consider applying an electric field. It forces the charged particle along a particular direction. To move there, the charged particle knocks aside fluid particles, which slow it down. That is, the fluid particles exert a drag force, which dissipates energy from the charged particle. This dissipation shares its origin with the equilibrium fluctuations---collisions between the charged particle and fluid particles---so the two phenomena must be related. The FDT quantifies this relationship.

This appendix concerns a quantum system of the following type. The system has observables $A$ and $B$. From time 0 to time $t'$, and from time $t' + \epsilon$ to time $t$, the system evolves under an unperturbed Hamiltonian $H$. $\epsilon$ is infinitesimally small. During the infinitesimal interval, the system evolves under the perturbed Hamiltonian $H' \coloneqq H - h B$. Later, we will consider the zero-field limit ($h = 0$). At time 0, the system begins in a state $\rho$. In App.~\ref{app_Kubo}, $\rho$ is arbitrary; afterward, it is thermal, as described in App.~\ref{app_Conventional_FDT}. We set Planck's reduced constant to $\hbar = 1$.

\subsection{Derivation of Kubo formula}
\label{app_Kubo}

This section addresses the following question about the time-$t$ expectation value of $A$. Consider the expectation value's rate of change with respect to $h \epsilon$ in the zero-field limit. This rate quantifies the system's response to the perturbation $-hB$. What is that response?

To answer this question, we introduce the expectation value. First, we define the unitary that encodes the full time evolution:
\begin{align}
   \label{eq_U_B_1}
   U_B (t, 0)
   \coloneqq  e^{-i H (t - t' - \epsilon) }  \,
   e^{-i (H - hB) \epsilon}  \,
   e^{-i H t' }  \, .
\end{align}
This unitary evolves $A$ in the Heisenberg picture to
$A(t)  \coloneqq  U_B^\dag (t, 0)  \,  A  \,  U_B (t, 0)$.
The time-evolved operator has the expectation value
\begin{align}
   \label{eq_A_Expval_1}
   \expval{ A(t) }_B
   \coloneqq  \Tr \left(  U_B^\dag (t, 0 )  \, A  \,  U_B (t, 0)  \,  \rho  \right) .
\end{align}
In terms of this expectation value, we define the response function:
\begin{align}
   \label{eq_R_Def}
   R_{AB} (t, t')
   \coloneqq 
   \frac{ \partial  \expval{ A(t) }_B }{ \partial (h \epsilon) }
   \Bigg\lvert_{h=0}
   \Theta (t - t') .
\end{align}
The Heaviside function encodes causality: the system can respond to perturbations applied in the past, not the future.

Let us calculate the response function. By the product rule,
\begin{align}
   \label{eq_R_2}
   R_{AB} (t, t')
   = \left[  \Tr  \left( \frac{ \partial U_B^\dag (t, 0) }{\partial (h \epsilon)}  \,
   A  \,  U_B (t, 0)  \right)  \Bigg\lvert_{h=0}
   +  \Tr  \left(  U_B^\dag (t, 0)  \,  A  \,
       \frac{ \partial U_B (t, 0) }{\partial (h \epsilon)}  \right)
       \Bigg\lvert_{h=0}  \,
   \right]
   \Theta (t - t') .
\end{align}
We approximate the derivatives to first order in $h \epsilon$. To do so, we approximate the central factor on the right-hand side of Eq.~\eqref{eq_U_B_1}. This exponential of a sum would equal a product of exponentials, if $H$ commuted with $B$. The two operators might not commute, however. According to the Baker--Campbell--Hausdorff formula, 
\begin{align}
   e^{-i H \epsilon }  \,  e^{i h \epsilon B }
   =  e^{- i (H - h B) \epsilon
   + \frac{i h \epsilon^2}{2}  \,  [ H, B]
   + \frac{i h \epsilon^3}{12}  \,  [H, [H, B]] + \ldots}
\end{align}
The largest correction is of $O(h \epsilon^2 [H, B])$, which we neglect.
Next, we Taylor-approximate the $e^{i h \epsilon B} \, .$ The unitary formula~\eqref{eq_U_B_1} becomes
\begin{align}
   \label{eq_U_B_2}
   U_B (t, 0)
   = e^{- i H (t - t') }
   \left\{ \id + i h \epsilon B + O \left( [h \epsilon]^2 \right)  \right\}
   e^{-i H t' } \, .
\end{align}
This result implies the unitary's rate of change with respect to $h \epsilon$ in the zero-field limit:
\begin{align}
   \label{eq_U_Deriv}
   \frac{\partial}{\partial (h \epsilon)}  \,
   U_B (t, 0)
   \bigg\lvert_{h=0}
   = i \, e^{-i H (t - t') }  \,  B  \,  e^{-i H t' }  \, . 
\end{align}
Calculating the Hermitian conjugate, too, is worthwhile:
\begin{align}
   \label{eq_UDag_Deriv}
   \frac{\partial}{\partial (h \epsilon)}  \,
   U_B^\dag (t, 0)
   \bigg\lvert_{h=0}
   =  - i \, e^{i H t'}  \,  B  \,  e^{i H (t - t') }  \, .
\end{align}
Let us substitute into Eq.~\eqref{eq_R_2}:
\begin{align}
   R_{AB} (t, t') 
   & =  \left[ -i \Tr \left( e^{i H t'} \, B \, e^{i H (t - t') }  \,  A  \, e^{-iHt}  \,  \rho  \right)
   + i \Tr \left( e^{iHt}  \,  A  \,  e^{-iH(t - t') }  \,  B  \,  e^{-iH t'}  \,  \rho  \right)
   \right]  \,  \Theta (t - t') \\
   & = i \left[  \Tr \LParen  A(t) \, B(t' )  \,  \rho \RParen
   -  \Tr \LParen B(t')  \, A(t)  \,  \rho  \right)  \RParen  \,
   \Theta (t - t')  \\
   & = i  \Tr \left( \left[ A(t),  B(t')  \right]  \rho  \right)  \,
   \Theta (t - t')  \\
   \label{eq_Kubo_app}
   & \equiv  i \expval{ \left[ A(t), B(t') \right]  }_\rho  \,
   \Theta (t - t' )  \, .
\end{align}
We have defined $\expval{X}_\rho  \coloneqq  \Tr ( X \rho )$. The final equality gives the Kubo formula for the response function.

\subsection{Derivation of conventional KMS relation}
\label{app_Conventional_KMS}

The following considerations motivate the KMS relation's derivation. The Kubo formula~\eqref{eq_Kubo_app} expresses the time-domain response function $R_{AB} (t, t')$ in terms of a time-domain correlator $\expval{ \left[ A(t), B(t') \right]  }_\rho$. The FDT will express a frequency-domain response function in terms of a frequency-domain correlator. The latter correlator obeys a symmetry property used to derive the FDT. That symmetry property is the KMS relation.

First, we introduce an assumption and notation. From now on, we assume that $\rho$ is the canonical thermal state, 
$\rho_\th  \coloneqq e^{- \beta H} / Z$.
The partition function is defined as $Z \coloneqq \Tr ( e^{-\beta H} )$.
The two-time thermal correlator has the form
\begin{align}
   \label{eq_C_AB_Def_App}
   C_{AB} (t, t') 
   \coloneqq  \expval{ A(t)  \, B(t')  }_{\rho_\th}  \, .
\end{align}

Now, we prove a time-domain version of the KMS relation,
\begin{align}
   \label{eq_KMS_time}
   C_{AB}(t, t') = C_{BA} (t', t + i \beta ) .
\end{align}
We substitute into the left-hand side from the definition~\eqref{eq_C_AB_Def_App}. Then, we substitute in the forms of $A(t)$ and $\rho_\th$:
\begin{align}
   C_{AB}(t, t')
   =  \frac{1}{Z}  \,  \Tr  \left(
   e^{i H t}  \,  A  \,  e^{-iH t}  \,  B(t')  \,  e^{-\beta H}  \right) .
\end{align}
Let us cycle the $B(t') \, e^{-\beta H}$ to the trace's lefthand side. The thermal exponential combines with the $e^{iHt}$ into $e^{iH (t + i \beta)} \, .$ Then, we insert 
$\id = e^{-\beta H}  \,  e^{\beta H}$ rightward of $A$. The $e^{\beta H}$ combines with the $e^{-iHt}$. The $e^{-\beta H}$ commutes through them, to form a new thermal exponential on the right-hand side of the trace's argument:
\begin{align}
   C_{AB}(t, t')
   =  \Tr \left( B(t')  \,  e^{i H (t + i \beta) }  \,  A  \,
   e^{-i H (t + i \beta) }  \,  e^{-\beta H}  \right)
   =  \Tr  \LParen  B (t')  \,  A (t + i \beta )  \,  \rho_\th  \RParen
   =  C_{BA} ( t', t + i \beta ) .
\end{align}
We have proved Eq.~\eqref{eq_KMS_time}.

We now take advantage of a property of the thermal state. $\rho_\th$ is stationary, or invariant under time evolution. $C_{AB} (t, t')$ therefore is stationary, depending on $t$ and $t'$ only through the difference $t - t'$. Abusing notation, we redefine $C_{AB} (t, t')$ as $C_{AB} (t - t')$. The time-domain KMS relation becomes
$C_{AB} ( t - t' ) = C_{BA} (t' - t - i \beta )$. Let us relabel $t - t'$ as $t$:
\begin{align}
   \label{eq_KMS_Time_2}
   C_{AB} (t)
   = C_{BA} (- t - i \beta ) .
\end{align}

From this relation, we derive the frequency-domain KMS relation. We define the Fourier transform and inverse Fourier transform as in solid-state physics and in~\cite{DAlessio_16_From,Noh_20_Numerical}: 
\begin{align}
   & \bar{f} (k) 
   = \int_{-\infty}^\infty  dx  \,  f(x)  \,  e^{i k x } 
   \, ,  \quad \text{and}  \quad
   f(x)  =  \frac{1}{2 \pi} 
   \int_{-\infty}^\infty  dk  \,  \bar{f}(k)  \, e^{- i k x }  \, .
\end{align}
Fourier-transforming each side of Eq.~\eqref{eq_KMS_Time_2} yields
\begin{align}
   \bar{C}_{AB} (\Omega)
   = \int_{-\infty}^\infty  dt  \,  
   C_{BA} (- t - i \beta )  \,  e^{i \Omega t } \, .
\end{align}
We multiply the right-hand side by 
$1 = e^{\beta \Omega}  \,  e^{-\beta \Omega}
= e^{\beta \Omega}   \,  e^{i \omega (i \beta) } \, ,$
incorporating the final factor into the $e^{i \Omega t }$:
\begin{align}
   \bar{C}_{AB} (\Omega)
   & =  e^{\beta \Omega} 
   \int_{-\infty}^\infty  dt  \,  
   C_{BA} (- t - i \beta )  \,  e^{i \Omega (t + i \beta) } \\
   \label{eq_Conv_KMS_app}
   & =  e^{\beta \Omega }  \,
   \bar{C}_{BA}  ( - \Omega ) .
\end{align}
We have derived the (frequency-domain) KMS relation.

\subsection{Derivation of conventional FDT}
\label{app_Conventional_FDT}

We derive the FDT using the Kubo formula~\eqref{eq_Kubo_app} and the KMS relation~\eqref{eq_Conv_KMS_app}. First, we introduce notation for antisymmetrized and symmetrized thermal correlators:
\begin{align}
   \label{eq_Def_Sym_Corr}
   C_{[A, B]}  (t, t')
   \coloneqq  \frac{1}{2}  \,  \expval{  \left[  A(t),  B(t')  \right]  }_{\rho_\th} 
   \, ,  \quad  \text{and}  \quad
   C_{ \{A, B \} } (t, t' )
   \coloneqq  \frac{1}{2}  \,  \expval{  \left\{  A(t),  B(t')  \right\}  }_{\rho_\th} \, .
\end{align}
The notation $\{ X, Y \}  \equiv  XY + YX$ represents the anticommutator. Why have we introduced the correlators~\eqref{eq_Def_Sym_Corr}? The Kubo relation expresses the response function in terms of $C_{[A, B]}$. We aim to replace this correlator with $C_{ \{A, B \} }$ for the following reason. Consider reinstating $\hbar \neq 1$. In the classical limit (as $\beta \hbar \omega \to 0$), we should recover the classical FDT from the quantum FDT.

We interrelate the correlators as follows. Both are stationary; so, abusing notation again, we redefine
$C_{[A, B]} (t, t')  \equiv  C_{[A, B]} (t - t')$ and
$C_{ \{ A, B \} } (t, t')  \equiv  C_{ \{ A, B \} } (t - t')$.
We then Fourier-transform the equations~\eqref{eq_Def_Sym_Corr}, beginning with the commutator equation. Since
$C_{[A, B]} (t - t')
= \frac{1}{2} [ \expval{ A(t) \, B(t')  }_{ \rho_\th }
   -  \expval{ B(t')  \, A(t)  }_{\rho_\th}  ]$,
\begin{align}
   \label{eq_FT_Comm_corr}
   \bar{C}_{[A, B]} (\Omega)
   & = \frac{1}{2}  \left[
   \bar{C}_{AB} (\Omega)  -  \bar{C}_{BA} (- \Omega)  \right] 
   =  \frac{1}{2}  \left( 1  -  e^{-\beta \Omega }  \right)  \,
   \bar{C}_{AB} (\Omega) .
\end{align}
The final equality follows from the KMS relation, Eq.~\eqref{eq_Conv_KMS_app}.
Similarly, the second equation in~\eqref{eq_Def_Sym_Corr} Fourier-transforms to
\begin{align}
   \label{eq_FT_Anticomm_corr}
   \bar{C}_{ \{ A, B \} } (\Omega)
   & = \frac{1}{2}  \left( 1  +  e^{-\beta \Omega }  \right)  \,
   \bar{C}_{AB} (\Omega) .
\end{align}
Let us divide Eq.~\eqref{eq_FT_Comm_corr} by Eq.~\eqref{eq_FT_Anticomm_corr} and rearrange factors:
\begin{align}
   \label{eq_Comm_Anticomm}
   \bar{C}_{[A, B]} (\Omega)
   & = \tanh \left( \beta \Omega / 2 \right)  
   \bar{C}_{ \{ A, B \} } (\Omega) .
\end{align}

To combine this equality with the Kubo formula~\eqref{eq_Kubo_app}, we must Fourier-transform the latter. We do so in multiple steps. First, we represent the Kubo formula's correlator as an inverse Fourier transform. The dummy $\Omega'$ serves as the variable conjugate to $t - t'$. Then, we substitute in from Eq.~\eqref{eq_Comm_Anticomm}:
\begin{align}
   R_{AB} (t - t')
   & = i 2 C_{ [A, B] } (t - t')  \,  \Theta (t - t')
   =  2 i \Theta (t - t' )  \times
   \frac{1}{2 \pi}  \int_{-\infty}^\infty  d \Omega'  \,
   \bar{C}_{[A, B]}  ( \Omega' )  \,
   e^{-i \Omega' (t - t') }   \\
   & =  \frac{ i \Theta (t - t') }{ \pi }  
   \int_{-\infty}^\infty  d \Omega'  \,
   \tanh \left( \frac{\beta \Omega'}{2}  \right)
   \bar{C}_{ \{A, B \} }  (\Omega')  \,
   e^{-i \Omega' (t - t') }  \, .
\end{align}
We now Fourier-transform each side, using the dummy $\Omega$ as the variable conjugate to $t - t'$. The left-hand side Fourier-transforms to $\tilde{R}_{AB} (\Omega)$. The right-hand side requires detailed evaluation:
\begin{align}
   \bar{R}_{AB} (\Omega)
   & = \frac{i}{\pi}
   \int_{-\infty}^\infty  dt  \,  e^{i \Omega (t - t') } \,  \Theta (t - t')
   \int_{-\infty}^\infty  d \Omega'  \,
   \tanh \left( \frac{\beta \Omega' }{2}  \right)
   \bar{C}_{ \{A, B \} } ( \Omega' )  \,
   e^{-i \Omega' (t - t') }  \,  \\
   \label{eq_conv_FDT_help1}
   & =  \frac{i}{\pi}  \int_{-\infty}^\infty d \Omega'  \,
   \tanh \left( \frac{\beta \Omega' }{2}  \right)
   \bar{C}_{ \{A, B \} } ( \Omega' )
   \int_{-\infty}^\infty  dt  \,  \Theta (t - t')  \,
   e^{i ( \Omega - \Omega' ) (t - t') }  \, .
\end{align}
We have interchanged the integrals. The Heaviside function cuts off the final integral from below at $t - t' = 0$. That integral becomes, via complex-analysis identities,
\begin{align}
   \int_0^\infty  dt  \,  e^{i (\Omega - \Omega') (t - t')}
   =  \lim_{\varepsilon \to 0^+}  
   \frac{i}{(\Omega - \Omega') + i \varepsilon}
   = \pi \, \delta (\Omega - \Omega')  +  i \frac{P}{\Omega - \Omega' } \, .
\end{align}
$P$ denotes the principal part. We substitute the final expression into Eq.~\eqref{eq_conv_FDT_help1} and evaluate the remaining integral. Then, we split up the response function as 
$\bar{R}_{AB}(\Omega)
=  \bar{R}^\1_{AB} (\Omega)  +  i  \bar{R}^\2_{AB} (\Omega)$, wherein
\begin{align}
   \label{eq_Conv_FDT_Im_app} &
   \bar{R}^\1_{AB} (\Omega)
   =  \tanh \left( \frac{\beta \Omega}{2}  \right)
   \bar{C}_{ \{A, B \} } (\Omega) 
   \quad \text{and} \\ 
   \label{eq_Conv_FDT_Re_app} &
   \bar{R}^\2_{AB} (\Omega)
   = - \frac{2P}{\pi}  \int_{-\infty}^\infty  \frac{d \Omega'}{\Omega - \Omega'}  \,
   \tanh \left( \frac{\beta \Omega'}{2}  \right)
    \bar{C}_{ \{A, B \} } (\Omega' )  .
\end{align}
Equation~\eqref{eq_Conv_FDT_Im_app} forms the conventional quantum FDT.

\section{Derivation of the fine-grained KMS relation for energy eigenstates of SU(2)-symmetric quantum many-body systems}
\label{sec_Derive_FDT_SU2}

Below, we detail the derivation of our fine-grained KMS relation from the non-Abelian ETH. In App.~\ref{sec_Thermo_props}, we introduce the modified NATS $\rho_\NATS$ in detail. In App.~\ref{sec_Thermo_props_2}, we review thermodynamic properties of $\rho_\NATS$. These thermodynamic properties will inform our analysis of energy eigenstates $\ket{\alpha, m}$, by the local similarity between $\ket{\alpha, m}$ and $\rho_\NATS$. In App.~\ref{app_Thermo_KMS}, we derive a fine-grained KMS relation satisfied by the modified NATS. We begin to derive the fine-grained KMS relation satisfied by an energy eigenstate in App.~\ref{app_Corr}, by introducing the correlator $\bar{C}_{AB}(\Omega)$. We also apply the non-Abelian ETH to the correlator there. In App.~\ref{app_Gen_correxn}, we calculate the general finite-size correction in $\bar{C}_{AB}(\Omega)$. In App.~\ref{app_KMS_0th}, we prove the fine-grained KMS relation satisfied by certain energy eigenstates $\ket{\alpha, m}$. We do not argue, there, that the correction can be anomalously large. We argue so in App.~\ref{app_KMS_Anom}, which focuses on a different parameter regime.

This appendix concerns the setup described in Sec.~\ref{sec_Review_NAETH}.
Certain parts of our argument involve extra assumptions, introduced below. Also, we define $\tilde{S}_\th(E, s)$ as the thermodynamic entropy at the energy $E$ and spin quantum number $s$.  Define $\tilde{D}_\tot(E, s)$ as the density of states at energy $E$ and spin quantum number $s$. This density of states obeys
$\tilde{S}_\th(E,s) = \ln \LParen \tilde{D}_\tot(E, s) \RParen$.
In contrast, we defined $S_\th(E, s)$ as the thermodynamic entropy at the energy $E$, the spin quantum number $s$, and any fixed magnetic spin quantum number 
(Sec.~\ref{sec_Review_NAETH} and~\cite{24_Lasek_Numerical}). The corresponding density of states is $D_\tot(E, s) = \tilde{D}_\tot(E, s) / (2s + 1)$: 
$S_\th(E, s) = \ln \LParen D_\tot(E, s) \RParen$.

\subsection{Modified NATS}
\label{sec_Thermo_props}

The multiqubit system's NATS is defined as follows~\cite{guryanova_2016_thermodynamics,Lostaglio_2017_MaxEnt,Halpern_2016_microcanonical,Majidy_23_Noncommuting}. Let $\beta$ denote an inverse temperature; and $\mu_a$, the effective chemical potential associated with $S_a$. Define 
$\mu \coloneqq \sqrt{ \sum_{a=x,y,z} \mu_a^2 } \, .$ The NATS has the form
\begin{align}
   \label{eq_NATS_app}
   \tilde{\rho}_\NATS
   & \coloneqq \exp \left( - \beta H - \sum_{a=x,y,z} \mu_a S_a \right) / \tilde{Z}
   = \exp \left( - \beta H - \mu S_z \right) 
   / \tilde{Z} .
\end{align}
The final equality follows from pointing the $z$-axis parallel to 
$\vec{\mu}=(\mu_x, \mu_y, \mu_z)$ [equivalently, parallel to $\Tr ( \tilde{\rho}_\NATS \vec{S} )$], as we do throughout these appendices. This orientation implies that $\mu \geq 0$. The partition function 
$\tilde{Z}
\coloneqq \Tr \LParen 
\exp \left( - \beta H - \mu S_z \right) \RParen$
normalizes the state.

Having reviewed the NATS, we define the modified NATS. First, we introduce the observable
\begin{align}
   \label{S_def_app}
   S \coloneqq \sum_{\alpha, m} s_\alpha \ketbra{\alpha, m}{\alpha, m} .
\end{align} 
Its eigenvalues---the spin quantum numbers $s_\alpha$---label the eigenvalues $s_\alpha (s_\alpha + 1)$ of $\vec{S}^2$. Denote by $\gamma \in \mathbb{R}$ the thermodynamic force conjugate to $S$. In terms of these quantities, the modified NATS has the form
\begin{align}
   \label{eq_NATS_mod_app}
   \rho_\NATS
   \coloneqq \exp \left( - \beta \left[ H - \mu S_z - \gamma S \right] \right) / Z .
\end{align}
The partition function
$Z \coloneqq \Tr \LParen
\exp ( - \beta [H - \mu S_z - \gamma S] ) \RParen$. Justifying the use of $\rho_\NATS$ is subtle: $\rho_\NATS$ does not follow from a common derivation of the thermal state's form~\cite{landau_1980_statistical}. In that derivation, we assume that the system of interest and its environment are jointly in equilibrium---in a microcanonical state. The total entropy therefore attains its maximal value. The total entropy depends straightforwardly on the total energy and on other total extensive charges. The system-of-interest charges contribute additively to these total charges. One can therefore trace out the environment from the microcanonical state, leaving the system-of-interest state expressed in terms of its extensive charges. However, $S$ is not extensive (additive).
The thermal state therefore appears not to depend on $S$, according to the microcanonical derivation.\footnote{
Hence, in much prior literature, the NATS has not depended on $S$ (e.g.,~\cite{guryanova_2016_thermodynamics,Lostaglio_2017_MaxEnt,Halpern_2016_microcanonical,NYH_20_Noncommuting,Kranzl_23_Experimental,MurthyNAETH,Majidy_23_Noncommuting}).}
This independence resembles the thermal state’s independence of $H^2$, $H^3$, $S_z^9$, etc. Yet $S$ is an observable whose expectation value we can control and have information about. Introducing $S$ into the system-of-interest density operator is therefore reasonable~\cite{Jaynes_57_Info_II}. 
Furthermore, one can infer empirically about $\expval{S}$ in two ways, as explained in the following two paragraphs.
One might debate whether the density operator deserves the label \emph{thermal state}, though. 
Having introduced $\rho_\NATS$, we define the thermal expectation value 
$\expval{ A } \coloneqq \Tr ( \rho_\NATS \, A )$ of arbitrary operators $A$.

One can infer empirically about $\expval{S}$ in two ways. First, consider measuring the fine-grained correlator experimentally. From the fine-grained KMS relation, one can infer $\gamma$'s value, which implies $\expval{S}$'s value~\cite{Jaynes_57_Info_II}.

Second, one can efficiently bound $\expval{S}$, using shadow tomography. We can express the spin-squared operator $\vec{S}^2$ in two ways. In terms of eigenvalues,
\begin{align}
   \label{eq_S2_1}
   \vec{S}^2  
   =  \sum_{\alpha, m}  s_\alpha (s_\alpha + 1)
   \ketbra{\alpha, m}{\alpha, m}
   =  \sum_{\alpha, m}  \left( s_\alpha^2 + s_\alpha \right)
   \ketbra{\alpha, m}{\alpha, m}
   =  \left( S^2  +  S  \right) .
\end{align}
Denote by $\vec{S}^{(j)}$ qubit $j$'s spin operator. In terms of these local operators,
\begin{align}
   \vec{S}^2
   = \left(  \sum_{j=1}^N  \vec{S}^{(j)}  \right)^2
   \label{eq_S2_2}
   = \sum_{j, k} \vec{S}^\JParen \cdot \vec{S}^\KParen .
\end{align}
Let us equate the rightmost expressions in~\eqref{eq_S2_1} and~\eqref{eq_S2_2}. Taking the expectation value of each yields $\expval{S^2} + \expval{S} 
= \sum_{j, k} 
\langle \vec{S}^\JParen \cdot \vec{S}^\KParen \rangle$.
Since $\expval{S^2} \geq 0$,
\begin{align}
   \expval{S} 
   \leq \sum_{j, k}
   \expval{ \vec{S}^\JParen \cdot \vec{S}^\KParen } .
\end{align}
The right-hand side equals a sum of two-local terms. One can therefore estimate it efficiently, using shadow tomography~\cite{20_Huang_Predicting,21_Paini_Estimating}.

\subsection{Thermodynamic properties of the modified NATS}
\label{sec_Thermo_props_2}

Here, we calculate thermodynamic properties of $\rho_\NATS$. The main results are formulae for $\tilde{S}_\th$ derivatives [Eqs.~\eqref{eq_E_deriv_help1},~\eqref{eq_S_bm0}, and~\eqref{eq_Big_params_thermo_1}]. One formula is general; the second applies in just one parameter regime; and the third, just in another parameter regime. These formulae will help us prove the fine-grained KMS relation for energy eigenstates $\ket{\alpha, m}$, by the local resemblance between $\rho_\NATS$ and $\ket{\alpha, m}$. To derive the formulae, we calculate a finite-size system's $Z$, $\expval{S_z}$, and $\expval{S}$. Then, we approximate the results in the thermodynamic limit. Finally, we specialize to certain parameter regimes.

\subsubsection{Thermodynamic properties of a finite-size system in $\rho_\NATS$}

Let us calculate $Z$, $\expval{S_z}$, and $\expval{S}$:
\begin{enumerate}

   \item $\bm{Z}$:
We compute the partition function's trace using $\{ \ket{\alpha, m} \}$:
   \begin{align}
      \label{eq_Z_1_app}
      Z
      \coloneqq \Tr \left( e^{ - \beta H - \mu S_z - \gamma S } \right)
      = \sum_\alpha  \sum_{m = -s_\alpha}^{s_\alpha}
      e^{-\beta (E_\alpha - \mu m - \gamma s_\alpha) }
      = \sum_\alpha e^{-\beta (E_\alpha - \gamma s_\alpha) }
      \sum_{m = -s_\alpha}^{s_\alpha}
      e^{\beta \mu m} \, .
   \end{align}
Let us define a function $G_s(x)$ similar to the $m$-dependent sum. By the formula for a finite geometric series,
   \begin{align}
      \label{eq_G_Def}
      G_s(x)
      & \coloneqq \frac{1}{2s+1} 
      \sum_{m = -s}^s  e^{xm} 
      =\frac{e^{-xs}}{2s+1} \sum_{m=0}^{2s} e^{xm} \\
      \label{eq_G_form_app}
      & = \frac{1}{2s+1} \,
      \frac{\sinh \left(x \left[s + \frac{1}{2} \right] \right) }{ \sinh(x/2) } \, .
   \end{align}
In terms of $G_s(x)$, the partition function has the form
   \begin{align}
      \label{eq_Z_app}
      Z = \sum_\alpha (2s_\alpha + 1) \, 
      G_{s_\alpha} (\beta \mu) \,
      e^{-\beta (E_\alpha - \gamma s_\alpha) } \, .
   \end{align}

   \item $\bm{\expval{S_z}}$: By definition,
   $\expval{S_z}
      = \Tr ( S_z \, \rho_\NATS)
      = \Tr \left( S_z \, e^{ - \beta ( H - \mu S_z - \gamma S) } / Z  \right)
      = \frac{1}{Z}  \sum_\alpha  \sum_{m = -s_\alpha}^{s_\alpha}
      m  \,  e^{ - \beta (E_\alpha - \mu m - \gamma s_\alpha) } \, .$
   The sum over $m$ has the form 
   $\sum_{m=-s_\alpha}^{s_\alpha} m \, e^{\beta \mu m}
   = (2s_\alpha + 1) \, G'_{s_\alpha} (\beta \mu)$. Hence
   \begin{align}
      \label{eq_Sz_Expval_app}
      \expval{S_z}
      = \frac{1}{Z} \sum_\alpha (2s_\alpha + 1) \,
      G'_{s_\alpha} (\beta \mu) \,
      e^{-\beta (E_\alpha - \gamma s_\alpha) } \, .
   \end{align}

   \item $\bm{\expval{S}}$: We evaluate $\expval{S}$ analogously to $\expval{S_z}$:
      $\expval{S}
      = \frac{1}{Z} \sum_\alpha s_\alpha (2s_\alpha + 1) \,
      G_{s_\alpha} (\beta \mu) \,
      e^{-\beta (E_\alpha - \gamma s_\alpha) } \, .$

\end{enumerate}

\subsubsection{Thermodynamic properties in the large-system regime}

Let us begin to evaluate the partition function. In the large-system regime, we approximate $E_\alpha$ as the continuous variable $E$. We replace the $\sum_\alpha (2s_\alpha + 1)$ with an integral $\int dE$ and the density of states, 
$\tilde{D}_\tot (E, s_\alpha) 
= \exp \LParen \tilde{S}_\th (E, s_\alpha) \RParen$.\footnote{
\label{foot_alpha}
Before we introduced this continuum approximation, $\alpha$ served as an index that participated in a one-to-one mapping with eigenenergies $E_\alpha$. Furthermore, each $\alpha$ could be mapped to one spin quantum number $\alpha$. This structure ceases to exist once we approximate $E_\alpha$ as continuous, however. $\alpha$ is now merely an artifact in the notation for $s_\alpha$.}
For convenience, we move the $G_{s_\alpha}(\beta \mu)$ into an exponential and a logarithm: 
$G_{s_\alpha}(\beta \mu)
= e^{\ln \LParen G_{s_\alpha}(\beta \mu) \RParen} \, .$
The partition function assumes the form
\begin{align}
   \label{eq_Z_large_app}
   Z = \sum_{s_\alpha}  \int dE \;
   e^{ \tilde{S}_\th(E, s_\alpha)
   + \ln \LParen G_{s_\alpha} (\beta \mu) \RParen
   - \beta (E - \gamma s_\alpha) } \, .
\end{align}

To evaluate $Z$ further, we would invoke Laplace's method. (Laplace's method specializes the saddle-point approximation to real variables.) The exponent peaks sharply, so we would typically Taylor-approximate its argument about the peak to second order. At the peak, the argument's first derivatives vanish (and second derivatives are negative). However, we will not approximately calculate $Z$ fully.

Instead, we calculate the first derivatives at the peak, as well as $\expval{S_z}$. In Eq.~\eqref{eq_Z_large_app}, the exponent has an argument that we define as
\begin{align}
   \label{eq_F_Def}
   \mathscr{F} (E, s_\alpha)
   \coloneqq \tilde{S}_\th (E, s_\alpha)
   + \ln \LParen G_{s_\alpha} (\beta \mu) \RParen
   - \beta (E - \gamma s_\alpha) .
\end{align}
Denote by $E_*$ and $s_*$ the $E$ and $s_\alpha$ values at which 
$\mathscr{F} (E, s_\alpha)$ maximizes. 
These values approximately equal expectation values (App.~\ref{app_Peak_vals}): $E_* \approx \expval{H}$, and $s_* \approx \expval{S}$. 

Let us evaluate the derivatives of $\mathscr{F} (E, s_\alpha)$ where the function peaks. We begin with the energy derivative:
\begin{align}
   \label{eq_F_E_deriv}
   & 0  \approx  \frac{\partial}{\partial E} \mathscr{F} (E, s_\alpha) 
   \Big\lvert_{\expval{H}, \expval{S}}
   = \frac{\partial}{\partial E} \tilde{S}_\th (E, s_\alpha) 
   \Big\lvert_{\expval{H}, \expval{S}}
   - \beta 
   \qquad \Rightarrow \\
   \label{eq_E_deriv_help1}
   & \frac{\partial}{\partial E} \tilde{S}_\th (E, s_\alpha) 
   \Big\lvert_{\expval{H}, \expval{S}}
   \approx \beta .
\end{align}
Equation~\eqref{eq_E_deriv_help1} is the first main result in this subsection.

The second main result concerns an $s_\alpha$-derivative of $\tilde{S}_\th$ at $\beta \mu = 0$. The $s_\alpha$-derivative of $\mathscr{F}$ vanishes reminiscently of Eq.~\eqref{eq_F_E_deriv}:
\begin{align}
   0  & \approx  \frac{\partial}{\partial s_\alpha} 
   \mathscr{F} (E, s_\alpha) \Bigg\lvert_{\expval{H}, \expval{S}}
   \label{eq_s_deriv_help1}
   = \frac{\partial}{\partial s_\alpha} 
   \tilde{S}_\th (E, s_\alpha) \Bigg\lvert_{\expval{H}, \expval{S}}
   + \frac{1}{ G_{s_\alpha} (\beta \mu)}
   \frac{\partial}{\partial s_\alpha}  G_{s_\alpha} (\beta \mu) \Bigg\lvert_{\expval{S}}
   + \beta \gamma .
\end{align}
The final derivative's form follows from the $G_s(x)$ formula~\eqref{eq_G_form_app}:
\begin{align}
   \frac{\partial}{\partial s_\alpha}  G_{s_\alpha} (\beta \mu)  \Big\lvert_{\expval{S}}
   & = \frac{1}{2 \expval{S} + 1} \, \frac{1}{ \sinh(\beta \mu / 2) } \left\{
   \beta \mu \cosh \left( \frac{\beta \mu}{2} \left[ 2 \expval{S} + 1 \right] \right)
   - \frac{2}{2 \expval{S} + 1} 
     \sinh \left( \frac{\beta \mu}{2} \left[ 2 \expval{S} + 1 \right] \right) \right\}
\end{align}
We factor out $\sinh \big( \frac{\beta \mu}{2} [2 \expval{S} + 1]  \big)$, then apply Eq.~\eqref{eq_G_form_app} to the prefactor:
\begin{align}
   \label{eq_G_deriv_help1}
   \frac{\partial}{\partial s_\alpha}  G_{s_\alpha} (\beta \mu)  \Big\lvert_{\expval{S}}
   & = G_{\expval{S}} (\beta \mu) \left\{
   \beta \mu \coth \left( \frac{\beta \mu}{2} \left[2 \expval{S} + 1 \right] \right)
   - \frac{2}{2 \expval{S} + 1} \right\} .
\end{align}
Upon substituting into Eq.~\eqref{eq_s_deriv_help1}, we isolate the $\tilde{S}_\th$ term:
\begin{align}
   \label{eq_s_deriv_help2}
   \frac{\partial}{\partial s_\alpha}   \tilde{S}_\th (E, s_\alpha) 
   \Big\lvert_{\expval{H}, \expval{S}}
   \approx - \beta \mu \coth \left( \frac{\beta \mu}{2} 
   \left[2 \expval{S} + 1 \right] \right)
   + \frac{2}{2 \expval{S} + 1}  -  \beta \gamma .
\end{align}
We will evaluate this formula in each of two parameter regimes.

We must also calculate $\expval{S_z}$ in these regimes. In App.~\ref{app_S_z_calcn}, we derive the following general formula from Eq.~\eqref{eq_Z_large_app}:
\begin{align}
   \label{eq_S_z_app}
   \expval{S_z}
   \approx G'_{ \expval{S} } (\beta \mu) / G_{ \expval{S} } (\beta \mu) .
\end{align}
Calculating the right-hand side from Eq.~\eqref{eq_G_form_app} is straightforward. The result echoes a result in the theory of magnetism~\cite{04_Kittel_Solid}. Consider a paramagnet of spins, each associated with the spin quantum number $s$. Denote by $S_z^\JParen$ the $j^\th$ spin's Pauli-$z$ operator. Suppose that the spins do not interact, evolving under the Hamiltonian $-\mathcal{B} \sum_j S_z^\JParen$. The magnetization depends on the \emph{Brillouin function},
\begin{align}
   \label{eq_Brillouin}
   B_s(x)
   \coloneqq \frac{2s + 1}{2s} \,
   \coth \left( \frac{x}{2s} \, [2s+1] \right)
   - \frac{1}{2s} \, \coth \left( \frac{x}{2s} \right) .
\end{align}
At the inverse temperature $\beta$, the paramagnet's average magnetization has the form
$\langle S_z^\JParen \rangle  =  s \, B_s(\beta s)$.
Our system has a similar magnetization:
\begin{align}
   \label{eq_S_z_Brill_app}
   \expval{S_z}
   \approx  \expval{S} \, B_{ \expval{S} } \left( \beta \mu \expval{S} \right) .
\end{align}
Two useful limiting behaviors of $B_s(x)$ derive from limiting behaviors of the hyperbolic tangent. First, suppose each function's argument is small:
\begin{align}
   \label{eq_Approx_coth_small}
   \coth(x \approx 0) 
   = \frac{1}{x} + \frac{x}{3} + O \left( x^3  \right)
   \qquad \Rightarrow \qquad
   B_s (x \approx 0)
   = \frac{x (s + 1)}{3s}  +  O \left( x^3 \right) .
\end{align}
Now, suppose each function's argument is large:
\begin{align}
   \label{eq_Approx_coth_large}
   \coth(x \gg 1) = 1 +  O \left( e^{-2 x} \right)
   \qquad \Rightarrow \qquad
   \lim_{x \to \infty} B_s(x) = 1 .
\end{align}

Let us apply the foregoing results, with the first-derivative conditions~\eqref{eq_E_deriv_help1} and~\eqref{eq_s_deriv_help2}, to ascertain how 
$\partial_{s_\alpha} \tilde{S}_\th (E, s_\alpha)$ and $\expval{S_z}$ behave under certain conditions on $\beta$ and $\mu$ (and, in one case, $\expval{S}$):
\begin{enumerate}

   \item \textbf{If} $\bm{\beta \mu = 0}$: We infer about 
   $\partial_{s_\alpha} \tilde{S}_\th (E, s_\alpha)$ from Eq.~\eqref{eq_s_deriv_help2}, using the small-argument formula~\eqref{eq_Approx_coth_small}:
   \begin{align}
      \label{eq_S_bm0}
      \frac{\partial}{\partial s_\alpha} \tilde{S}_\th (E, s_\alpha) \Big\lvert_{ \expval{H}, \expval{S} }
   \approx - \beta \gamma .
   \end{align}
   Next, we calculate $\expval{S_z}$. By the $\expval{S_z}$ formula~\eqref{eq_S_z_Brill_app} and the small-argument approximation~\eqref{eq_Approx_coth_small},
   \begin{align}
      \label{eq_Sz_bm0}
      \expval{S_z} \approx 0.
   \end{align}

   \item \textbf{If $\bm{\beta \mu > 0}$ and $\bm{\expval{S}  =  O (N^\zeta)}$,
   wherein $\bm{\zeta  \in  (0, 1]}$}:
   Again, we infer about $\partial_{s_\alpha}  \tilde{S}_\th (E, s_\alpha)$, from Eq.~\eqref{eq_s_deriv_help2}. To the hyperbolic tangents therein, we apply the large-argument approximation~\eqref{eq_Approx_coth_large}:
   \begin{align}
      \label{eq_Big_params_thermo_1}
      \frac{\partial \tilde{S}_\th (E, s_\alpha) }{\partial s_\alpha}
      \Bigg\lvert_{ \expval{H}, \expval{S} }
      \approx - \beta (\mu + \gamma) + O \left( \expval{S}^{-1} \right) .
   \end{align}
   To infer about $\expval{S_z}$, we apply the large-argument approximation~\eqref{eq_Approx_coth_large} to Eq.~\eqref{eq_S_z_Brill_app}:
   \begin{align}
      \label{eq_Sz_bm_large}
      \expval{S_z}
      \approx \expval{S} - \frac{1}{2} 
      \left[ \coth \left( \frac{\beta \mu}{2} \right) - 1 \right]
      + O \left( e^{-2 \beta \mu \expval{S} }  \right) 
      =  \expval{S} + O(1) .
   \end{align}

\end{enumerate}

\subsubsection{Proof that, at the peak of $\mathscr{F} (E, s_\alpha)$, 
$\expval{H} \approx E_*$ and $\expval{S} \approx s_*$}
\label{app_Peak_vals}


Let us prove that $\expval{H} \approx E_*$ and $\expval{S} \approx s_*$ where 
$\mathscr{F} (E, s_\alpha)$ [Eq.~\eqref{eq_F_Def}] peaks. 
To do so, we calculate $\expval{S}$, then $\expval{H}$. We substitute into the first expectation value from the $\rho_\NATS$ definition~\eqref{eq_NATS_app}:
\begin{align}
   \expval{S}
   = \Tr \left( S  \rho_\NATS \right)
   = \frac{1}{Z}  \Tr \left( S \, e^{-\beta (H - \mu S_z - \gamma S) } \right)
   \label{eq_S_expval_help1}
   = \frac{1}{\beta Z}  \frac{\partial Z}{\partial \gamma} \, .
\end{align}
Applying Laplace's method to $Z$, one Taylor-approximates to second order. By Eqs.~\eqref{eq_Z_large_app} and~\eqref{eq_F_Def}, the partition function has the form
\begin{align}
   Z
   & = \sum_{s_\alpha} \int dE \, 
   \exp \LParen \mathscr{F} (E, s_\alpha)  \RParen
   \approx \sum_{s_\alpha} \int dE \, 
   \exp \LParen \mathscr{F} (E \approx E_*, s_\alpha \approx s_*)  \RParen \\
   & \approx  \sum_{s_\alpha \approx s_*}  \int  dE  \,
   \exp \Big(  \mathscr{F} (E_*, s_*) 
   + \vec{\nabla}  \mathscr{F} (E, s_\alpha)  \Big\lvert_{E_*, s_*}  \cdot
      (E - E_*,  s_\alpha - s_* ) \\
    & \qquad \qquad \qquad \qquad \quad
    + \frac{1}{2} \, (E - E_* , s_\alpha - s_* )^{\rm T}  \cdot
       \vec{\nabla}_{\rm H}^2  \mathscr{F} (E, s_\alpha) \Big\lvert_{E_*, s_*}
       \cdot  (E - E_*,  s_\alpha - s_* ) \Big) .
\end{align}
$\vec{\nabla}_{\rm H}^2$ denotes the Hessian. The gradient (first-order term) vanishes by the definition of $E_*$ and $s_*$. If the system is very large, the second-order correction vanishes. We therefore approximate the partition function to zeroth order:
\begin{align}
   \label{eq_Z_approx_0th}
   Z  
   \approx  \exp \Big( \tilde{S}_\th ( E_*, s_* )  
                                 +  \ln \LParen G_{s_*} (\beta \mu)  \RParen
                                 - \beta [E_* - \gamma s_* ]  \Big) .
\end{align}
Substituting into Eq.~\eqref{eq_S_expval_help1} yields
$\expval{S} \approx s_*$, the first claim we aimed to prove.

Now, we prove that $\expval{H} \approx E_*$. If $\rho_\NATS$ were a canonical state, rather than a modified NATS, then $\expval{H}$ would equal 
$- \frac{1}{Z} \frac{\partial Z}{\partial \beta}$~\cite[Sec.~16.1]{Callen_book}.
Guided by this observation, we calculate $- \frac{1}{Z} \frac{\partial Z}{\partial \beta}$, using the NATS formula~\eqref{eq_NATS_app}:
\begin{align}
   - \frac{1}{Z} \frac{\partial Z}{\partial \beta}
   = - \frac{1}{Z} \frac{\partial}{\partial \beta}
   \Tr \left( e^{-\beta (H - \mu S_z - \gamma S) }  \right)
   = \expval{H}  -  \mu \expval{S_z}  -  \gamma \expval{S} .
\end{align}
Solving for $\expval{H}$ yields
\begin{align}
   \label{eq_H_expval_1_app}
   \expval{H}
   = - \frac{1}{Z}  \frac{\partial Z}{ \partial \beta}
   + \mu \expval{S_z}  +  \gamma \expval{S} .
\end{align}

Let us approximate the first term to zeroth order. We substitute into Eq.~\eqref{eq_Z_approx_0th} for $G_{s_\alpha} (\beta \mu)$ from the definition~\eqref{eq_G_Def}:
\begin{align}
   \frac{1}{Z}  \frac{\partial Z}{ \partial \beta}
   & \approx \frac{1}{Z}  \frac{\partial}{\partial \beta}
   \sum_{m = - s_*}^{s_*}  (2 s_* + 1) \,
   \exp \left( \tilde{S}_\th (E_*, s_*)  -  \beta [E_* - \mu m - \gamma s_*] \right)  \\
   & = \frac{1}{Z}  \sum_{m = - s_*}^{s_*}  (2 s_* + 1)
   (E_* - \mu m - \gamma s_*)
   \exp \left( \tilde{S}_\th (E_*, s_*)  -  \beta [E_* - \mu m - \gamma s_*] \right) \\
   & \approx E_*  
   - \frac{\mu}{Z}  \sum_{m = - s_*}^{s_*}  (2s_* + 1) m \,
   \exp \left( \tilde{S}_\th (E_*, s_*)  -  \beta [E_* - \mu m - \gamma s_*] \right)
   - \gamma s_* \\
   & \approx  E_* - \mu \expval{S_z}  -  \gamma \expval{S} .
\end{align}
Again, we have approximated $Z$ to zeroth order. Substituting into Eq.~\eqref{eq_H_expval_1_app} yields
$\expval{H} \approx E_*$, the second claim we aimed to prove.

\subsubsection{Calculation of $\expval{S_z}$}
\label{app_S_z_calcn}

Let us derive the formula~\eqref{eq_S_z_app} for the modified-NATS expectation value $\expval{S_z}$. By textbook thermodynamics, reviewed in App.~\ref{app_Peak_vals},
$\expval{S_z} = \frac{1}{Z} \, \frac{\partial Z}{\partial (\beta \mu)}$. 
We approximated $Z$ to zeroth order in Eq.~\eqref{eq_Z_approx_0th}. Substituting into the $\expval{S_z}$ formula, we pull the $G$ outside the exponential. Across App.~\ref{app_Peak_vals}, we showed that $E_* \approx \expval{H}$ and $s_* \approx \expval{S}$. Combining these results yields
\begin{align}
   \expval{S_z}
   & \approx \exp \left(  - \tilde{S}_\th ( \expval{H}, \expval{S} )
   + \beta \left[ \expval{H} - \gamma \expval{S} \right]  \right)
   \left[ G_{ \expval{S} } (\beta \mu) \right]^{-1}
   \cdot  \exp \left(  \tilde{S}_\th ( \expval{H}, \expval{S} )
   - \beta \left[ \expval{H} - \gamma \expval{S} \right]  \right)
   \frac{\partial G_{ \expval{S} } (\beta \mu) }{ \partial (\beta \mu) } \, .
\end{align}
The exponentials cancel; Eq.~\eqref{eq_S_z_app} results.

\subsection{Fine-grained KMS relation satisfied by the modified non-Abelian thermal state}
\label{app_Thermo_KMS}

Here, we derive the fine-grained KMS relation obeyed by $\rho_\NATS$. This relation forms a model for that derived in the rest of App.~\ref{sec_Derive_FDT_SU2}: the fine-grained KMS relation for an eigenstate of an SU(2)-symmetric Hamiltonian. The derivation here is less straightforward than that in App.~\ref{app_Conventional_KMS}, where we proved the conventional KMS relation for a quantum system in a canonical state. The reason is, $\rho_\NATS$ contains an $S$ that does not participate in simple commutation relations with spherical tensor operators, to our knowledge. 

This subsection is organized as follows. In App.~\ref{app_Corr_NATS_H}, we introduce a time-domain two-point correlator $C_{AB}^\NATS (t)$ and a frequency-domain correlator $C_{AB}^\NATS (\Omega)$. We introduce a variation on the latter, the \emph{fine-grained correlator} 
$\hat{\bar{C}}_{AB}^\NATS (\Omega, \Delta m, \Delta s)$, in App.~\ref{app_Finegrain}. There, we also derive the fine-grained KMS relation for the fine-grained correlator. 

We focus on correlators of spherical tensor operators, reviewed in Sec.~\ref{sec_Review_NAETH}.
For convenience, we address 
$A  \coloneqq  A^{(k')}_{-q}$ and $B  \coloneqq  B^{(k)}_q \, .$

\subsubsection{Correlators defined in terms of the modified NATS}
\label{app_Corr_NATS_H}

This subsubsection bridges the conventional KMS relation (App.~\ref{app_Conventional_KMS}) to the fine-grained KMS relation. 
Define the time-domain connected two-point correlator
\begin{align}
   \label{eq_Def_Mixed_Corr_app}
   C_{AB}^\NATS (t)
   & \coloneqq  \expval{ A(t)  \,  B }  -  \expval{ A(t) }  \expval{B} .
\end{align}
Section~\ref{sec_FDT_Thermo} concerned a disconnected correlator, for consistency with the conventional KMS relation (App.~\ref{app_Review_FDT}). Here, we analyze a connected correlator because it more reliably encodes correlations between variables, by definition. However, the proof below (of the fine-grained KMS relation) extends directly to disconnected correlators; one simply eliminates certain terms.

To evaluate $C_{AB}^\NATS (t)$, we replace the Heisenberg-picture operator $A(t)$ with its definition, $e^{iHt}  \,  A  \,  e^{-iHt} \, .$ We also calculate the traces using the energy eigenbasis. Finally, we insert an
$\id = \sum_{\alpha', m'}  \ketbra{\alpha', m'}{\alpha', m'}$ rightward of the first $A$:
\begin{align}
   C_{AB}^\NATS (t)
   & =  \Tr  \left(  e^{iHt}  \,  A  \,  e^{-iHt}  \,  B  \,  
           \frac{ e^{-\beta (H - \mu S_z - \gamma S) } }{Z}  \right)
   -  \Tr  \left(  e^{i H t}  \,  A  \,  e^{-iHt}  \,  
       \frac{ e^{-\beta (H - \mu S_z - \gamma S) } }{Z}  \right)  
   \Tr  \left( B  \, 
                 \frac{ e^{-\beta (H - \mu S_z - \gamma S) } }{Z}  \right)  \\
   & =  \sum_{\alpha, m, \alpha', m'}  \Bigg(
   \bra{\alpha, m}  A  \ketbra{\alpha', m'}{\alpha', m'}  B  \ket{\alpha, m}  \,
   \frac{ e^{-\beta \left( E_\alpha - \mu m - \gamma s_\alpha  \right) } }{Z}  \,
   e^{-i \left(  E_{\alpha'}  -  E_\alpha  \right) t }
   \\ & \nonumber \qquad \qquad \qquad
   -  \bra{\alpha, m}  A  \ketbra{\alpha, m}{\alpha', m'}  B  \ket{\alpha', m'}  \,
   \frac{ e^{-\beta \left( E_\alpha - \mu m - \gamma s_\alpha  \right) } }{Z}  \,
   \frac{ e^{-\beta \left( E_{\alpha'} - \mu m' - \gamma s_{\alpha'}  \right) } }{Z}
   \Bigg)  .
\end{align}

Fourier-transforming the time-domain correlator yields the frequency-domain correlator. We follow the Fourier-transform convention in App.~\ref{app_Review_FDT}:
\begin{align}
   \bar{C}_{AB}^\NATS (\Omega)
   & \coloneqq  \int_{-\infty}^\infty  dt  \:  C_{AB} (\Omega)  \,
   e^{i \Omega t} \\
   & = \sum_{\alpha, m, \alpha', m'}  \Bigg(
   \bra{\alpha, m} A \ketbra{\alpha', m'}{\alpha', m'}  B  \ket{\alpha, m}  \,
   \frac{ e^{-\beta \left( E_\alpha - \mu m - \gamma s_\alpha \right) } }{Z}
   \int_{-\infty}^\infty  dt  \:  
   e^{i \left[ \Omega - \left(  E_{\alpha'}  -  E_\alpha  \right)  t  \right]  }
   \\ \nonumber & \qquad \qquad \qquad
   - \bra{\alpha, m}  A  \ketbra{\alpha, m}{\alpha', m'}  B  \ket{\alpha',  m'}  \,
   \frac{ e^{-\beta \left( E_\alpha - \mu m - \gamma s_\alpha \right) } }{Z}  \,
   \frac{ e^{-\beta \left( E_{\alpha'} - \mu m' - \gamma s_{\alpha'} \right) } }{Z}
   \int_{-\infty}^\infty  dt  \:  e^{i \Omega t}
   \Bigg)
\end{align}
The Dirac  delta function has the integral representation
$\delta (k)  =  \frac{1}{2 \pi}  \int_{-\infty}^\infty  dx  \,  e^{i k x} \, .$  Therefore,
\begin{align}
   \label{eq_S_NATS_app}
   \bar{C}_{AB}^\NATS (\Omega)
   & =  2 \pi  \sum_{\alpha, m, \alpha', m'}
   \Bigg\{  \bra{\alpha, m} A \ketbra{\alpha', m'}{\alpha', m'}  B  \ket{\alpha, m}  \,
   \frac{ e^{- \beta \left( E_\alpha - \mu m - \gamma s_\alpha \right) } }{Z}  \,
   \delta \left( \Omega - \left[ E_{\alpha'} - E_\alpha \right] \right)
   \\ \nonumber & \qquad \qquad \qquad \qquad
   - \bra{\alpha, m} A \ketbra{\alpha, m}{\alpha', m'}  B  \ket{\alpha', m'}  \,
   \frac{ e^{- \beta \left( E_\alpha - \mu m - \gamma s_\alpha \right) } }{Z}  \,
   \frac{ e^{- \beta \left( E_{\alpha'} - \mu m - \gamma s_{\alpha'} \right) } }{Z}  \,
   \delta (\Omega) \Bigg\} .
\end{align}
Whether $\bar{C}_{AB}^\NATS (\Omega)$ obeys a KMS relation is unclear, so we fine-grain the correlator.

\subsubsection{Fine-grained correlator and KMS relation for the modified NATS}
\label{app_Finegrain}

To motivate the fine-grained correlator, we highlight the Dirac delta function in Eq.~\eqref{eq_S_NATS_app}. That delta function singles out an energy difference, $E_{\alpha'} - E_\alpha$. Energy plays a role analogous to the magnetic spin quantum number and to the spin quantum number throughout the rest of Eq.~\eqref{eq_S_NATS_app}. Hence we put the latter two quantities on completely equal footing with energy: we introduce a $\Delta m$ and a $\Delta s$ analogous to $\Omega$. We do so using Kronecker delta functions $\delta_{m' \, (m + \Delta m)}$ and 
$\delta_{s_{\alpha'} \, (s_\alpha + \Delta s)}$, rather than Dirac delta functions, because the spacings between consecutive $m$ values remains constant in the thermodynamic limit, as do the spacings between consecutive $s_\alpha$ values. We define the \emph{fine-grained (frequency-domain connected) correlator for the modified NATS} as
\begin{align}
   \label{eq_S_Fine_app}
  & \hat{\bar{C}}_{AB}^\NATS (\Omega, \Delta m, \Delta s)
  \coloneqq  2 \pi  \sum_{\alpha, m, \alpha', m'}  \Bigg\{
  \bra{\alpha, m} A \ketbra{\alpha', m'}{\alpha', m'}  B  \ket{\alpha, m}  \,
  \frac{ e^{- \beta \left( E_\alpha - \mu m - \gamma s_\alpha \right) } }{Z}  \,
  \delta \left( \Omega - \left[ E_{\alpha'} - E_\alpha \right] \right) \,
  \\ \nonumber & \times
  \delta_{m' \, (m + \Delta m)}  \,
  \delta_{s_{\alpha'} \, (s_\alpha + \Delta s)}
  - \bra{\alpha, m} A \ketbra{\alpha, m}{\alpha', m'} B \ket{\alpha', m'}  \,
  \frac{ e^{- \beta \left( E_\alpha - \mu m - \gamma s_\alpha \right) } }{Z}  \,
  \frac{ e^{- \beta \left( E_{\alpha'} - \mu m' - \gamma s_{\alpha'} \right) } }{Z}  \,
  \delta (\Omega )  \,
  \delta_{ ( \Delta m) \, 0}  \,
  \delta_{ (\Delta s) \, 0} \, .
\end{align} 

Summing the fine-grained correlator over $\Delta m$ and $\Delta s$ yields the conventional correlator for the modified NATS [Eq.~\eqref{eq_S_NATS_app}]:
\begin{align}
   \bar{C}_{AB}^\NATS  (\Omega)
   & =  \sum_{\Delta m, \Delta s}
   \hat{\bar{C}}_{AB}^\NATS ( \Omega, \Delta m, \Delta s ) .
\end{align}
The first sum must run over $\Delta m = q$, and the second sum must run over
$\Delta s = - \min \{k, k' \}, - \min \{k, k' \} + 1 , \ldots, \min \{ k, k' \}$.
Otherwise, the sums' limits are arbitrary.

One can measure $\hat{\bar{C}}_{AB}^\NATS (\Omega, \Delta m, \Delta s)$ experimentally by, for example, eigendecomposing the $A$ and $B$. The correlator reveals itself as an average with respect to an extended Kirkwood--Dirac quasiprobability~\cite{23_Lostaglio_KD,Gherardini_24_Quasi,24_DRMAS_Properties}. One can infer such a quasiprobability from each of multiple experimental protocols~\cite{23_Lostaglio_KD,Gherardini_24_Quasi,24_DRMAS_Properties}. Granted, such protocols involve weak or strong measurements of the energy eigenbasis. Such measurements are impractical if the system is large. Three points weigh against this criticism, however: (i) The same criticism applies to direct experimental tests of the ETH, which is used widely nonetheless. 
(ii) $\hat{\bar{C}}_{AB}^\NATS (\Omega, \Delta m, \Delta s)$ is measurable in principle and therefore is well-defined physically. (iii) Experimentalists' control over quantum many-body systems has been advancing rapidly~\cite{24_Fauseweh_Quantum}.

The fine-grained correlator obeys the \emph{fine-grained KMS relation for the modified NATS},
\begin{align}
   \label{eq_Detailed_KMS_app}
   \hat{\bar{C}}_{AB}^\NATS (\Omega, \Delta m, \Delta s)
   =  e^{\beta \left( \Omega - \mu \Delta m - \gamma \Delta s \right) }  \,
   \hat{\bar{C}}_{BA}^\NATS (- \Omega, - \Delta m, - \Delta s)  .
\end{align}
To prove this equation, we substitute into the right-hand side from the definition~\eqref{eq_S_Fine_app}:
\begin{align}
   \label{eq_Fine_KMS_help1}
   & e^{\beta \left( \Omega - \mu \Delta m - \gamma \Delta s \right) }  \,
   \hat{\bar{C}}_{BA}^\NATS (- \Omega, - \Delta m, - \Delta s)
   = 2 \pi \sum_{\alpha, m, \alpha', m'}  \bigg\{  
   \bra{\alpha, m} B \ketbra{\alpha', m'}{\alpha', m'} A \ket{\alpha, m} \,
   e^{ \beta ( \Omega - \mu \Delta m - \gamma \Delta s ) }  \,
   \\ \nonumber & \qquad \qquad \qquad \qquad \qquad \qquad \qquad \qquad \qquad \qquad \qquad \; \; \times
   \frac{ e^{- \beta \left( E_\alpha - \mu m - \gamma s_\alpha \right) } }{Z}  \,
   \delta \left( - \Omega - \left[ E_{\alpha'} - E_\alpha \right] \right) \,
   \delta_{m' \, (m - \Delta m) }  \,
   \delta_{ s_{\alpha'}  \,  (s_\alpha - \Delta s) }
   \\ & \nonumber \qquad \quad 
   - \bra{\alpha, m} B \ketbra{\alpha, m}{\alpha', m'} A \ket{\alpha', m'}  \,
   e^{ \beta ( \Omega - \mu \Delta m - \gamma \Delta s ) }  \,
   \frac{ e^{- \beta \left( E_{\alpha} - \mu m - \gamma s_{\alpha} \right) } }{Z}
    \frac{ e^{- \beta \left( E_{\alpha'} - \mu m' - \gamma s_{\alpha'} \right) } }{Z}  \,
    \delta (- \Omega )  \,
    \delta_{ (- \Delta m) \, 0}  \,
    \delta_{ (-\Delta s) \, 0 }   \bigg\} .
\end{align}
The first Dirac delta function simplifies as
$\delta ( - \Omega - [ E_{\alpha'} - E_\alpha ] )
= \delta ( \Omega + [ E_{\alpha'} - E_\alpha ] )
= \delta ( \Omega - [E_\alpha - E_{\alpha'} ] )$.
It is nonzero if and only if $\Omega = E_\alpha - E_{\alpha'}$. Therefore, we replace the first thermal exponential's $\Omega$ with $E_\alpha - E_{\alpha'}$ while keeping the Dirac delta function present. Similarly, first term's Kronecker deltas are nonzero if and only if $\Delta m = m - m'$ and $\Delta s = s_\alpha - s_{\alpha'} \, .$
In the final term of Eq.~\eqref{eq_Fine_KMS_help1}, we can remove the negative signs without changing the formula's value:
\begin{align}
   \label{eq_Fine_KMS_help2}
   & e^{\beta \left( \Omega - \mu \Delta m - \gamma \Delta s \right) }  \,
   \hat{\bar{C}}_{BA}^\NATS (- \Omega, - \Delta m, - \Delta s)
   = 2 \pi \sum_{\alpha, m, \alpha', m'}  \bigg\{ 
   \bra{\alpha, m} B \ketbra{\alpha', m'}{\alpha', m'} A \ket{\alpha, m}  \,
   \frac{ e^{- \beta \left( E_{\alpha'} - \mu m' - \gamma s_{\alpha'} \right) } }{Z}  
   \nonumber \\ & \qquad \qquad \qquad \qquad \qquad \qquad \qquad \qquad \qquad \qquad \qquad \qquad \quad \times
   \delta \left( \Omega - \left[ E_\alpha - E_{\alpha'} \right] \right) \,
   \delta_{m' \, (m - \Delta m) }  \,
   \delta_{ s_{\alpha'}  \, ( s_\alpha - \Delta s ) }
   \\ & \nonumber \qquad \qquad \qquad \qquad \qquad \qquad 
   - \bra{\alpha, m} B \ketbra{\alpha, m}{\alpha', m'} A \ket{\alpha', m'}  \,
   \frac{ e^{- \beta \left( E_{\alpha} - \mu m - \gamma s_{\alpha} \right) } }{Z}  
   \frac{ e^{- \beta \left( E_{\alpha'} - \mu m' - \gamma s_{\alpha'} \right) } }{Z}    \,
   \delta (\Omega)  \,   \delta_{(\Delta m) \, 0}  \,  \delta_{(\Delta s)  \,  0}  \, .
\end{align}
We can interchange dummy indices: $\alpha \leftrightarrow \alpha'$ and $m \leftrightarrow m'$. By the commutativity of scalar multiplication, the right-hand side of Eq.~\eqref{eq_Fine_KMS_help2} equals
$\hat{\bar{C}}_{AB}^\NATS (\Omega)$.

\subsection{Fine-grained correlator in an energy eigenstate of an SU(2)-symmetric quantum many-body system}
\label{app_Corr}

This appendix introduces the correlator in our fine-grained KMS relation.
Appendix~\ref{app_Corr_Context} specifies the setting. Appendix~\ref{app_Corr_Time} introduces the time-domain correlator 
$C_{AB}(t)$. Appendix~\ref{app_Corr_Freq} introduces the frequency-domain correlator 
$\bar{C}_{AB}(\Omega)$ and the fine-grained (dynamical) correlator 
$\hat{\bar{C}}^\dyn_{AB}(\Omega,  \Delta m, \Delta s)$.  We rewrite 
$\hat{\bar{C}}_{AB}^\dyn(\Omega,  \Delta m, \Delta s)$, using the Wigner–Eckart theorem and non-Abelian ETH, in App.~\ref{app_Corr_WE_NAETH}. To simplify the resulting expression, we package factors together into coarse-grained functions that exhibit symmetries. The symmetries will help us prove the fine-grained KMS relation.

\subsubsection{Setting}
\label{app_Corr_Context}

We specified much of our setup at the beginning of App.~\ref{sec_Derive_FDT_SU2}. Here, we sharpen our focus to certain energy eigenstates $\ket{\alpha, m}$ and to certain local operators. 

The following reasoning motivates or focus on certain $\ket{\alpha, m}$s. To derive the fine-grained KMS relation, we will evaluate correlators $\bar{C}_{AB}(\Omega,  \Delta m, \Delta s)$ on eigenstates $\ket{\alpha, m}$. We expect these eigenstates to locally resemble modified NATSs [Eq.~\eqref{eq_NATS_mod_app}], whose thermodynamic properties we can therefore apply (App.~\ref{sec_Thermo_props_2}). However, we expect a $\ket{\alpha, m}$ to resemble a $\rho_\NATS$ sufficiently only if the two share certain properties:
\begin{align}
   \label{eq_NATS_eigen_correspondence}
   E_\alpha = \expval{H} , \quad
   m = \expval{S_z} , 
   \quad \text{and} \quad
   s_\alpha = \expval{S} .
\end{align}
We focus on eigenstates that obey these relations. Furthermore, given any $\ket{\alpha, m}$, one can construct a $\rho_\NATS$ that satisfies Eqs.~\eqref{eq_NATS_eigen_correspondence}: one fixes the equations' left-hand sides and solves for $(\beta, \mu, \gamma)$~\cite{Jaynes_57_Info_I,Jaynes_57_Info_II,06_Liu_Gibbs,guryanova_2016_thermodynamics,25_Liu_Quantum}.

We focus on correlators of spherical tensor operators, as in App.~\ref{app_Thermo_KMS}. For convenience, we analyze
\begin{align}
   \label{eq_AB}
   A  \coloneqq  A^{(k')}_{-q}
   \quad \text{and} \quad
   B  \coloneqq  B^{(k)}_q .
\end{align}
We assume that $A$ and $B$, with $H$, obey the non-Abelian ETH. Local operators tend to obey ETH statements (although other operators can)~\cite{18_Garrison_Does}, so we assume that $k, k' = O(1)$. Consequently, $q = O(1)$.

\subsubsection{Time-domain correlator in an energy eigenstate} 
\label{app_Corr_Time}

The connected two-time correlator depends on the Heisenberg-picture operator 
$A(t)  \coloneqq  e^{iHt} \, A \, e^{-iHt} \, :$
\begin{align}
   \label{eq_S_t_Def_app}
   C_{AB} (t)
   \coloneqq  C_{AB} (t; \alpha, m)
   \coloneqq  \bra{\alpha, m} A(t) \, B \ket{\alpha, m} 
   - \bra{\alpha, m} A(t) \ketbra{\alpha, m}{\alpha, m} B \ket{\alpha, m} .
\end{align}
Section~\ref{sec_FDT_Analytics} concerned a disconnected correlator, for consistency with the conventional KMS relation (App.~\ref{app_Review_FDT}). In the appendices, we analyze connected correlators because they more reliably encode correlations between variables, by definition (App.~\ref{app_Corr_NATS_H}). Yet whether one begins with a connected or a disconnected correlator, here, does not matter: we will soon form a dynamical correlator, subtracting off the static component, to which the 
$- \bra{\alpha, m} A(t) \ketbra{\alpha, m}{\alpha, m} B \ket{\alpha, m}$
above contributes.

To evaluate $C_{AB} (t)$, we substitute in the definition of $A(t)$. Leftward of the first term's $B$, we introduce an
$\id = \sum_{\alpha', m'}  \ketbra{\alpha', m'}{\alpha', m'}$. Invoking the Hamiltonian eigenvalue equation yields
\begin{align}
   C_{AB}(t)
   & = \sum_{\alpha', m'}  \bra{\alpha, m} A \ketbra{\alpha', m'}{\alpha', m'} B \ket{\alpha, m}
   e^{i (E_\alpha - E_{\alpha'} ) t } 
   - \bra{\alpha, m} A \ketbra{\alpha, m}{\alpha, m} B \ket{\alpha, m} \\
   \label{eq_S_t_2_app}
   & = \sum_{\alpha'} \bra{\alpha, m} A \ketbra{\alpha', m+ q}{\alpha', m + q} B \ket{\alpha, m}  e^{i (E_\alpha - E_{\alpha'}) (t) }
   - \bra{\alpha, m} A \ketbra{\alpha, m}{\alpha, m} B \ket{\alpha, m} \, .
\end{align}
The final equality follows from Eq.~\eqref{eq_AB} and a selection rule (encoded in the Wigner–Eckart theorem's Clebsch–Gordan coefficient). The $\sum_{\alpha'}$ represents a sum over 
the energy eigenstates $\ket{\alpha', m+q}$ labeled by the magnetic spin quantum number $m+q$.

\subsubsection{Frequency-domain and (dynamical) fine-grained correlators in an energy eigenstate} 
\label{app_Corr_Freq}

We Fourier-transform the time-domain correlator as in App.~\ref{app_Conventional_KMS}:
\begin{align}
   \label{eq_S_w_Def_app}
   \bar{C}_{AB}(\Omega)
   & \coloneqq \bar{C}_{AB} (\Omega; \alpha, m)
   \coloneqq  \int_{-\infty}^\infty  dt  \;
   C_{AB} (t; \alpha, m)  \, e^{i \Omega t} \, .
\end{align}
Upon substituting in from Eq.~\eqref{eq_S_t_2_app}, we invoke the Dirac delta function's integral representation, 
$\delta(x) = \frac{1}{2 \pi} \int_{-\infty}^\infty dk \; e^{ikx}$:\footnote{
This Dirac delta function is an idealization approximated in our numerics (Sec.~\ref{sec_Num}). There, we average over the energy eigenstates $\ket{\alpha', m+q}$ associated with energies $E_{\alpha'}$ within a window: 
$| E_{\alpha'} - E_\alpha - \omega | < \Delta \omega / 2$.
The average assumes the form
$\frac{1}{\Delta \omega}
\sum_{\alpha' \, : \, | E_{\alpha'} - E_\alpha - \omega | < \Delta \omega / 2} \, .$}
\begin{align}
   \label{eq_S_w_app_help1}
   \bar{C}_{AB} (\Omega)
   & = 2 \pi  \left\{
   \sum_{\alpha'}  \bra{\alpha, m} A \ketbra{\alpha', m+q}{\alpha', m+q} B \ket{\alpha, m} \,
   \delta \left( \Omega - \left[ E_{\alpha'} - E_\alpha \right] \right) 
   - \bra{\alpha, m} A \ketbra{\alpha, m}{\alpha, m} B \ket{\alpha, m} \, \delta(\Omega)
   \right\} .
\end{align}

Let us separate the static correlator, which multiplies $\delta (\Omega)$, from the dynamical correlator, in which $E_{\alpha'} \neq E_\alpha$:
\begin{align}
   \label{eq_S_w_app_help1a}
   \bar{C}_{AB} (\Omega)
   & =  \bar{C}_{AB}^\stat  \,  \delta (\Omega)
   +  \bar{C}_{AB}^\dyn (\Omega)  .
\end{align}
The static correlator has the form 
\begin{align}
   \label{eq_Define_stat_1st}
   \bar{C}_{AB}^\stat
   & \coloneqq 2 \pi \left(  \bra{\alpha, m} A \ketbra{\alpha, m + q}{\alpha, m + q}
   B \ket{\alpha, m}
   - \bra{\alpha, m} A \ketbra{\alpha, m}{\alpha, m} B \ket{ \alpha, m}  \right)   .
\end{align}
It can be nonzero, suggestively of quantum memory, we explain in upcoming work~\cite{Static_corr}. The dynamical correlator has the form
\begin{align}
   \label{eq_S_w_app_help1bi}
   \bar{C}_{AB}^\dyn (\Omega)
   & \coloneqq 2 \pi \sum_{\alpha' \neq \alpha}
   \bra{\alpha, m} A \ketbra{\alpha', m+ q}{\alpha', m + q}  B  \ket{\alpha, m}  \,
   \delta \left( \Omega - \left[ E_{\alpha'} - E_\alpha \right] \right) .
\end{align}

Let us fine-grain the dynamical correlator. As in App.~\ref{app_Thermo_KMS}, we introduce Kronecker deltas $\delta_{m'  \, (m + \Delta m)}$ and 
$\delta_{s_{\alpha'} \, (s_\alpha + \Delta s)} \, .$ 
A selection rule has mapped $m'$ to $m + q$, so the first Kronecker delta becomes $\delta_{(\Delta m) \, q} \, .$
The rest of this appendix concerns the \emph{fine-grained dynamical correlator},
\begin{align}
   \label{eq_S_w_app_help1b}
   \hat{\bar{C}}_{AB}^\dyn (\Omega, \Delta m, \Delta s)
   & \coloneqq 2 \pi \sum_{\alpha' \neq \alpha}
   \bra{\alpha, m} A \ketbra{\alpha', m+ q}{\alpha', m + q}  B  \ket{\alpha, m}  \,
   \delta \left( \Omega - \left[ E_{\alpha'} - E_\alpha \right] \right)  \,
   \delta_{(\Delta m) \, q}  \,
   \delta_{s_{\alpha'} \, (s_\alpha + \Delta s)} \, .
\end{align}

\subsubsection{Application of Wigner–Eckart theorem and non-Abelian ETH to the fine-grained correlator} 
\label{app_Corr_WE_NAETH}

The Wigner–Eckart theorem appears in the main text as Eq.~\eqref{eq_WE}; and the non-Abelian ETH, as Eq.~\eqref{eq_NAETH}. We apply these equations to the dynamical correlator~\eqref{eq_S_w_app_help1b}:
\begin{align}
   \label{eq_S_w_dis_help1} &
   \hat{\bar{C}}_{AB}^\dyn (\Omega, \Delta m, \Delta s)  
   = 2 \pi \sum_{\alpha' \neq \alpha} 
   \braket{ s_\alpha \, , m }{ s_{\alpha'} \, , m + q ; k' , -q }
   \braket{ s_{\alpha'} \, , m+ q}{ s_\alpha \, , m ; k , q } 
   \\ & \nonumber \times
   \left[ \mathcal{T}^{(A)} \left( E_\alpha \, , s_\alpha  \right) \,
           \delta_{\alpha \alpha'}
           + e^{- S_\th  \left(  \frac{E_\alpha + E_{\alpha'} }{2} \, , 
                                         \frac{ s_\alpha + s_{\alpha'} }{2}  \right) / 2 }  \,
              f^{(A)}_{-\nu_{\alpha \alpha'} }  
              \left(  \frac{E_\alpha + E_{\alpha'} }{2} \, , 
                       \frac{ s_\alpha + s_{\alpha'} }{2} \, , 
                        - \omega_{\alpha \alpha'}  \right)
              R^{(A)}_{\alpha \alpha'}  \right]
   \\ & \nonumber \times
   \left[ \mathcal{T}^{(B)} \left( E_\alpha \, , s_\alpha  \right) \,
           \delta_{\alpha \alpha'}
           + e^{- S_\th  \left(  \frac{E_\alpha + E_{\alpha'} }{2} \, , 
                                         \frac{ s_\alpha + s_{\alpha'} }{2}  \right) / 2 }  \,
              f^{(B)}_{\nu_{\alpha \alpha'} }  
              \left(  \frac{E_\alpha + E_{\alpha'} }{2} \, , 
                                \frac{ s_\alpha + s_{\alpha'} }{2} \, , 
                                \omega_{\alpha \alpha'}  \right)
              R^{(B)}_{\alpha' \alpha}  \right]  \,
   \\ & \nonumber \times
   \delta \left( \Omega - \omega_{\alpha' \alpha} \right)  \,
   \delta_{(\Delta m) \, q}  \,
   \delta_{s_{\alpha'} \, (s_\alpha + \Delta s)}  \, .
\end{align}
We have replaced the Dirac delta function's $E_{\alpha'} - E_\alpha$ with $\omega_{\alpha' \alpha}$. This replacement motivated the main-text definitions
$\omega_{\alpha' \alpha} \coloneqq E_{\alpha'} - E_\alpha$ and
$\nu_{\alpha' \alpha} \coloneqq  s_{\alpha'} - s_\alpha \, ,$
despite the right-hand sides' equaling the negatives of the $\omega$ and $\nu$ definitions in~\cite{MurthyNAETH}.

Let us multiply out the $(\ldots + \ldots) \times (\ldots + \ldots)$ in Eq.~\eqref{eq_S_w_dis_help1}. Four terms result; three vanish due to the summand's $\alpha' \neq \alpha$. To neaten what remains, we package factors into two auxiliary functions. One is a product of Clebsch–Gordan coefficients:
\begin{align}
   \label{eq_CG_prod_app}
   \mathcal{C} ( \tilde{\nu} | s_{\tilde{\alpha}} \, , \tilde{m} , \tilde{k} , \tilde{k'}, \tilde{q} )
   \coloneqq \braket{ s_{\tilde{\alpha}} \, , \tilde{m} }{ s_{\tilde{\alpha}} + \tilde{\nu} , \tilde{m} + \tilde{q} ; \tilde{k'} , - \tilde{q} }
   \braket{ s_{\tilde{\alpha}} + \tilde{\nu} , \tilde{m} + \tilde{q} }{ s_{\tilde{\alpha}} \, , \tilde{m} ; \tilde{k} , \tilde{q} } .
\end{align}
We have singled out the $\tilde{\nu}$ in the left-hand side because we will sum over it later. The second auxiliary function is a product of $f$s:
\begin{align}
   \label{eq_f_prod_app}
   \mathcal{F}_{T T'} \left( \tilde{E} , \tilde{s} ; \tilde{\omega}, \tilde{\nu}  \right)
   \coloneqq f^{(T)}_{ - \tilde{\nu} }  \left( \tilde{E}, \tilde{s}, - \tilde{\omega} \right)
   f^{(T')}_{ \tilde{\nu} } \left( \tilde{E}, \tilde{s}, \tilde{\omega} \right) .
\end{align}
Equation~\eqref{eq_S_w_dis_help1} simplifies to
\begin{align}
   \label{eq_S_w_dis_help1b} 
   \hat{\bar{C}}_{AB}^\dyn (\Omega, \Delta m, \Delta s)  
   & = 2 \pi  \sum_{\alpha'  \neq  \alpha} 
   \mathcal{C} ( \nu_{\alpha' \alpha} | s_\alpha \, , m , k , k' , q )  \,
   e^{-S_\th  \left( \frac{E_\alpha + E_{\alpha'} }{2} \, , 
                            \frac{ s_\alpha + s_{\alpha'} }{2}  \right) }  
   \mathcal{F}_{AB}   \left( \frac{E_\alpha + E_{\alpha'} }{2} \, , 
                                          \frac{ s_\alpha + s_{\alpha'} }{2} \, ;
                                          \omega_{\alpha \alpha'} , 
                                          \nu_{\alpha \alpha'}  \right) \,
   \nonumber \\ & \qquad \qquad \quad \times
   R^{(A)}_{\alpha \alpha'}  \,  R^{(B)}_{\alpha' \alpha} \,   
   \delta \left( \Omega - \omega_{\alpha' \alpha} \right)  \,
   \delta_{(\Delta m) \, q}  \,
   \delta_{s_{\alpha'} \, (s_\alpha + \Delta s)}  \, .
\end{align}

We define a third auxiliary function in terms of the product $R^{(A)} R^{(B)}$. The motivation is, $R^{(A)} R^{(B)}$ is the only random variable in Eq.~\eqref{eq_S_w_dis_help1b}; the other factors are smooth functions. We expect the average of  $R^{(A)} R^{(B)}$  to vary smoothly, too~\cite{Noh_20_Numerical}. Therefore, we decompose $R^{(A)} R^{(B)}$ into its average and its deviation therefrom. First, we define the average over the eigenstate pairs 
$(\ket{ \tilde{\alpha}, \tilde{m} },  \ket{ \tilde{\alpha}',  \tilde{m}' })$ whose energy and spin quantum numbers participate in appropriate sums and differences:
\begin{align}
   \label{eq_R_Def_app}
   \mathcal{R}_{TT'} \left( \tilde{E}, \tilde{s}; \tilde{\omega}, \tilde{\nu} \right)
   \coloneqq  \expval{ R^{(T)}_{ \tilde{\alpha} , \tilde{\alpha}' }
                                   R^{(T')}_{ \tilde{\alpha}', \tilde{\alpha} }  }_{
   \frac{ E_{ \tilde{\alpha} } + E_{ \tilde{\alpha}' } }{2}  =  \tilde{E} , \:
   \frac{ s_{ \tilde{\alpha} } + s_{ \tilde{\alpha}' } }{2}  =  \tilde{s} , \:
   E_{ \tilde{\alpha}' } - E_{ \tilde{\alpha} }  =  \tilde{\omega} , \:
   s_{ \tilde{\alpha}' } - s_{ \tilde{\alpha} }  =  \tilde{\nu}
   } \, .
\end{align}
Now, we estimate how much
$R^{(A)}_{\alpha \alpha'}  \,  R^{(B)}_{\alpha' \alpha}$ 
deviates from its average.
$R^{(A/B)}_{\alpha \alpha'}$, analyzed alone, resembles a Gaussian-distributed random variable of zero mean and unit standard deviation across a small energy window at fixed spin quantum numbers~\cite{24_Lasek_Numerical}. However, 
$R^{(A)}_{\alpha \alpha'}$ may share correlations with 
$R^{(B)}_{\alpha' \alpha} \, .$
We therefore expect the deviation to be a random variable of zero mean and $O(1)$ standard deviation across a small window. The relative deviation (the deviation divided by $\mathcal{R}_{AB}$), summed over $\sum_{\alpha'}$, obeys the central limit theorem: it scales as 
(number of terms)$^{-1/2}$
$= D_\tot^{-1/2} (E_{\alpha'} \, s_{\alpha'} )  \, .$ 
($D_\tot$ is defined at the beginning of App.~\ref{sec_Derive_FDT_SU2}.)
We expect this scaling to characterize also the summed deviation divided by (i) $\mathcal{R}_{AB}$ and (ii) the prefactors of the $R^{(A)}_{\alpha \alpha'}  \,  R^{(B)}_{\alpha' \alpha}$ in Eq.~\eqref{eq_S_w_dis_help1b}.\footnote{
One might be concerned because the prefactors include not only smooth functions, but also $e^{-S_\th}$. Exponentials of entropies are infamous for peaking sharply. However, the $S_\th$ remains constant as $\alpha'$ varies: $S_\th$ is a function of $E_\alpha$, which is fixed implicitly in the equation's left-hand side; $s_\alpha$, which is fixed likewise; $E_{\alpha'}$, which is fixed by $E_\alpha$, the Dirac delta function, and the $\Omega$ in the equation's left-hand side; and $s_{\alpha'}$, which is fixed analogously.}
Substituting into Eq.~\eqref{eq_S_w_dis_help1b} therefore yields
\begin{align}
   \label{eq_S_w_dis_help2} &
   \hat{\bar{C}}_{AB}^\dyn (\Omega, \Delta m, \Delta s) 
   = 2 \pi  \sum_{\alpha'  \neq  \alpha}  
   \mathcal{C} ( \nu_{\alpha' \alpha} | s_\alpha \, , m , k , k' , q )  \,
   e^{-S_\th  \left( \frac{E_\alpha + E_{\alpha'} }{2} \, , 
                            \frac{ s_\alpha + s_{\alpha'} }{2}  \right) }  \,
   \mathcal{F}_{AB}   \left( \frac{E_\alpha + E_{\alpha'} }{2} \, , 
                                          \frac{ s_\alpha + s_{\alpha'} }{2} \, ;
                                          \omega_{\alpha \alpha'} , 
                                          \nu_{\alpha \alpha'}  \right) \,
   \nonumber \\ & \quad \times 
   \mathcal{R}_{AB} \left(  \frac{E_\alpha + E_{\alpha'}}{2} \, ,
   \frac{s_\alpha + s_{\alpha'}}{2} \, ; 
           \omega_{\alpha' \alpha} \, , 
           \nu_{\alpha' \alpha}  \right)  \,
   \delta \left( \Omega - \omega_{\alpha' \alpha} \right)  \,
   \delta_{(\Delta m) \, q}  \,
   \delta_{s_{\alpha'} \, (s_\alpha + \Delta s)}  \, 
   \left[ 1 +  D_\tot^{-1/2}  \left(  E_{\alpha'} \, , s_{\alpha'}  \right)  \right]  \, .
\end{align}

We eliminate the Dirac delta function as follows. Let us replace the
$\sum_{\alpha' \neq \alpha}$  with  $\sum_{\alpha'}  (1 - \delta_{\alpha' \alpha})$.
As explained below Eq.~\eqref{eq_S_t_2_app}, 
$\sum_{\alpha'}$ represents a sum over the energy eigenstates $\ket{\alpha', m+q}$ labeled by the magnetic spin quantum number $m+q$. Denote the set of these states by
$\mathcal{S} \coloneqq \{ \ket{\alpha' , m+q} \}$.
The set is a union of subsets: if $N$ is an even integer,
$\mathcal{S} = \mathcal{S}_0  \cup  \mathcal{S}_1  \cup  \ldots$\footnote{
If $N$ is an odd number,
$\mathcal{S} = \mathcal{S}_{1/2}  \cup  \mathcal{S}_{3/2}  \cup  \ldots$}
Each $\mathcal{S}_{s_{\alpha'}}$ consists of the energy eigenstates labeled by the spin quantum number $s_{\alpha'}$ and the magnetic spin quantum number $m+q$.
Hence $\sum_{\alpha'}  
=  \sum_{s_{\alpha'}}  
\sum_{ \ket{\alpha', m+q} \in \mathcal{S}_{s_{\alpha'}} } \, $. In the thermodynamic limit, the second sum becomes an integral over eigenenergies, weighted by the density of states within $\mathcal{S}_{s_{\alpha'}}$:\footnote{
The significances of the symbols $\alpha'$ and $s_{\alpha'}$ change slightly, as explained in footnote~\ref{foot_alpha}.}
\begin{align}
   \sum_{\alpha'} 
   =  \sum_{s_{\alpha'}}  
   \sum_{ \ket{\alpha', m+q} \in \mathcal{S}_{s_{\alpha'}} }
   \to  \sum_{s_{\alpha'}}  
   \int  dE  \;  e^{ S_\th (E, s_{\alpha'} ) } \, .
\end{align}
Accordingly, each $E_{\alpha'}$ in Eq.~\eqref{eq_S_w_dis_help2} becomes an $E$. The $\delta( \Omega - \omega_{\alpha' \alpha} )$ becomes a
$\delta( \Omega - [E - E_\alpha] )$. When we integrate, $E$ becomes $E_\alpha + \Omega$. :
%
\begin{align}
   & \hat{\bar{C}}_{AB}^\dyn (\Omega, \Delta m, \Delta s) 
   \label{eq_S_w_app_help3}
   = 2 \pi  \sum_{s_{\alpha'}}  
   \left( 1 - \delta_{\alpha \alpha'}  \right)
   e^{ S_\th ( E_\alpha + \Omega , s_{\alpha'} ) }  
   \mathcal{C} ( \nu_{\alpha' \alpha} | s_\alpha \, , m , k , k' , q ) \,
   e^{ - S_\th \left( E_\alpha + \frac{\Omega}{2} \, , 
                             \frac{s_\alpha + s_{\alpha'} }{2}  \right) }  \,
   \\ & \nonumber \qquad \qquad  \;  \times
   \mathcal{F}_{AB}  \left( E_\alpha + \frac{\Omega}{2} \, , 
                                          \frac{s_\alpha + s_{\alpha'} }{2}  \, ; 
                                         -\Omega , \nu_{\alpha \alpha'}  \right) \,
   \mathcal{R}_{AB}  \left(  E_\alpha + \frac{\Omega}{2}  \, ,
     \frac{s_\alpha + s_{\alpha'} }{2} \, ;  
     \Omega  \, ,
     \nu_{\alpha' \alpha}  \right)
   \delta_{(\Delta m) \, q}  \,
   \delta_{s_{\alpha'} \, (s_\alpha + \Delta s)}  
   \\ & \nonumber \qquad \qquad  \;  \times
   \left[ 1 +  D_\tot^{-1/2} \left( E_\alpha + \frac{\Omega}{2} \, , s_{\alpha'}  \right)   \right]   .
\end{align}

For ease of evaluation, we rewrite Eq.~\eqref{eq_S_w_app_help3} in six steps detailed in the next several paragraphs. First, we express the sum over $s_{\alpha'}$ as a sum over $\nu_{\alpha' \alpha} \, .$ Second, we perform the sum. Third, we replace $s_{\alpha'}$ with $s_\alpha + \nu_{\alpha' \alpha}$ everywhere else. Fourth, we eliminate the $1 - \delta_{\alpha \alpha'} \, .$ Fifth, we replace the thermodynamic entropies $S_\th$ with densities $D$ of states. 
Sixth, we package $\mathcal{F}_{AB}$ and $\mathcal{R}_{AB}$ into a variable $\mathcal{G}_{AB}$ that exhibits useful symmetries. 

Let us express the sum over $s_{\alpha'}$, in Eq.~\eqref{eq_S_w_app_help3}, as a sum over $\nu_{\alpha' \alpha}$. Equation~\eqref{eq_S_w_app_help1b} bounds $\nu_{\alpha' \alpha}$, implicitly containing Clebsch–Gordan coefficients subject to selection rules:
\begin{align}
   \label{eq_Bound_nu}
   \begin{cases}
   \bra{\alpha, m}  A^{(k')}_{-q}  \ket{\alpha', m + q }
   \quad \Rightarrow \quad
   s_\alpha \in \left\{ | s_{\alpha'} - k' | ,  | s_{\alpha'} - k' | + 1 , 
   \ldots , s_{\alpha'} + k \right\}
   \quad \Rightarrow \quad
   - k' \leq s_{\alpha'} - s_\alpha  \leq  k'  \\
   \bra{\alpha' , m + q } B^{(k)}_q \ket{\alpha, m}
   \quad \Rightarrow \quad
   s_{\alpha'}  \in  \left\{  | s_\alpha - k | ,  | s_\alpha - k | +  1 , 
   \ldots ,  s_\alpha + k  \right\}
   \quad  \Rightarrow  \quad
   - k  \leq  s_{\alpha'} - s_\alpha  \leq  k .
   \end{cases}
\end{align}
These inequalities condense into
$- \min \{k, k' \} \leq \nu_{\alpha' \alpha} \leq \min \{k, k' \}$.
Now, we perform the sum over $\nu_{\alpha' \alpha} \, .$ This variable becomes $\Delta s$, due to the $\delta_{s_{\alpha'} \, (s_\alpha + \Delta s)}$ in Eq.~\eqref{eq_S_w_app_help3}. 

Once we eliminate $\alpha'$ from Eq.~\eqref{eq_S_w_app_help3}, the
$1 - \delta_{\alpha \alpha'}$ has no meaning. Conveniently, its effect on Eq.~\eqref{eq_S_w_app_help3} is exponentially small in $N$: the 
$1 - \delta_{\alpha \alpha'}$ excludes only one term from the sum, which contains $O(2^N)$ other terms. We therefore eliminate the $1 - \delta_{\alpha \alpha'} \, ,$ incurring an exponentially small correction.

Let us replace the thermodynamic entropies $S_\th$ in Eq.~\eqref{eq_S_w_app_help3}. $S_\th (E, s)$ equals the logarithm of the density $D(E, s) / (2s+1)$
of states at energy $E$, spin quantum number $s$, and magnetic spin quantum number $m = 0$ 
(see Sec.~\ref{sec_Review_NAETH}, the beginning of App.~\ref{sec_Derive_FDT_SU2}, and~\cite{24_Lasek_Numerical}). 
In our final rewriting, we package two smooth functions together:
\begin{align}
   \label{eq_G_Def_app}
   \mathcal{G}_{T T'}  \left(  \tilde{E}, \tilde{s} ; \tilde{\Omega} , \tilde{\nu}  \right)
   \coloneqq  \mathcal{F}_{T T'}  \left(  \tilde{E}, \tilde{s} ; -\tilde{\Omega} , -\tilde{\nu}  \right) \,
   \mathcal{R}_{T T'}  \left(  \tilde{E}, \tilde{s} ; \tilde{\Omega} , \tilde{\nu}  \right) .
\end{align}
We have introduced three calligraphic-font smooth functions that obey a symmetry relation: 
\begin{align}
   \label{eq_Sym_app}
   \text{If} \;  \chi  \in  \{ \mathcal{F}, \mathcal{R}, \mathcal{G} \}, \;
   \text{then}  \;
   \chi_{T T'} \left(  \tilde{E}, \tilde{s} ; \tilde{\Omega} , \tilde{\nu}  \right)
   =  \chi_{T' T}  \left(  \tilde{E}, \tilde{s} ; - \tilde{\Omega} , - \tilde{\nu}  \right) .
\end{align}
Let us apply all six rewritings to Eq.~\eqref{eq_S_w_app_help3}: to within an exponentially-small-in-$N$ correction,
\begin{align}
   \hat{\bar{C}}_{AB}^\dyn (\Omega, \Delta m, \Delta s)
   \label{eq_S_w_dynamic_app_1}
   & = 2 \pi  \,
   \mathcal{C} ( \Delta s | s_\alpha \, , m , k , k' , q ) \,
   \frac{ D_\tot \left( E_\alpha + \Omega , 
                               s_\alpha + \Delta s  \right) }{ 
              D_\tot \left( E_\alpha + \frac{\Omega}{2} \, , 
                                 s_\alpha + \frac{\Delta s}{2}  \right)  }  \,
     \mathcal{G}_{AB}  \left(  E_\alpha + \frac{\Omega}{2}  \, ,
     s_\alpha + \frac{\Delta s}{2} \, ;  
     \Omega  \, ,  \Delta s  \right)  \,
   \delta_{(\Delta m) \, q} \, .
\end{align}

\subsection{General finite-size corrections in the fine-grained correlator}
\label{app_Gen_correxn}

From now on, we assume that $s_\alpha \, , E_\alpha  \gg  1$.
Let us Taylor-approximate the $\mathcal{G}_{AB}$ in Eq.~\eqref{eq_S_w_dynamic_app_1}. We will Taylor-approximate the density-of-states ratio, too; however, different approximation methods will prove useful in different subsections below.

Let us Taylor-approximate $\mathcal{G}_{AB}$ about $(E_\alpha, s_\alpha)$. We take the natural log of $\mathcal{G}_{AB}$, Taylor-approximate to first order, and exponentiate. This strategy yields a simple polynomial correction that is easy to analyze. To simplify notation, we define derivatives of $\ln \LParen \mathcal{G}_{AB} (\ldots) \RParen$:
\begin{align}
   \label{eq_Gamma_Def} &
   \text{If} \; \;  X \in \{ E, s \} ,  \; \;  \text{then}  \; \;
   \Gamma_X  
   \coloneqq \partial_X  \ln \LParen
   \mathcal{G}_{AB} \left( E, s; \Omega, \Delta s \right)
   \RParen  \big\rvert_{E_\alpha \, , \, s_\alpha} \, . 
\end{align}
We define derivatives $\Gamma_{XY}$ and $\Gamma_{XYZ}$ analogously. In terms of these derivatives,
\begin{align}
   \label{eq_Approx_G}
   \mathcal{G}_{AB} 
   \left( E_\alpha + \Omega / 2 , s_\alpha + \Delta s / 2 ; \Omega , \nu_{\alpha' \alpha} \right) 
   & = \mathcal{G}_{AB} \left( E_\alpha \, , s_\alpha \, ; \Omega , \Delta s \right) 
   \\ \nonumber & \; \times
   \exp  \left( \frac{1}{2}  \,  \Gamma_E \,  \Omega
                    + \frac{1}{2}  \,  \Gamma_s \,  \Delta s
                    + \frac{1}{8}  \left\{  \Gamma_{EE}  \Omega^2
                    + 2 \Gamma_{Es}  \Omega \, \Delta s
                    + \Gamma_{ss}  [\Delta s]^2  \right\} 
                    +  O \left( \Gamma_{XYZ}  \right) \right)  .
\end{align}

Let us substitute from Eq.~\eqref{eq_Approx_G} into Eq.~\eqref{eq_S_w_dynamic_app_1}:
\begin{align}
   \label{eq_S_w_dynamic_app_2}
   \hat{\bar{C}}_{AB}^\dyn (\Omega, \Delta m, \Delta s)
   & = 2 \pi  \,
   \mathcal{C} ( \Delta s | s_\alpha \, , m , k , k' , q ) \,
   \frac{ D_\tot \left( E_\alpha + \Omega , 
                               s_\alpha + \Delta s  \right) }{ 
              D_\tot \left( E_\alpha + \frac{\Omega}{2} \, , 
                                 s_\alpha + \frac{\Delta s}{2}  \right)  }  \,
   \mathcal{G}_{AB} \left( E_\alpha \, , s_\alpha \, ; \Omega , \Delta s \right) \,
   \delta_{(\Delta m) \, q}  
   \\ \nonumber & \quad \; \times
   \exp  \left( \frac{1}{2}  \,  \Gamma_E \,  \Omega
                    + \frac{1}{2}  \,  \Gamma_s \,  \Delta s
                    + \frac{1}{8}  \left\{  \Gamma_{EE}  \Omega^2
                    + 2 \Gamma_{Es}  \Omega \, \Delta s
                    + \Gamma_{ss}  [\Delta s]^2  \right\} 
                    +  O \left( \Gamma_{XYZ}  \right) \right) .
\end{align}

\subsection{Fine-grained KMS relation, not necessarily with any anomalous correction, for certain energy eigenstates}
\label{app_KMS_0th}

In this subsection, we prove the fine-grained KMS relation for energy eigenstates $\ket{\alpha, m}$, under two sets of conditions. We will not argue, here, that this KMS relation contains any anomalously large correction.

First, we Taylor-approximate the density-of-states ratio in the correlator~\eqref{eq_S_w_dynamic_app_2}. Define the first derivatives of $S_\th$ as
\begin{align}
   \label{eq_Sigma_Def}
   \text{If} \; \;  X \in \{ E, s \} , \; \;  \text{then} \; \;
   \Sigma_{X}  
   \coloneqq  \partial_X  S_\th (E, s) 
   \big\rvert_{E_\alpha, s_\alpha} \,  .
\end{align}
Define $\Sigma_{XY}$ and $\Sigma_{XYZ}$ analogously. By the beginning of App.~\ref{sec_Derive_FDT_SU2},
$D_\tot (E, s) = \frac{ \tilde{D}_\tot (E, s)}{2s + 1}
= \frac{\exp \LParen \tilde{S}_\th (E, s) \RParen}{2s + 1} \, .$
We Taylor-approximate the entropies:
\begin{align}
   \label{eq_Approx_ratio2_help1}
   & \frac{ D_\tot ( E_\alpha + \Omega , s_\alpha + \Delta s ) }{
            D_\tot ( E_\alpha + \Omega / 2 , s_\alpha + \Delta s / 2 ) }
   = \exp \left( \tilde{S}_\th \left( E_\alpha + \Omega , s_\alpha + \Delta s \right)
   - \tilde{S}_\th \left( E_\alpha + \Omega / 2 , s_\alpha + \Delta s / 2 \right)
   \right)  
   \\ \nonumber & \qquad \qquad \qquad \qquad \qquad \qquad \quad \; \times
   \frac{2 s_\alpha + \Delta s + 1}{ 2 s_\alpha + 2 \Delta s + 1} \\
   & = \exp \Bigg(
   \frac{1}{2} \partial_E  \tilde{S}_\th ( E, s_\alpha ) \big\lvert_{E_\alpha} \Omega
   + \frac{1}{2}  \partial_s  \tilde{S}_\th ( E_\alpha, s )  
      \big\lvert_{s_\alpha}  \Delta s
   + \frac{3}{8} \left[ \Sigma_{EE} \Omega^2  
                               +  2 \Sigma_{Es} \Omega (\Delta s)
                               + \Sigma_{ss} (\Delta s)^2  \right]
      \nonumber \\ & \qquad \qquad
      + O \left(  \Sigma_{XYZ}  \right)
   \Big)  \,
   \frac{2 s_\alpha + \Delta s + 1}{ 2 s_\alpha + 2 \Delta s + 1} ,
\end{align}
The final ratio approximates to 
$1  +  O( s_\alpha^{-1} ) \approx  \exp \LParen O( s_\alpha^{-1} ) \RParen$.
We package all the corrections into their own exponential.  Equation~\eqref{eq_E_deriv_help1}, with the similarity between $\rho_\NATS$ and $\ket{\alpha, m}$, implies
$\partial_E  \tilde{S}_\th (E, s) \lvert_{E_\alpha, \, s_\alpha}
      \approx  \beta$.
(The $\approx$ signs come from the zeroth-order approximation of $Z$ in App.~\eqref{app_Peak_vals}. We assume that corrections of this type are negligible.) Equation~\eqref{eq_Approx_ratio2_help1} becomes
\begin{align}
   \label{eq_Approx_ratio2} 
   \frac{ D_\tot ( E_\alpha + \Omega , s_\alpha + \Delta s ) }{
            D_\tot ( E_\alpha + \Omega / 2 , s_\alpha + \Delta s / 2 ) }
   & = \exp \left(  \frac{1}{2}  \left[  \beta \Omega
   + \partial_s  \tilde{S}_\th ( E, s )  
      \big\lvert_{E_\alpha \, , s_\alpha}  \Delta s  \right]  \right)
      \\ \nonumber & \quad \; \times
   \exp  \left(  \frac{3}{8} \left[ \Sigma_{EE} \Omega^2  
                               +  2 \Sigma_{Es} \Omega (\Delta s)
                               + \Sigma_{ss} (\Delta s)^2  \right]
      +  O \left( s_\alpha^{-1} \right)
      + O \left( \Sigma_{XYZ}  \right)  \right) .
\end{align}
Substituting into Eq.~\eqref{eq_S_w_dynamic_app_2} yields
\begin{align}
   \label{eq_S_w_dynamic_app_3} &
   \hat{\bar{C}}_{AB}^\dyn (\Omega, \Delta m, \Delta s)
   = 2 \pi  \,
   \mathcal{C} ( \Delta s | s_\alpha \, , m , k , k' , q ) \,
   \exp \left(  \frac{1}{2}  \left[  \beta \Omega
   + \partial_s  \tilde{S}_\th ( E, s )  
      \big\lvert_{E_\alpha \, , s_\alpha}  \Delta s  \right]  \right)
   \mathcal{G}_{AB} \left( E_\alpha \, , s_\alpha \, ; \Omega , \Delta s \right) \,
   \delta_{(\Delta m) \, q}  
   \nonumber \\ &  \times
   \exp  \Bigg( \frac{1}{2}  \,  \Gamma_E \,  \Omega
                    + \frac{1}{2}  \,  \Gamma_s \,  \Delta s
                    + \frac{1}{8}  \left\{  
                        \left[ \Sigma_{EE}  +  \Gamma_{EE}  \right]  \Omega^2
                       + 2  \left[ \Sigma_{Es}  +  \Gamma_{Es}  \right]  \Omega \, \Delta s
                       +  \left[ \Sigma_{ss}  +  \Gamma_{ss}  \right]  [\Delta s]^2  \right\} 
                    \nonumber \\ & \qquad \qquad
                    +  O \left( s_\alpha^{-1} \right)
                    +  O \left( \Gamma_{XYZ}  \right) 
                    + O \left( \Sigma_{XYZ}  \right)  \big) .
\end{align}

We must identify how $\hat{\bar{C}}_{AB}^\dyn (\Omega, \Delta m, \Delta s)$ [Eq.~\eqref{eq_S_w_dynamic_app_2}] transforms under
\begin{align}
   \label{eq_Trans_compare}
   B \leftrightarrow A, \quad
   \Omega  \mapsto - \Omega,  \quad
   \Delta m  \mapsto  - \Delta m,  \quad
   \Delta s  \mapsto - \Delta s .
\end{align}
$S_\th$ does not depend on any of the quantities in Eq.~\eqref{eq_Trans_compare}. Therefore, all derivatives of $S_\th$ remain constant. So do $\mathcal{G}$ and its derivatives, by Eq.~\eqref{eq_Sym_app}. Also under the transformation~\eqref{eq_Trans_compare}, 
$k \leftrightarrow k'$, $q \leftrightarrow -q$, and 
$\delta_{(\Delta m) q}  \mapsto  \delta_{(-\Delta m) (-q)}  =  \delta_{(\Delta m) q} \, .$ Hence the transformed correlator is
\begin{align}
   \label{eq_C_Trans_Help1} &
   \hat{\bar{C}}_{BA}^\dyn ( - \Omega, - \Delta m, - \Delta s)
   = 2 \pi \mathcal{C} ( - \Delta s | s_\alpha , m , k', k, - q)  \,
   \exp \left(  - \frac{1}{2}  \left[  \beta \Omega
   + \partial_s  \tilde{S}_\th ( E, s )  
      \big\lvert_{E_\alpha \, , s_\alpha}  \Delta s  \right]  \right)
   \mathcal{G}_{AB} \left( E_\alpha \, , s_\alpha \, ; \Omega , \Delta s \right) \,
   \nonumber \\ &  \times
   \delta_{(\Delta m) \, q}  
   \exp  \Bigg( - \frac{1}{2}  \,  \Gamma_E \,  \Omega
                    - \frac{1}{2}  \,  \Gamma_s \,  \Delta s
                    + \frac{1}{8}  \left\{  
                        \left[ \Sigma_{EE}  +  \Gamma_{EE}  \right]  \Omega^2
                       + 2  \left[ \Sigma_{Es}  +  \Gamma_{Es}  \right]  \Omega \, \Delta s
                       +  \left[ \Sigma_{ss}  +  \Gamma_{ss}  \right]  [\Delta s]^2  \right\} 
                    \nonumber \\ & \qquad \qquad \qquad \quad
                    +  O  \left( s_\alpha^{-1}  \right)
                    +  O \left( \Gamma_{XYZ}  \right) 
                    + O \left( \Sigma_{XYZ}  \right)  \big) .
\end{align}

We must compare the correlators~\eqref{eq_S_w_dynamic_app_3} and~\eqref{eq_C_Trans_Help1}. To do so, we introduce the logarithmic ratio
\begin{align}
   \label{eq_Logratio_Def_app}
    \tilde{\mathcal{{L}}}_{AB} (\Omega, q, \Delta s;  \alpha,m)
    \coloneqq \ln  \left(  \frac{  
    \hat{\bar{C}}^{\rm dyn}_{AB} (\Omega, \Delta m {=} q, \Delta s; \alpha, m)}{   \hat{\bar{C}}^{\rm dyn}_{BA} (-\Omega, -\Delta m {=} q, -\Delta s;
\alpha, m)}  \right) .
\end{align}
The correlators' $\delta_{(\Delta m) q}$s enforce the $\Delta m  {=} q$ condition on the formula's right-hand side. Once we substitute in, many factors cancel exactly:
\begin{align}
   \label{eq_Logratio_app_help2}
    \tilde{\mathcal{{L}}}_{AB} (\Omega, q, \Delta s;  \alpha,m)
    & = \beta \Omega  
    + \partial_s  \tilde{S}_\th ( E, s )  \big\lvert_{E_\alpha \, , s_\alpha}  \Delta s 
    + \ln \left(  \frac{ \mathcal{C} ( \Delta s | s_\alpha \, , m , k , k' , q ) }{ 
                               \mathcal{C} ( - \Delta s | s_\alpha , m , k', k, - q) }  \right) 
   \\ \nonumber & \quad  \;
   +  \Gamma_E \Omega  
   +  \Gamma_s  \,  \Delta s
    +  O  \left(  s_\alpha^{-1}  \right)
    + O \left( \Sigma_{XYZ}  \right) .
\end{align}

We wish to prove that the right-hand side has the form 
$\beta (\Omega - \mu m - \gamma s_\alpha) + (\text{correction})$.
To analyze the second and third terms in Eq.~\eqref{eq_Logratio_app_help2}, we make further assumptions about the parameter regime in Apps.~\ref{app_KMS_s_Large_m_Large} and~\ref{app_KMS_s_Large_m_0}. Here, we estimate the explicit corrections in Eq.~\eqref{eq_Logratio_app_help2}. We cannot know \emph{a priori} how $\Gamma_{X}$ and $\Sigma_{XYZ}$ scale (as emphasized in App.~\ref{app_KMS_Anom}). The exact scalings do not matter here, as emphasized at the beginning of this subsection. 

Therefore, we estimate these corrections crudely. First, $\Gamma_X$ is a function of $O(1)$ numbers. Hence we estimate
$\Gamma_{X}  =  O ( X^{-1} )$.
Second, $\Sigma_{XYZ}$ is a third derivative of $S_\th$, which often scales extensively. Hence we estimate
$\Sigma_{XYZ}  =  O ( N X^{-1} Y^{-1} Z^{-1} )$.  
Third, Eqs.~\eqref{eq_Bound_nu} bound $\Delta s = s_{\alpha'} - s_\alpha$ in terms of $k$ and $k'$, which are $O(1)$. Fourth, the non-Abelian ETH's $f$ function is sizable---and so the correlator 
$\hat{\bar{C}}^\dyn_{AB}$ is---only when $\Omega = O(1)$~\cite{24_Lasek_Numerical,DAlessio_16_From,Srednicki_99_Approach,Khatami_13_Fluctuation}. Therefore, we estimate the explicit correction in Eq.~\eqref{eq_Logratio_app_help2} to be
\begin{align}
   \label{eq_Estimate_Exp_Correxn}
   O \left( s_\alpha^{-1}  \right)
   + O \left( E_\alpha^{-1}  \right)
   + O \left( N E_\alpha^{-2}  \right)
   + O  \left( N s_\alpha^{-2}  \right)
   + O \left( N E_\alpha^{-1}  s_\alpha^{-1}  \right) .
\end{align}
The $O(s_\alpha^{-1})$ correction that appears explicitly in Eq.~\eqref{eq_Logratio_app_help2} [whose source we explained below Eq.~\eqref{eq_Approx_ratio2_help1}] may cancel between the log-ratio's numerator and denominator. ETHs often hold when additive conserved quantities are $O(N)$~\cite{DAlessio_16_From}, so the overall correction may often be $O(N^{-1})$. Regardless, the correction vanishes in the thermodynamic limit.

\subsubsection{Fine-grained KMS relation for energy eigenstates whose $s_\alpha, m = O \left( N^\zeta  \right)$, wherein $\zeta \in (0, 1]$, and $s_\alpha - m = O(1)$}
\label{app_KMS_s_Large_m_Large}

In the above-named parameter regime, we estimate the Clebsch–Gordan product $\mathcal{C}$. Then, we estimate the $\partial_s \tilde{S}_\th$ in the log-ratio~\eqref{eq_Logratio_app_help2}.

Let us estimate the Clebsch–Gordan product $\mathcal{C}$ in the log-ratio~\eqref{eq_Logratio_app_help2}. We leverage the following claim, labeled as Eqs.~(B11)--(B12) in~\cite{24_Lasek_Numerical}: 
   \begin{align}
      \label{eq_CG_prop_24a}  &
      \text{If} \; \tilde{s}, \tilde{m} = O \left(  N^\zeta  \right), 
      \;  \text{wherein}  \;   \zeta \in (0, 1],  \;  \text{and}  \;
      \;  \tilde{\nu}, \tilde{k}, \tilde{q}, \tilde{s} - \tilde{m} = O(1),
      \; \text{then} \\ &
      \label{eq_CG_prop_24b}
      \braket{ \tilde{s} + \tilde{\nu} , \tilde{m} + \tilde{q} }{
                   \tilde{s}, \tilde{m} ; \tilde{k} , \tilde{q}  }
      =  c \left( \tilde{\nu} , \tilde{k} , \tilde{q}, \tilde{s} - \tilde{m}  \right)  \,
      (\tilde{s})^{- | \tilde{\nu} - \tilde{q} | / 2}  \,
      \left[ 1 + O \left( \tilde{s}^{-1} \right)  \right] ,
      \; \text{wherein} \\ &
      \label{eq_CG_prop_24c}
      c \left( \tilde{\nu} , \tilde{k} , \tilde{q}, \tilde{s} - \tilde{m}  \right)
      =  \delta_{ \tilde{\nu} \tilde{q} }  
          +  \left( 1 - \delta_{ \tilde{\nu}  \tilde{q} }  \right)  \,  O(1) .
   \end{align}
   Our parameters satisfy the conditions~\eqref{eq_CG_prop_24a}.\footnote{
   Equations~\eqref{eq_Bound_nu} bound $\Delta s = s_{\alpha'} - s_\alpha$ in terms of $k$ and $k'$, which are $O(1)$.}
   We can therefore apply Eqs.~\eqref{eq_CG_prop_24b} and~\eqref{eq_CG_prop_24c}, with the $\mathcal{C}$ definition~\eqref{eq_CG_prod_app}, to the Clebsch–Gordan product in Eq.~\eqref{eq_Logratio_app_help2}:
   \begin{align}
      \label{eq_KMS_thermo_case1_help1}
      \mathcal{C} ( \Delta s | s_\alpha, m , k , k' , q ) 
      & = \left\{  \delta_{ (\Delta s) \, q }
      + O \left( \left[ 1 - \delta_{ (\Delta s) \, q}  \right]  \,
                     s_\alpha^{- | \Delta s - q | }  \right)  \right\}
      \left[ 1 + O \left( s_\alpha^{-1}  \right)  \right]  \\
      & =  \delta_{ (\Delta s)  q}  \,
            \left[ 1 + O \left( s_\alpha^{-1}  \right)  \right]
            +  \left( 1 - \delta_{ (\Delta s) q }  \right)  \,
                O  \left(  s_\alpha^{- O(1)}  \right) .
   \end{align}
This function remains invariant under the transformation~\eqref{eq_Trans_compare}. Hence the two Clebsch–Gordan products in~\eqref{eq_Logratio_app_help2} cancel. Yet consider substituting for the $\mathcal{C}$s in Eqs.~\eqref{eq_S_w_dynamic_app_3} and~\eqref{eq_C_Trans_Help1}. Each $\mathcal{C}$'s leading-order term is proportional to $\delta_{(\Delta s) q} \, ,$ which multiplies a $\delta_{(\Delta m) q} \, .$ The following consequence will affect our result:
\begin{align}
   \label{eq_Delta_s_m}
   \Delta s = \Delta m .
\end{align}

Now, we can apply the $\rho_\NATS$ property~\eqref{eq_Big_params_thermo_1} to energy eigenstates $\ket{\alpha, m}$. This application is justified only if $s_\alpha$, $m$, and $E_\alpha$ satisfy Eqs.~\eqref{eq_NATS_eigen_correspondence}. $s_\alpha$ and $m$ do by assumption [by the title of this subsubsection and by the boldface text above Eq.~\eqref{eq_Big_params_thermo_1}] and by Eq.~\eqref{eq_Sz_bm_large}.
Hence
\begin{align}
   \label{eq_partial_s_help1}
   \partial_s \tilde{S}_\th (E, s) \lvert_{E_\alpha, \, s_\alpha}
   =  - \beta (\mu + \gamma )  +  O( s_\alpha^{-1} ) .
\end{align}
   
Let us substitute from Eqs.~\eqref{eq_KMS_thermo_case1_help1},~\eqref{eq_Delta_s_m},~\eqref{eq_partial_s_help1}, and~\eqref{eq_Estimate_Exp_Correxn} into the log-ratio~\eqref{eq_Logratio_app_help2}:
\begin{align}
   \label{eq_Logratio_app_help3}
    \tilde{\mathcal{{L}}}_{AB} (\Omega, q, \Delta s;  \alpha,m)
    & = \beta ( \Omega  -  \mu \, \Delta m  -  \gamma \,  \Delta s )  
    + O \left( s_\alpha^{-O(1)}  \right)
   + O \left( E_\alpha^{-1}  \right)
   \nonumber \\ & \quad \; 
   + O \left( N E_\alpha^{-2}  \right)
   + O  \left( N s_\alpha^{-2}  \right)
   + O \left( N E_\alpha^{-1}  s_\alpha^{-1}  \right) .
\end{align}
We have proved the fine-grained KMS relation.

\subsubsection{Fine-grained KMS relation for energy eigenstates whose 
$s_\alpha = O(N^\zeta)$, wherein $\zeta \in (0, 1]$, and $m = 0$}
\label{app_KMS_s_Large_m_0}

Again, we approximate the $\mathcal{C}$ in the log-ratio~\eqref{eq_Logratio_app_help2}. Then, we approximate the $\tilde{S}_\th$ derivative.

Equation~\eqref{eq_CG_prod_app} defines $\mathcal{C}$ in terms of two Clebsch--Gordan coefficients, approximated as follows in App.~\ref{app_CG_property}:
   \begin{align}
      \text{If} \; 
      \tilde{s} = O \left( N^\zeta \right) , \;
      \zeta \in (0, 1 ] ,
      \; \text{and} \;
      \tilde{k}, \tilde{q}, \tilde{m}  =  O(1) ,
      \; \text{then}  \;
      \braket{ \tilde{s} + \tilde{\nu} , \tilde{m} + \tilde{q} }{  
                   \tilde{s} , \tilde{m} ; \tilde{k} , \tilde{q} }
      =  \bar{c} \left(  \tilde{\nu} ; \tilde{k} , \tilde{q}  \right)  \,
      \left[ 1 + O \left( \tilde{s}^{-1}  \right)  \right] .
   \end{align} 
   Appendix~\ref{app_CG_property} specifies the form of 
   $\bar{c}  (  \tilde{\nu} ; \tilde{k} , \tilde{q}  )$.
   Therefore, the Clebsch--Gordan product in Eq.~\eqref{eq_S_w_dynamic_app_2} approximates to
   \begin{align}
      \mathcal{C} ( \Delta s | s_\alpha , m , k , k' , q )
      & = \bar{c} ( -\Delta s ; k' , -q )  \,
      \bar{c} ( \Delta s; k, q )  \,
      \left[ 1 + O \left( s_\alpha^{-1}  \right)  \right] 
      \label{eq_C_case2_help1}
      =  \mathcal{C} ( - \Delta s | s_\alpha, m, k', k, -q)
      \left[ 1 + O \left( s_\alpha^{-1}  \right)  \right] .
   \end{align}
   
Now, we estimate the $\tilde{S}_\th$ derivative in the log-ratio~\eqref{eq_Logratio_app_help2}.
Since $m = 0$ by assumption, $\mu = 0$. Therefore, by the similarity between $\rho_\NATS$ and $\ket{\alpha, m}$, Eq.~\eqref{eq_S_bm0} implies
   $\partial_s  \tilde{S}_\th (E, s) |_{ E_\alpha, \, s_\alpha}  \approx  - \beta \gamma$.

Let us substitute from the foregoing equation, Eq.~\eqref{eq_C_case2_help1}, and Eq.~\eqref{eq_Estimate_Exp_Correxn} into Eq.~\eqref{eq_Logratio_app_help2}:
\begin{align}
   \label{eq_L_case2_app}
   \tilde{\mathcal{{L}}}_{AB} (\Omega, q, \Delta s;  \alpha, m {=} 0)
    = \beta ( \Omega  -  \gamma \,  \Delta s )  
    + O \left( s_\alpha^{-1}  \right)
   + O \left( E_\alpha^{-1}  \right)
   + O \left( N E_\alpha^{-2}  \right)
   + O  \left( N s_\alpha^{-2}  \right)
   + O \left( N E_\alpha^{-1}  s_\alpha^{-1}  \right) .
\end{align}
Again, we have proved the fine-grained KMS relation.

\subsection{Fine-grained KMS relation, with anomalously large correction, for certain energy eigenstates}
\label{app_KMS_Anom}

We argue in App.~\ref{app_KMS_Anom_If_Then} that, if $s_\alpha = O (N^{ \zeta \in (0, 1) })$ and $m = q = \beta = 0$, the fine-grained KMS relation contains an anomalously large correction. In App.~\ref{app_KMS_Anom_argue_for_s_scaling}, we argue that $s_\alpha$ can be of $O (N^{ \zeta \in (0, 1) })$. In fact, $s_\alpha = O( N^{1/2} )$ is typical.

\subsubsection{If $s_\alpha = O (N^{ \zeta \in (0, 1) })$ and $m = q = \beta = 0$, the fine-grained KMS relation contains an anomalously large correction}
\label{app_KMS_Anom_If_Then}

Let us return to the relatively general formula~\eqref{eq_S_w_dynamic_app_2} for the correlator $\hat{\bar{C}}_{AB}^\dyn$. Again, we approximate the ratio of $D_\tot$s. In App.~\ref{app_KMS_0th}, we expressed the $D_\tot$s in terms of $\tilde{S}_\th$, which we Taylor-approximated. Here, we invoke
$D_\tot (E, s)  
=  \frac{ \exp \LParen S_\th (E, s) \RParen }{ 2s + 1 }$
and Taylor-approximate the $S_\th$s:
\begin{align}
   & \frac{ D_\tot ( E_\alpha + \Omega , s_\alpha + \Delta s ) }{
            D_\tot ( E_\alpha + \Omega / 2 , s_\alpha + \Delta s / 2 ) }
   = \exp \LParen S_\th \left( E_\alpha + \Omega , s_\alpha + \Delta s \right)
   - S_\th \left( E_\alpha + \Omega / 2 , s_\alpha + \Delta s / 2 \right)
   \RParen \\
   \label{eq_Approx_ratio2b}
   & = \exp \Bigg(
   \frac{1}{2}  \Sigma_E  \Omega
   + \frac{1}{2}  \Sigma_s  \,  \Delta s
   + \frac{3}{8} \left[ \Sigma_{EE} \Omega^2  
                               +  2 \Sigma_{Es} \Omega (\Delta s)
                               + \Sigma_{ss} (\Delta s)^2  \right]
      + O \left( \Sigma_{XYZ}  \right)
   \Big)  .
\end{align}
Substituting into Eq.~\eqref{eq_S_w_dynamic_app_2} yields
\begin{align}
   \label{eq_C_Anom_help1} &
   \hat{\bar{C}}_{AB}^\dyn (\Omega, \Delta m, \Delta s)
   = 2 \pi  \,
   \mathcal{C} ( \Delta s | s_\alpha \, , m , k , k' , q ) \,
   \exp \left(
   \frac{1}{2}  \left[ \Sigma_E + \Gamma_E \right]  \Omega
   + \frac{1}{2}  \left[ \Sigma_s + \Gamma_s \right]  \,  \Delta s   \right)
   \mathcal{G}_{AB} \left( E_\alpha \, , s_\alpha \, ; \Omega , \Delta s \right) \,
   \delta_{(\Delta m) \, q}  
   \\ \nonumber &  \times
   \exp \left(  \frac{1}{8} \left\{  
           \left[ 3 \Sigma_{EE} + \Gamma_{EE}  \right]  \Omega^2
           + 2 \left[ 3 \Sigma_{Es} + \Gamma_{Es}  \right]  \Omega  \,  \Delta s
           + \left[ 3 \Sigma_{ss} + \Gamma_{ss}  \right]  [\Delta s]^2  \right\}
           +  O \left( \Sigma_{XYZ}  \right)
                    +  O \left( \Gamma_{XYZ}  \right)
   \right) .
\end{align}

To progress further, we focus on a parameter regime:
\begin{align}
   \label{eq_Param_regime_app}
   s_\alpha = O \left( N^\zeta \right) ,
   \quad \text{wherein} \quad
   \zeta \in (0, 1],
   \quad \text{and} \quad
   m = q = 0.
\end{align}
Under these conditions [and $k, k' = O(1)$, as assumed earlier], the Clebsch–Gordan product $\mathcal{C}$ exhibits the symmetry
\begin{align}
   \label{eq_C_symm_appB}
   \mathcal{C} (\Delta s | s_\alpha, 0, k, k', 0)
   =  \mathcal{C} (- \Delta s | s_\alpha, 0, k', k, 0 )  \, 
   \left[ 1  +  O \left(  s_\alpha^{-2}  \right)  \right] .
\end{align}
This property follows from Eqs.~(C12),~(C22), and~(C23b) of~\cite{Messiah_book}. [If $m$ or $q$ is nonzero, the correction is $O(s_\alpha^{-1})$, by App.~\eqref{app_CG_property}.]

The logarithmic ratio~\eqref{eq_Logratio_Def_app} quantifies the correction in the fine-grained KMS relation. Let us substitute in from Eq.~\eqref{eq_C_Anom_help1}, from below Eq.~\eqref{eq_Param_regime_app}, and from Eq.~\eqref{eq_C_symm_appB}:
\begin{align}
   \label{eq_L_eval_help1_app}
    \tilde{\mathcal{{L}}}_{AB} (\Omega, 0, \Delta s;  \alpha, 0)
    =  \left( \Sigma_E + \Gamma_E \right) \Omega  
    +  \left( \Sigma_s + \Gamma_s \right)  \Delta s
    +  O \left( s_\alpha^{-2}  \right)
    +  O  \left(  \Sigma_{XYZ}  \right)
    +  O  \left(  \Gamma_{XYZ}  \right) .
\end{align}

We aim to estimate how each term scales as $N$ grows. To do so, we must stipulate how two intermediate variables grow: 
\begin{enumerate}

   \item $s_\alpha$: Consistently with~\eqref{eq_Param_regime_app}, we assume that 
   \begin{align}
      \label{eq_s_scale_app}
      s_\alpha = \lfloor \mathscr{S} N^\zeta \rfloor ,
   \end{align}
   for some constant $\mathscr{S}$. 

   \item $E_\alpha$ or $\beta$: We can address these options in two ways:
   \begin{enumerate}
   
      \item  \label{item_E}
      We can assume that $E_\alpha = \mathscr{E} N$, for some constant $\mathscr{E}$.
      
      \item \label{item_beta}
      We can assume that
   \begin{align}
      \label{eq_Sigma_s_beta_app}
      \partial_E  S_\th (E, s) |_{E_\alpha, s_\alpha}
      =  \partial_E  \tilde{S}_\th (E, s) |_{E_\alpha, s_\alpha}
      =  \beta ,
   \end{align}
   to within a correction of $O(2^{-N})$.
   The first equality follows from $S_\th(E, s) = \tilde{S}_\th (E, s) - \ln(2s + 1)$ 
   (beginning of App.~\ref{sec_Derive_FDT_SU2}). The final equality follows from (i) Eq.~\eqref{eq_E_deriv_help1} and (ii) the local similarity between $\rho_\NATS$ and $\ket{\alpha, m}$ under the conditions~\eqref{eq_Correspond}. However, the system does not necessarily have an $\ket{\alpha, m}$ that satisfies Eq.~\eqref{eq_Sigma_s_beta_app} at arbitrary $\beta$ values, because $N$ is finite. But as $N$ grows, however, the system's energy gaps shrink exponentially. Hence the correction is expected to be exponentially small.
   
   \end{enumerate}
   Option~\ref{item_E} resembles the microcanonical ensemble, which models a system of fixed energy; and option~\ref{item_beta} resembles the canonical ensemble, which models a system of fixed temperature~\cite{Callen_book}. We adopt option~\ref{item_beta}, as the KMS relation contains a $\beta$.
   
\end{enumerate}

Now, we can estimate how each term in Eq.~\eqref{eq_L_eval_help1_app} scales with $N$:
\begin{enumerate}

   \item $\Sigma_E$: By definition, $\Sigma_E$ equals the left-hand side of Equation~\eqref{eq_Sigma_s_beta_app}. Therefore, to within an exponentially small correction, $\Sigma_E  =  \beta$.

   \item $\Gamma_E$: One cannot know the form of $\Gamma_E$ \emph{a priori}. Yet ETHs' components tend to depend on energy smoothly through an argument $E_\alpha / N$. We posit that $\mathcal{G}_{AB}$ depends so, such that
   $\Gamma_E = O (N^{-1})$.

   \item $\Sigma_s$: We estimate $\Sigma_s$ in five steps. First, we Legendre-transform $S_\th (E_\alpha, s_\alpha)$ into $\bar{S}_\th (\beta, s_\alpha)$. Second, we assume $\beta = 0$. Under this condition, the $s_\alpha$-dependence of $\bar{S}_\th (\beta, s_\alpha)$ is known. Third, we approximate the $s_\alpha$ derivative with a finite difference. A polynomial in $s_\alpha$ and $N$ results. Fourth, we interrelate this polynomial with $\beta \gamma$, using the thermodynamic results in App.~\ref{sec_Thermo_props_2}. Fifth, we calculate $\beta \gamma$ in the thermodynamic limit, defined as $(\beta \gamma)_\infty$. Sixth, we estimate $\Sigma_s$ in terms of $(\beta \gamma)_\infty$ and $N$-dependent corrections.
   
   We wish to assess the $s_\alpha$-derivative of $S_\th (E_\alpha, s_\alpha)$. We do not know this function's $s_\alpha$-dependence. However, we know the $s_\alpha$ dependence of a Massieu function, which follows from Legendre-transforming the entropy~\cite[p.~151]{Callen_book}: 
   $\bar{S}_\th (\beta, s_\alpha)  
   \coloneqq  S_\th (E_\alpha, s_\alpha)  -  \beta E_\alpha$.
   Conveniently, the Massieu function's $s_\alpha$ derivative equals the entropy's (equals $\Sigma_s$).
   Let $\beta = 0$. The Massieu function has the form~\cite[Eq.~(15)]{JaeDongXXZ}\footnote{
   The formula here contains only one $2s_\alpha + 1$ factor, whereas Eq.~(15) of~\cite{JaeDongXXZ} contains two such factors. The reason is, we have fixed $m$ to 0.}
   \begin{align}
      \label{eq_Massieu}
      \bar{S}_\th (0, s_\alpha)
      = \ln \left(  \frac{ N! \, (2s_\alpha + 1) }{  
                       \left( \frac{N}{2} - s_\alpha  \right) !  \,  
                       \left(  \frac{N}{2} + s_\alpha + 1  \right) ! }
              \right)
      = \ln \left(  \frac{2s_\alpha + 1}{ \frac{N}{2} + s_\alpha + 1}  \right)
      +  \ln  \left( \frac{N!}{ \left( \frac{N}{2} - s_\alpha \right)!  \, 
                                       \left( \frac{N}{2} + s_\alpha  \right) ! }  \right) \, .
   \end{align}
   We approximate the partial derivative with the finite difference:\footnote{
   We estimated this approximation's error by analytically continuing the factorial to the Gamma function. The estimated error is subleading, so we omit it.}
   \begin{align}
      \Sigma_s
      & \approx  \bar{S}_\th (0, s_\alpha + 1)  -  \bar{S}_\th (0, s_\alpha )
      \label{eq_Sigma_s_help0}
      = \ln \left( \frac{1 + \frac{3}{2 s_\alpha} }{ 1 + \frac{1}{2 s_\alpha} }  \right)
      + \ln \left( 1 - \frac{2 s_\alpha}{N}  \right)
      - \ln  \left( 1 + \frac{2 (s_\alpha + 2)}{N}  \right) \\
      \label{eq_Sigma_s_help1}
      & = \frac{1}{s_\alpha}  -  \frac{4 s_\alpha}{N}
      + O \left( s_\alpha^{-2}  \right)
      + O \left(  \left[  s_\alpha / N  \right]^2  \right)
      + O \left( N^{-1}  \right) .
   \end{align}
   The final equality holds only if $s_\alpha \ll N$.
   
   Let us interrelate this expression with $\beta \gamma$. By definition [Eq.~\eqref{eq_Sigma_Def}] 
   $\Sigma_s \coloneqq  \partial_s S_\th (E, s) |_{E_\alpha, s_\alpha} \, .$ 
   By the notation introduced at the beginning of App.~\ref{sec_Derive_FDT_SU2},
   $S_\th(E, s)  =  \tilde{S}_\th (E, s)  -  \ln (2 s + 1)$.
   Hence 
   \begin{align}
      \label{eq_Sigma_s_help2}
      \Sigma_s 
      = \partial_s  \tilde{S}_\th (E, s) \big\lvert_{E_\alpha, s_\alpha}
      - \frac{2}{2 s_\alpha + 1}
      \quad \Rightarrow \quad
      \partial_s  \tilde{S}_\th (E, s) \big\lvert_{E_\alpha, s_\alpha}
      =  \Sigma_s
      +  \frac{2}{2 s_\alpha + 1}  \, .
   \end{align}
   We can evaluate this derivative using App.~\ref{sec_Thermo_props_2}. Under the conditions~\eqref{eq_Correspond}, $\ket{\alpha, m}$ locally resembles a $\rho_\NATS$. The $\ket{\alpha, m}$ is labeled by $m = 0$, which translates into $\mu = 0$. Hence Eq.~\eqref{eq_S_bm0} implies
   $\partial_s  \tilde{S}_\th (E, s) |_{E_\alpha, s_\alpha}  \approx  - \beta \gamma$.
   [The $\approx$ sign comes from the zeroth-order approximation of $Z$ in App.~\eqref{app_Peak_vals}. We assume that corrections of this type are negligible.]
   Let us solve for $\beta \gamma$, substitute in from the right-hand side of Eq.~\eqref{eq_Sigma_s_help2}, and substitute in from~\eqref{eq_Sigma_s_help0} (if $\zeta = 1$) or~\eqref{eq_Sigma_s_help1} (if $\zeta < 1$):
   \begin{align}
      \label{eq_bg_help1_app}
      \beta \gamma
      = - \left( \Sigma_s  +  \frac{2}{2 s_\alpha + 1}  \right)
      \approx \begin{cases}
      - \ln \left( \frac{1 + \frac{3}{2 s_\alpha} }{ 1 + \frac{1}{2 s_\alpha} }  \right)
      - \ln \left( 1 - \frac{2 s_\alpha}{N}  \right)
      + \ln  \left( 1 + \frac{2 (s_\alpha + 2)}{N}  \right)
      - \frac{2}{2 s_\alpha + 1}  
      &  \zeta = 1  \\
      - \frac{2}{s_\alpha}  +  \frac{4 s_\alpha}{N}
      + O \left( s_\alpha^{-2}  \right)
      + O \left( N^{-1}  \right)
      + O \left( \left[ s_\alpha / N \right]^2  \right)
      &  \zeta < 1 .
      \end{cases}
   \end{align}
   
   Define the infinite-size parameter product
   $(\beta \gamma)_\infty
      \coloneqq  \lim_{N \to \infty} \beta \gamma$.
   To calculate this product, we apply the scaling relation~\eqref{eq_s_scale_app} to Eq.~\eqref{eq_bg_help1_app}:
   \begin{align}
      \label{eq_bg_inf}
      ( \beta \gamma )_\infty
      = \begin{cases}
         \ln \left( \frac{1 + 2 \mathscr{S} }{1 - 2 \mathscr{S} }  \right) ,  &  \zeta = 1  \\
         0,  &  \zeta \in (0, 1) .
      \end{cases}
   \end{align}
   
   Finally, we calculate $\Sigma_s$ in terms of $(\beta \gamma )_\infty$ and finite-size ($N$-dependent) corrections. We replace $s_\alpha$ with $\mathscr{S} N^\zeta$ in Eqs.~\eqref{eq_Sigma_s_help0} and~\eqref{eq_Sigma_s_help1}, then compare with~\eqref{eq_bg_inf}. The $\zeta = 1$ case is especially simple. If $\zeta \in (0, 1)$, then 
   $\Sigma_s = O \left( N^{- \zeta } \right)  +  O \left( N^{\zeta - 1}  \right)  + \ldots
   = (\beta \gamma)_\infty  +  O  \left( N^{- \min \{ \zeta, 1 - \zeta \} }  \right)$.
   In summary,
   \begin{align}
      \label{eq_Sig_s_anom}
      \Sigma_s
      = \begin{cases}
         - (\beta \gamma)_\infty  +  O \left( N^{-1}  \right) ,  
         &  \zeta = 1 \\
         - (\beta \gamma)_\infty  +  O  \left( N^{- \min \{ \zeta, 1 - \zeta \} }  \right) ,
         &  \zeta \in (0, 1) .
      \end{cases}
   \end{align}
   If $\zeta = 1$, the finite-size correction is as large as usual (in the absence of any non-Abelian symmetry)~\cite{Noh_20_Numerical}. If $\zeta \in (0, 1)$, the correction is polynomially larger in the system size---what we call anomalous.

   \item  $\Gamma_s$: We cannot know \emph{a priori} how $\Gamma_s$ varies with $s_\alpha$. If $\Gamma_s$ depends on $s_\alpha$ through $s_\alpha / N$, then $\Gamma_s = O (N^{-1})$; and, if through $s_\alpha / \sqrt{N}$, then 
   $\Gamma_s = O(N^{-1/2})$. In any case, the $\Gamma_s$ scaling will not eliminate the anomalous correction from $\Sigma_s$.
   
\end{enumerate}
   
Let us substitute from the list items above into Eq.~\eqref{eq_L_eval_help1_app}. We estimate the fine-grained KMS relation's finite-size correction to be, when $\beta = 0$,
\begin{align}
   \label{eq_L_Anom_app}
   \tilde{\mathcal{L}}_{AB} (\Omega, 0, \Delta s; \alpha, 0)
   = \begin{cases}
      \beta ( \Omega - \gamma \, \Delta s )  
      +  O  \left( N^{-1}  \right)
      +  O  \left(  \Gamma_s  \right)  ,
      &  \zeta = 1  \\
      \beta ( \Omega - \gamma \, \Delta s )  
      +  O  \left( N^{- \min \{ \zeta, 1 - \zeta \} }  \right)
      +  O  \left(  \Gamma_s  \right)  ,
      &  \zeta  \in  (0, 1)  .
   \end{cases}
\end{align}

Does the anomalous correction extend beyond $\beta = 0$? We expect so, for three reasons. First, the $\beta {=} 0$ anomalous correction relies on $\tilde{S}_\th(\beta, s_\alpha)$ [Eq.~\eqref{eq_Massieu}]. If $\tilde{S}_\th$ changed drastically as $\beta$ changed from 0, the system would exhibit an infinite-temperature phase transition, which seems unlikely. Second, one can regard the correction's $\tilde{S}_\th$ dependence as reliance on the Hilbert space's structure. The Hilbert space has this structure due to the non-Abelian symmetry (by Schur's lemma~\cite{Bouchard_20_Group,Gour_08_Resource}), 
which does not depend on temperature. Third, the high-temperature ($\beta {=} 0$) limit generally behaves relatively classically. One should therefore expect more-quantum behavior (perhaps exaggerated effects of noncommutation) at lower temperatures.

\subsubsection{$s_\alpha$ can be $O (N^{1/2})$ and is typically $O(N^{1/2})$ 
}
\label{app_KMS_Anom_argue_for_s_scaling}

In the previous subsubsection, we argued that an energy eigenstate's fine-grained KMS relation can contain an anomalous correction if $s_\alpha = O (N^{ \zeta \in (0, 1) })$ and $m = q = \beta = 0$. Here, we argue that $s_\alpha$ can satisfy this condition---can be of $O(N^{1/2})$. The argument relies on (i) the resemblance between $\ket{\alpha, m}$ and $\rho_\NATS$ and (ii) the equivalence of thermodynamic ensembles. We invoked these two principles in the previous subsubsection, arguing that an $\ket{\alpha, m}$ can resemble a $\rho_\NATS$ if the two share their $\beta$ values, rather than satisfying $E_\alpha = \expval{H}$. That is, one can construct a $\rho_\NATS$-like $\ket{\alpha, m}$ by specifying $(E_\alpha, m, s_\alpha)$ or by specifying $(\beta, m, s_\alpha)$. Similarly, one can construct a $\rho_\NATS$-like $\ket{\alpha, m}$ by specifying $(\beta, \beta \mu, \beta \gamma)$. We specify the latter here and prove that the modified NATS's $\expval{S}$ can be of $O(N^{1/2})$ while $m = 0$. Then, we ascribe this scaling to an $\ket{\alpha, m}$'s $s_\alpha$ via (i) and (ii). Afterward, we argue that $s_\alpha = O(N^{1/2})$ typically.

First, we specify the modified NATS under consideration. For convenience, we assume that $N$ is even. Let $\beta = 0$, as in the previous subsubsection. Let $\beta \mu = 0$, for consistency with that subsubsection's assumption $m = 0$. As $N$ grows toward the thermodynamic limit, $\gamma$ grows such that $\beta \gamma$ remains constant. The modified NATS assumes the form
\begin{align}
   \label{eq_NATS_Anom}
   \rho'_\NATS (\beta \gamma)
   = e^{\beta \gamma S} / \tilde{Z} (\beta \gamma) .
\end{align}
The partition function $\tilde{Z} (\beta \gamma)$ depends implicitly on $N$.

To calculate thermodynamic properties, we must calculate the partition function. We invoke its definition, then the formula for the total Hilbert space's dimensionality~\cite[Eq.~(15)]{JaeDongXXZ}:
\begin{align}
   \tilde{Z} (\beta \gamma)
   \coloneqq  \Tr  \left(  e^{\beta \gamma S}  \right)
   \label{eq_Z_Anom_help1}
   =  \sum_{s = 0}^{N/2}
   \frac{N! \, (2s + 1)^2 }{ 
           \left( \frac{N}{2} - s \right)!  \,  \left( \frac{N}{2} + s + 1  \right)! }  \,
   e^{\beta \gamma s}  \, .
\end{align}

To interpret this expression physically, we rewrite the summand as
\begin{align}
   \frac{N!}{ \left( \frac{N}{2} - s \right) ! \, \left( \frac{N}{2} + s \right) ! }  \,
   \frac{ \left(2 s + 1 \right)^2 }{ \frac{N}{2} + s + 1 }  \,
   e^{\beta \gamma s}
   \label{eq_walker1}
   = {N  \choose  N/2 + s}  
   \exp \left( \beta \gamma s 
                   + 2 \ln (2s + 1)  -  \ln \left( \frac{N}{2} + s + 1  \right)  \right) .
\end{align}
The right-hand side is proportional to a random-walk probability distribution. Consider a random walker who begins at $x = 0$. An infinitely hard reflecting wall keeps the walker at $x \geq 0$. The ground tilts away from the wall if $\beta \gamma > 0$ and toward the wall if $\beta \gamma < 0$. The slope is $\beta \gamma$. In addition to the tilt, the landscape exhibits a potential
$\sim \ln(2s) + O(s^0) + O(s/N)$.

Having described the random walker's setup, we translate between Eq.~\eqref{eq_walker1} and a walker-related probability.
Suppose the walker takes $N$ steps. What is the walker's probability of ending at $x = N/2 + 2s$ (of taking exactly $N/2 + s$ steps rightward, or of having a displacement $2s$)? This probability is proportional to~\eqref{eq_walker1}. The lower limit $s=0$, in Eq.~\eqref{eq_Z_Anom_help1}, encodes the wall and the walker's confinement to $x \geq 0$. This interpretation will illuminate our results below.

To estimate $\tilde{Z} (\beta \gamma)$ [Eq.~\eqref{eq_Z_Anom_help1}], we assume that $N \gg s, 1$. The large-$s$ terms in~\eqref{eq_Z_Anom_help1} contribute most to the sum. We therefore assume that $s \gg 1$ to Stirling-approximate the integrand:
\begin{align}
   \tilde{Z} (\beta \gamma)
   =  \frac{2^N}{ \sqrt{\pi (N/2)^3 } }
   \int_0^\infty  ds  \;   s^2  \:
   e^{- 2 s^2 / N + \beta \gamma s}  \,  .
\end{align}
We expect the large-system approximation to incur an exponentially small error, which we neglect. Let us change variables from $s$ to 
$x \coloneqq \sqrt{2 / N}  \, s$ and from $\beta \gamma$ to
\begin{align}
   \label{eq_Def_tilde_g}
   \tilde{\gamma}  \coloneqq  \sqrt{N / 8}  \,  \beta \gamma .
\end{align}
The partition function becomes
\begin{align}
   \label{eq_Z_Anom_help2}
   \tilde{Z}(\beta \gamma)  
   =  2^{N+1}  \tilde{\mathcal{Z}} \left( \tilde{\gamma}  \right) .
\end{align}
The reduced partition function is
\begin{align}
   \tilde{\mathcal{Z}} \left( \tilde{\gamma} \right)
   \coloneqq  \frac{2}{ \sqrt{\pi} }
   \int_0^\infty  dx  \;  x^2  \:
   e^{-x^2 + 2 \tilde{\gamma} x }
   \label{eq_Z_Anom_help3}
   =  \frac{ \tilde{\gamma} }{ \sqrt{\pi} }
   + \frac{1 + 2 \tilde{\gamma}^2 }{2}  
   \left[ 1 + {\rm erf} \left( \tilde{\gamma}  \right)  \right]  \,
   e^{ \tilde{\gamma}^2 }  \, .
\end{align}
The ${\rm erf}$ denotes the error function.

From the partition function, we can calculate $\expval{S}$. By Eq.~\eqref{eq_NATS_Anom}, 
$\expval{S}
= \partial_{\beta \gamma}  \ln \LParen \tilde{Z} (\beta \gamma)  \RParen$.
By the partition-function formulae~\eqref{eq_Z_Anom_help2} and~\eqref{eq_Z_Anom_help3},
\begin{align}
   \label{eq_S_Anom_help1} &
   \expval{S}  
   = \sqrt{ \frac{N}{8} }  \:
   \tilde{s}  \left(  \tilde{\gamma}  \right) ,
   \quad \text{wherein} \\ &
   \label{eq_Def_tilde_s}
   \tilde{s} \left( \tilde{\gamma} \right)
   \coloneqq \frac{d}{d \tilde{\gamma}}  \,
   \ln \LParen \tilde{\mathcal{Z}} \left( \tilde{\gamma} \right)  \RParen
   = \frac{ 4 \left( 1 + \tilde{\gamma}^2 \right) 
               + 2 \sqrt{\pi} \, \tilde{\gamma} \left( 3 + 2 \tilde{\gamma}^2 \right)
               \left[ 1 + {\rm erf} \left( \tilde{\gamma} \right) \right]  \,
               e^{ \tilde{\gamma}^2 } }{
               2 \tilde{\gamma} 
               + \sqrt{\pi}  \,  \left( 1 + 2 \tilde{\gamma}^2  \right)  \,
               \left[ 1 + {\rm erf} \left( \tilde{\gamma} \right)  \right]  \,
               e^{ \tilde{\gamma}^2 }  }  \, .
\end{align}
The auxiliary function $\tilde{s} ( \tilde{\gamma} )$ scales as follows in various parameter regimes controlled by $\tilde{\gamma}$ (by $\beta \gamma$ and $N$):
\begin{align}
   \label{eq_tilde_s_scaling}
   \tilde{s}  \left(  \tilde{\gamma}  \right)
   = \begin{cases}
      2 \tilde{\gamma} + O \left( \tilde{\gamma}^{-1}  \right)  \,  &
      \tilde{\gamma}  \gg  1  \\
      O(1) ,  &
      \left\lvert \tilde{\gamma} \right\rvert  \lesssim  1 \\
      - \frac{3}{ \tilde{\gamma} }  
      +  O \left( \left\lvert \tilde{\gamma} \right\rvert^{-3}  \right) ,  &
      \tilde{\gamma}  \ll  -1 .
   \end{cases}
\end{align}
From Eqs.~\eqref{eq_S_Anom_help1},~\eqref{eq_tilde_s_scaling}, and~\eqref{eq_Def_tilde_g}, we infer how $\expval{S}$ scales with $N$ as $\beta \gamma$ changes:\footnote{
The middle line of Eq.~\eqref{eq_S_scale_Anom_final} does not follow from the middle line of Eq.~\eqref{eq_tilde_s_scaling} but follows from the other equations.}
\begin{align}
   \label{eq_S_scale_Anom_final}
   \expval{S}
   = \begin{cases}
      \beta \gamma \, N / 4 ,  &
      \beta \gamma  \gg  \sqrt{8 / N }  \\
      \sqrt{2 N / \pi }  \, ,  &
      \beta \gamma = 0 \\
      - 3 / (\beta \gamma)  ,  &
      \beta \gamma \ll - \sqrt{8 / N }  \, .
   \end{cases}
\end{align}
The middle line shows that $\expval{S}$ can be of $O(N^{1/2})$, consistently with an assumption in App.~\ref{app_KMS_Anom_If_Then}.

We can understand Eq.~\eqref{eq_S_scale_Anom_final} more fully through the random-walker story described below Eq.~\eqref{eq_walker1}. $\expval{S}$ equals half the random walker's position after exactly $N$ steps. In the first line of Eq.~\eqref{eq_S_scale_Anom_final}, $\beta \gamma > 0$. Hence the ground tilts away from the hard wall (from $x=0$). The walker therefore escapes the wall ballistically; the average displacement is proportional to the number of steps: $\expval{S} \propto N$. In the second line of Eq.~\eqref{eq_S_scale_Anom_final}, $\beta \gamma = 0$; the ground does not tilt. The random walker escapes the logarithmic potential with an average displacement, $\expval{S}$, sublinear in the number $N$ of steps: $\expval{S} \propto \sqrt{N}$. In the third line of Eq.~\eqref{eq_S_scale_Anom_final}, $\beta \gamma < 0$; the ground tilts toward the wall. Hence the random walker reaches a steady state by the wall; the average position, $\expval{S}$, does not depend on the number $N$ of steps taken. 

We can progress beyond this observation, by calculating the finite-size correction to $\beta \gamma$ when $\expval{S} = O (N^{1/2})$ and $\beta \gamma \approx 0$. These conditions are close to those in the middle line of Eq.~\eqref{eq_S_scale_Anom_final}. Equation~\eqref{eq_L_Anom_app}, displaying finite-size corrections (of $\tilde{\mathcal{L}}_{AB}$) to $(\beta \gamma)_\infty$, motivates this study. Let $r$ denote a constant such that 
$\expval{S} = \sqrt{N / 8} \, r .$ By Eq.~\eqref{eq_S_Anom_help1},
$r = \tilde{s} ( \tilde{\gamma} )$. By this equation and Eq.~\eqref{eq_Def_tilde_g},
$r = \tilde{s} ( \sqrt{N/8} \, \beta \gamma )$. Using Eq.~\eqref{eq_Def_tilde_s}, we solve for $\beta \gamma$: $\beta \gamma = O ( N^{-1/2} )$.

Having argued that $s_\alpha$ can be $O( N^{1/2} )$, we argue that $s_\alpha$ typically scales so. The $N$-qubit Hilbert space decomposes into sectors characterized by the total spin quantum number $s_\alpha \, ,$ by Schur's lemma~\cite{Bouchard_20_Group,Gour_08_Resource,NYH_20_Noncommuting}. The $s_\alpha$ sector has a dimensionality
$(2s_\alpha+1)^2 N! / [(N/2-s)!(N/2+s+1)!]$, by the discussion around Eq.~\eqref{eq_Massieu}. This dimensionality attains its maximum at $s_\alpha = O(N^{1/2})$. Therefore, the spin quantum number typically scales as $s_\alpha \sim \sqrt{N}$. Therefore, when evaluated NATSs whose $\gamma$s vanish, the fine-grained KMS relation typically has an anomalous correction.

\section{Proof of property of Clebsch--Gordan coefficient}
\label{app_CG_property}

In App.~\ref{app_KMS_0th}, we invoked a property of Clebsch--Gordan coefficients:
\begin{align}
   \label{eq_CG_prop_proof1} &
   \text{If} \; s = O \left( N^\zeta \right) , \;
   \zeta \in (0, 1] , \;  \text{and} \;
   k, q, m = O(1) ,  \;  \text{then}  \\
   \label{eq_CG_prop_proof2}
   & \braket{s + \nu, m + q}{ s, m ; k, q }
   = \bar{c} ( \nu ; k, q )  \,
   \left[ 1 + O \left( s^{-1} \right) \right] ,  \; \text{wherein} \; \\
   \label{eq_CG_prop_proof3}
   & \bar{c} (\nu; k, q)
   \coloneqq \frac{1}{2^k}  
   \sqrt{ \frac{ (k + \nu)! \, (k - \nu)! }{ (k + q)! \, (k - q)! } }  
   \sum_{ \ell = \max\{ 0, q - \nu \} }^{ \min \{ k + q , k - \nu \} }
   (-1)^\ell    {k + q  \choose  \ell}  {k - q  \choose  \nu - q + \ell}  .
\end{align}
The beginning of our argument echoes Eqs.~(B29) and~(B30) in~\cite{24_Lasek_Numerical}. There, the general formula for a Clebsch--Gordan coefficient is copied from Eq.~(2.41) of~\cite{Bohm_Quantum_Book}:
\begin{align}
   \label{eq_CG_app_1}
   & \braket{s + \nu, m + q}{ s, m; k, q} 
   \\ \nonumber
   & = \sqrt{ \frac{ [2(s + \nu) + 1]  \,  
   (2s + \nu - k)!  \,  (k + \nu) !  \,  (k - \nu)!  \,  (s + m + \nu + q)!  \,  (s - m + \nu - q)!  \,  (s + m)!  \,  (s - m)!  \,  (k + q)!  \,  (k - q)!  }{
                            (2s + \nu + k + 1)!  }  }
   \nonumber \\ \nonumber & \quad \; \times
   \sum_\ell  \frac{ (-1)^\ell }{
   \ell!  \,  (k - \nu - \ell)!  \,  (s - m - \ell)!  \,  (k + q - \ell)!  \,  (s + m + \nu - k + \ell)!  \,  (\nu - q + \ell)!  }  \, .
\end{align}
The sum runs over the integers $\ell$ that render every factorial's argument non-negative.

Let us begin to approximate the Clebsch--Gordan coefficient. First, we approximate a fairly general factorial. If $x \gg \Delta$, then
\begin{align}
   \label{eq_Approx_fact}
   (x + \Delta)!
   =  [(x + \Delta) (x + \Delta - 1)  \ldots  (x + 1) ]  x!
   =  x^\Delta  \,  x!  \left[ 1  +  O \left( x^{-1}  \right)  \right] .
\end{align}
By Eq.~\eqref{eq_CG_prop_proof1}, $s \gg k, q, m$. Because $s \gg k$ and by the selection rule obeyed by the Clebsch--Gordan coefficient in~\eqref{eq_CG_prop_proof2}, $s \gg \nu$. We apply these inequalities and the approximation~\eqref{eq_Approx_fact} to Eq.~\eqref{eq_CG_app_1}. Many factors cancel exactly:
\begin{align}
   \label{eq_CG_app_2}
   \braket{s + \nu, m + q}{s, m; k, q}
   & =  \frac{1}{2^k}  
   \sqrt{ \frac{ [2(s + \nu) + 1]  \,  (k + \nu)!  \,  (k - \nu)!  \,  (k + q)!  \,  (k - q)!  \,  [1 + O(s^{-1}) ] }{  (2s)  [1 + O(s^{-1}) ] }  }
   \nonumber \\ & \quad \; \times
   \sum_\ell  \frac{ (-1)^\ell }{
   \ell!  \,  (k - \nu - \ell)!  \,  (k + q - \ell)!  \,  (\nu - q + \ell)!  \,  [1 + O(s^{-1} )]  }  \, .
\end{align}
The corrections simplify as
\begin{align}
   \sqrt{ \frac{ [1 + O(s^{-1}) ] }{ [1 + O(s^{-1}) ] }  }  \,
   \frac{1}{ [1 + O(s^{-1}) ] }
   & =  \left[ 1 + \frac{1}{2} \, O \left( s^{-1} \right)  \right]
   \left[ 1 - \frac{1}{2} \, O \left( s^{-1} \right)  \right]
   \left[ 1 - O \left( s^{-1} \right)  \right]
   =  1 + O \left( s^{-1} \right) .
\end{align}
Also, the square-root in Eq.~\eqref{eq_CG_app_2} contains two factors that approximately cancel:
\begin{align}
   \frac{2 (s + \nu) + 1}{2s} 
   =  1 + O \left(s^{-1} \right) .
\end{align}
Hence Eq.~\eqref{eq_CG_app_2} simplifies to
\begin{align}
   \label{eq_CG_app_3}
   \braket{s + \nu, m + q}{s, m; k, q}
   & = \frac{1}{2^k}  \,
   \sqrt{ (k + \nu) !  \,  (k - \nu) !  \,  (k + q)!  \,  (k - q)!  }  \,
   \\ \nonumber & \quad \; \times
   \sum_\ell  \frac{ (-1)^\ell }{ \ell!  \,  (k - \nu - \ell)!  \,  (k + q - \ell)!  \,  (\nu - q + \ell)! }  \,
   \left[ 1 + O \left( s^{-1} \right)  \right] .
\end{align}

The proof's final steps center on the sum. The summand, with the sentence below Eq.~\eqref{eq_CG_app_1}, implies that $\ell \geq 0$, $k - \nu \geq \ell$, $k + q \geq \ell$, and $\nu - q \geq -\ell$. In summary,
$\max \{0, q - \nu\}  \leq  \ell  \leq  \min \{ k - \nu, k + q \}$.
Consider inserting these limits in Eq.~\eqref{eq_CG_app_3}. The result is equivalent to Eqs.~\eqref{eq_CG_prop_proof2}--\eqref{eq_CG_prop_proof3}, one can check by evaluating the binomial coefficients: if $a, b \geq 0$, then
${a \choose b}  =  \frac{a!}{ b! (a - b)! } \, .$

\section{Transformation property of fine-grained correlator evaluated in energy eigenstate}
\label{app_Sym_for_num}

This appendix introduces the transformation property invoked in the numerical analysis (Sec.~\ref{sec_Num}). By invoking the property, one can easily deduce general results from the examples analyzed. The property stems from the Wigner–Eckart theorem as follows. In Sec.~\ref{sec_Num}, we analyze correlators of spherical tensor operators $A^{(k')}_{-q}$ and $B^{(k)}_q \, .$ The correlators depend on elements of the matrices that represent the operators relative to the energy eigenbasis, $\{ \ket{\alpha, m} \}$. By the Wigner–Eckart theorem, Clebsch–Gordan coefficients encapsulate all the matrix elements' $m$- and $q$-dependences. For example, 
\begin{align}
    \langle \alpha',m+q|B^{(k)}_q | \alpha,m\rangle &= 
    \langle s_{\alpha'},m+q|s_\alpha,m;k,q\rangle \langle \alpha' \| B^{(k)}
    \| \alpha \rangle   
    =  \frac{\langle s_\alpha', m+q|s_\alpha,m;k,q\rangle}{\langle s_\alpha',0|s_\alpha,0;k,0\rangle}   
    \langle \alpha',0 | B^{(k)}_0 | \alpha,0\rangle  \\ &
    \equiv  
             \mathsf{M}_1 (s_\alpha, s_{\alpha'}, m, k, q) 
             \langle \alpha',0 | B^{(k)}_0 | \alpha,0\rangle 
             \label{eq:element_symmetry}
\end{align}
We have defined the Clebsch–Gordan ratio 
\begin{equation}
    \mathsf{M}_1(s, s', m, k, q) 
    \coloneqq \frac{\langle s', m+q|s,m;k,q\rangle}
    {\langle s',0|s,0;k,0\rangle} .
    \label{eq:M1coefficient}
\end{equation}
In Eq.~\eqref{eq:element_symmetry}, the prefactor depends on $\alpha$ and $\alpha'$ only through $s_\alpha$ and $s_{\alpha'}$. This property allows us to transform one fine-grained correlator into another easily.

The matrix-element symmetry~\eqref{eq:element_symmetry} extends to the fine-grained dynamical correlator, 
$\hat{\bar{C}}^{\rm dyn}_{A^{(k')}_{-q},  B^{(k)}_q}
(\Omega, \Delta m{=}q; \Delta s; \alpha, m)$ [Eq.~\eqref{eq_Corr_help1}]. If 
\begin{align}
    & \mathsf{M}_2(s, \Delta s, m, k, k', q) \coloneqq 
        \mathsf{M}_1(s+ \Delta s, s, m+q, k', -q)
        \mathsf{M}_1(s, s+ \Delta s, m, k, q) ,
        \label{eq:M2coefficient}
        \quad \text{then} \\ &
    \hat{\bar{C}}^{\rm dyn}_{A^{(k')}_{-q},B^{(k)}_q}
    (\Omega, \Delta m{=}q, \Delta s; 
    \alpha, m) = 
    \mathsf{M}_2(s_\alpha, \Delta s, m, k, k', q)  \,
    \hat{\bar{C}}^{\rm dyn}_{A^{(k')}_{0},B^{(k)}_0}
    (\Omega, \Delta m{=}0, \Delta s;
    \alpha,0) .
    \label{eq:Stransform}
\end{align}
Therefore, we can deduce about general dynamical fine-grained correlators upon evaluating a correlator between spherical tensor operators' $q{=}0$ components on $\ket{\alpha, m{=}0}$.

To test the fine-grained KMS relation, we define the logarithmic ratio
\begin{equation}
    \tilde{\mathcal{{L}}}_{A^{(k')}_{-q} B^{(k)}_q} (\Omega, q, \Delta s;  \alpha,m)
    \coloneqq \ln  \left(  \frac{ 
    \hat{\bar{C}}^{\rm dyn}_{A^{(k')}_{-q} B^{(k)}_q}
                (\Omega, \Delta m{=}q, \Delta s;
\alpha, m)}
{ \hat{\bar{C}}^{\rm dyn}_{B^{(k)}_{q} A^{(k')}_{-q}}
               (-\Omega, \Delta m{=}{-q}, -\Delta s;
\alpha, m)}  \right) .
\label{eq:LogRatioFunc2}
\end{equation}
By the transformation rule~\eqref{eq:Stransform}, the log-ratio transforms as
\begin{equation}
   \tilde{\mathcal{L}}_{A^{(k')}_{-q} B^{(k)}_q} (\Omega, q, \Delta s;  \alpha,m) 
   = \mathsf{M}_3(s_\alpha, \Delta s, m; k, k', q)  
   +  \tilde{\mathcal{L}}_{A^{(k')}_{0}  B^{(k)}_0} (\Omega, 0, \Delta s;  \alpha,0) .
   \label{eq:Ltransform}
\end{equation}
We have defined the function
\begin{equation}
    \mathsf{M}_3(s, \Delta s, m; k, k', q) = \ln  \left( \frac{
    \mathsf{M}_2(s, \Delta s, m, k, k', q)}{
    \mathsf{M}_2(s, -\Delta s, m, k', k, -q)}   \right)
\end{equation}
of 8 Clebsch–Gordan coefficients.

\section{Additional numerics supporting the fine-grained KMS relation}
\label{app_Add_num}

This appendix provides further numerical evidence for the fine-grained KMS relation. The data describe a Heisenberg model that lacks symmetries other than SU(2), unlike the main text's Heisenberg model. First, we describe the setup. Then, we present results analogous to those in the main text.

We simulated a quantum many-body system that has SU(2) symmetry but, unlike the main-text model, lacks any other symmetry [apart from the U(1) symmetry equivalent to energy conservation]. The system is a chain of $N \in [14,16, 18]$ qubits subject to closed boundary conditions.
The chain evolves under a next-nearest-neighbor Heisenberg Hamiltonian. The Hamiltonian lacks translational symmetry due to the boundary conditions and a defect: 
    qubit  3 couples to its neighbors more strongly than any other qubit does.
   The defect also breaks spatial  inversion symmetry. The Hamiltonian has the form 
\begin{equation}
    H_2 = \sum^{N-1}_{j=1} J_j \, \vec{\sigma}^{(j)} \cdot \vec{\sigma}^{(j+1)} 
    + \frac{1}{2} \sum^{N-2}_{j=1} J_j \, \vec{\sigma}^{(j)} \cdot \vec{\sigma}^{(j+2)} \, .
\end{equation}
The coupling strength is $J_j = 1+0.3\delta_{j3}$. We solved for the Hamiltonian's eigenvalues and eigenstates exactly.

Our analysis focuses on the operator $ T^{(0)}_0$, defined in the main text's Eq.~\eqref{eq_T00_T20}. As $k = q =0$, $\Delta s= \Delta m=0$. From $T^{(0)}_0$, we calculated quantities that are defined in the main text but that we partially review here for convenience.
We evaluated the fine-grained correlator~\eqref{eq_Corr_help1} on energy eigenstates $\ket{\alpha,m{=}0}$. From the correlator, we calculated the logarithmic ratio 
$\mathcal{L}_{k,q=0}(\Omega, \Delta s {=} 0; \alpha, m {=} 0)$ [Eq.~\eqref{eq:LogRatioFunc0}]. As described in Sec.~\ref{sec_Num}, $\mathcal{L}$ fluctuates from eigenstate to eigenstate. Therefore, we averaged $\mathcal{L}$ over appropriate eigenstates, to produce $\overline{\mathcal{L}}$. We averaged over energy eigenstates from a narrower window than in the main text---an energy window of width $\Delta E = 0.1$. The reason is, the present system has a greater density of states than the main-text system. The average $\overline{\mathcal{L}}$ is calculated over energy eigenstates $\ket{\alpha,m}$ labeled by eigenvalues $s_\alpha{=}s$, $m{=}0$, and $E_{\alpha} \approx E(\beta)$.  $E(\beta)$ denotes the eigenenergy closest to the Hamiltonian expectation value in the canonical average at the inverse temperature $\beta$. By $E_{\alpha} \approx E(\beta)$, we mean that $E_\alpha$ lies within a width-$(\Delta E)$ window centered on $E(\beta)$. The window is of width $\Delta E =0.1$ (and so is narrower than in the main text).
In summary, $\overline{\mathcal{L}}$ is the average over the set $\{\mathcal{L}_{k,0} (\Omega, 0;\alpha,0) : s_{\alpha} = s, |E_{\alpha}-E(\beta)| < \Delta E /2 \}$.

\begin{figure}[H]
    \centering
    \begin{subfigure}[t]{0.45\textwidth}
        \centering
        \includegraphics[width=0.9\textwidth]{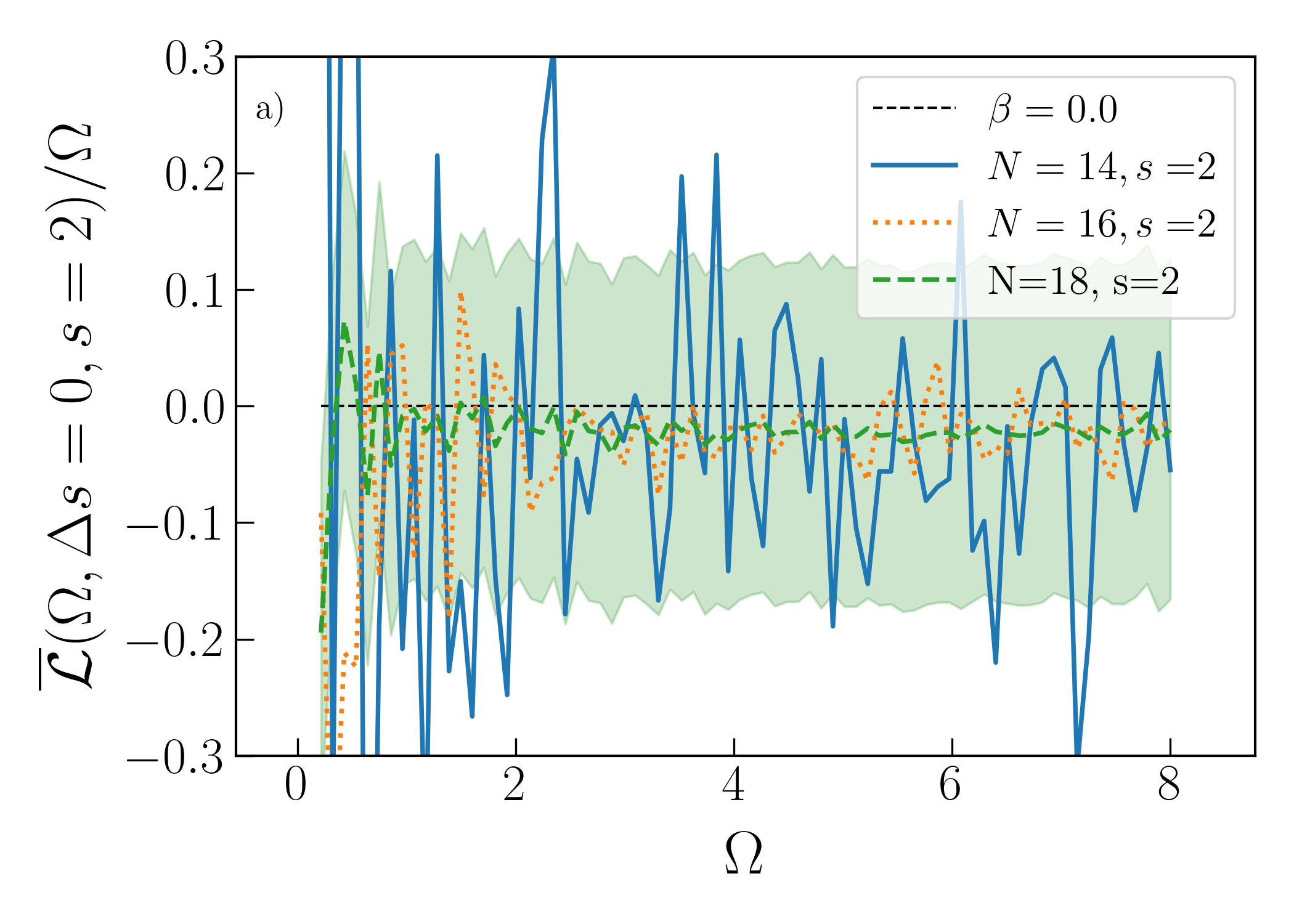}
    \end{subfigure}\hspace{2em}%
    \begin{subfigure}[t]{0.45\textwidth}
        \centering
        \includegraphics[width=0.9\textwidth]{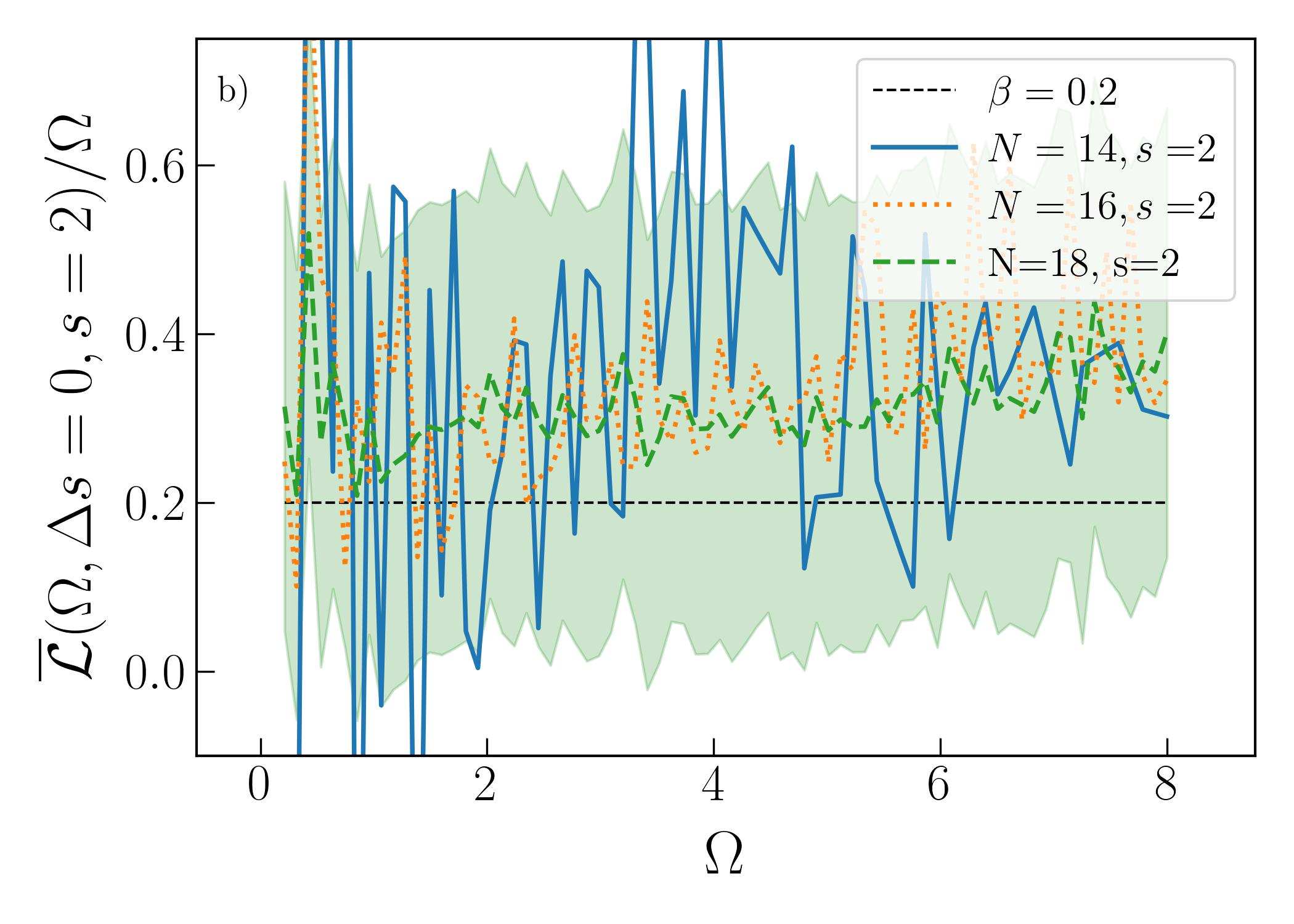}
    \end{subfigure}
    \caption{Normalized average logarithmic ratio $\overline{\mathcal{L}} (\Omega, \Delta s {=} 0, s{=}2) / \Omega$ versus $\Omega$. The dashed black lines represent the inverse temperature, $\beta$, which equals 0.0 in (a) and 0.2 in (b). Each of the other curves corresponds to one system size. The shaded area shows one standard deviation at $N=18$. }
    \label{fig:AppE_fig1}
\end{figure}

\begin{figure}[H]
    \centering
    \begin{subfigure}[t]{0.45\textwidth}
        \centering
        \includegraphics[width=0.9\textwidth]{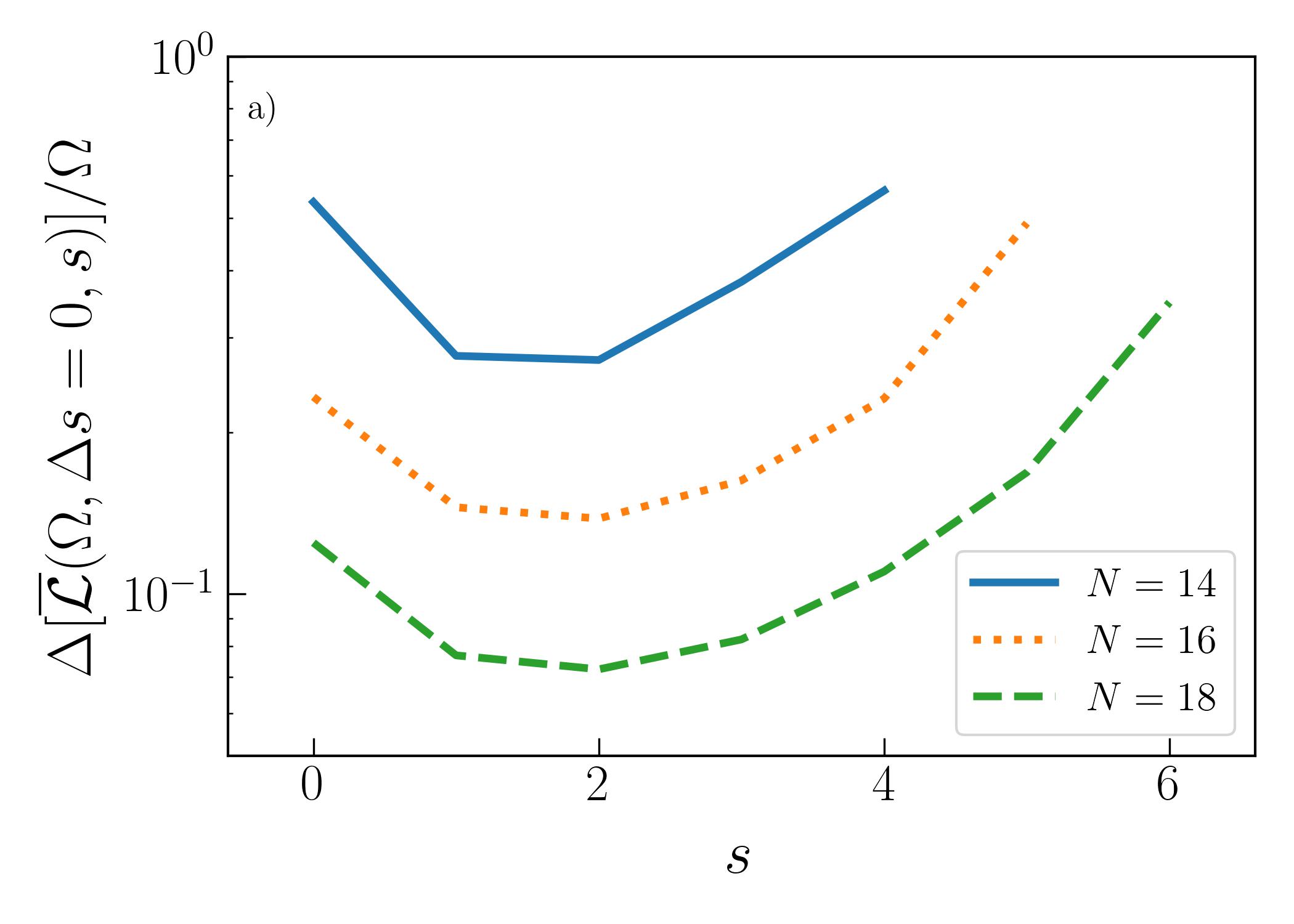}
    \end{subfigure}\hspace{2em}%
    \begin{subfigure}[t]{0.45\textwidth}
        \centering
        \includegraphics[width=0.9\textwidth]{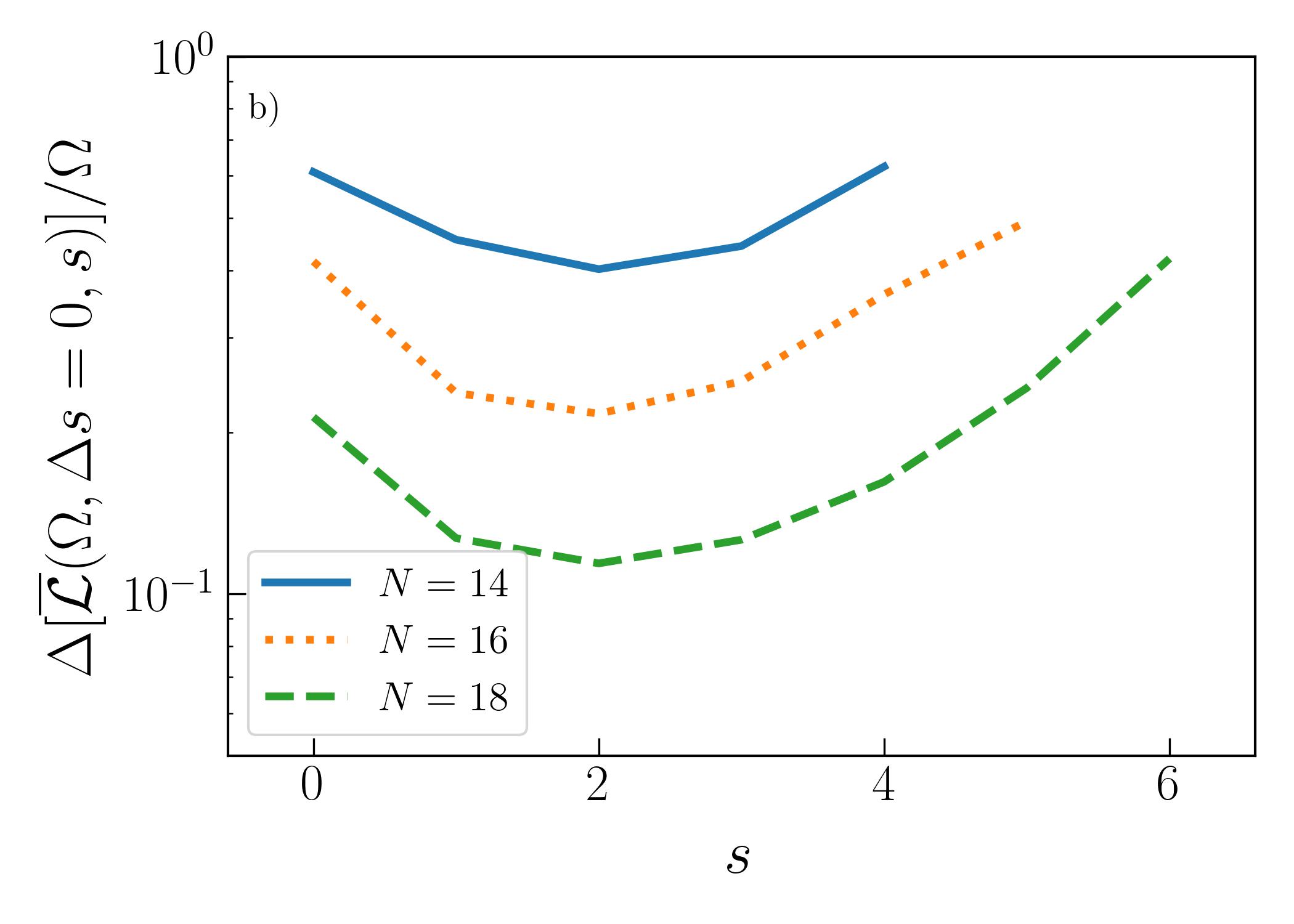}
    \end{subfigure}
    \caption{Standard deviation in the normalized average logarithmic ratio  $\Delta \overline{\mathcal{L}}/ \Omega$ versus the spin quantum number $s$, for $N=14,16,18$. The inverse temperature $\beta$ is $0.0$ in (a), and $0.2$ in (b). This data quantifies the width of the shaded area in Fig.~\ref{fig:AppE_fig1}, which shows $\overline{\mathcal{L}}(\Omega)$ for $s$ that minimizes  $\Delta \overline{\mathcal{L}}/ \Omega$. Different curves correspond to different system sizes. Each curve terminates when the $s$ subspace becomes too small to provide enough data to produce high-quality results.}
    \label{fig:AppE_fig2}
\end{figure}

Figure~\ref{fig:AppE_fig1} shows the average logarithmic ratios $\overline{\mathcal{L}}$, normalized by $\Omega$, versus $\Omega$. The inverse temperature $\beta=0.0$ in Fig.~\ref{fig:AppE_fig1}(a), and  $\beta=0.2$ in Fig.~\ref{fig:AppE_fig1}(b).  
The spin quantum number is $s=2$, at which the fluctuations 
$\Delta  \overline{\mathcal{L}}/ \Omega $ (the standard deviation of $\mathcal{L}/ \Omega $ within a small energy window $\Delta E$) are minimized [Fig.~\ref{fig:AppE_fig2}].
According to the KMS relation~\eqref{eq_KMS_eigen}, 
$\overline{\mathcal{L}}/ \Omega$ equals $\beta$, to within finite-size corrections. We define the effective inverse temperature $\beta_{\rm eff}$ as the average of 
$\overline{\mathcal{L}}/\Omega$ across $\Omega \in [2, 5]$. The figure bears out this expectation: $\overline{\mathcal{L}}/ \Omega$ lies within one standard deviation of $\beta$ throughout the $\Omega$ range, at each $\beta$ value. However, $\overline{\mathcal{L}}/ \Omega$  tends to overestimate $\beta$ when $\beta=0.2$. The reason is likely that, when $\beta = 0.2$, $\overline{\mathcal{L}}$ is calculated from a relatively small subspace. The main text's Fig.~\ref{fig:FSSnonzeroNu} displays the same effect.

Figure~\ref{fig:AppE_fig2} shows the fluctuations (the standard deviation of $\mathcal{L}/ \Omega $ within a small energy window $\Delta E$) $\Delta \overline{\mathcal{L}}/ \Omega$ in $\overline{\mathcal{L}}/ \Omega$. At each $s$ value, 
$\Delta \overline{\mathcal{L}}/ \Omega$ decreases exponentially as $N$ increases. This finding is consistent with the non-Abelian ETH~\eqref{eq_NAETH}, whose off-diagonal term is exponentially small in $N$. As $s{=}2$ minimizes
$\Delta \overline{\mathcal{L}}/ \Omega$, we use this $s$ value in Fig.~\ref{fig:AppE_fig1}.

\begin{figure}[H]
    \centering
    \begin{subfigure}[t]{0.45\textwidth}
        \centering
        \includegraphics[width=0.9\textwidth]{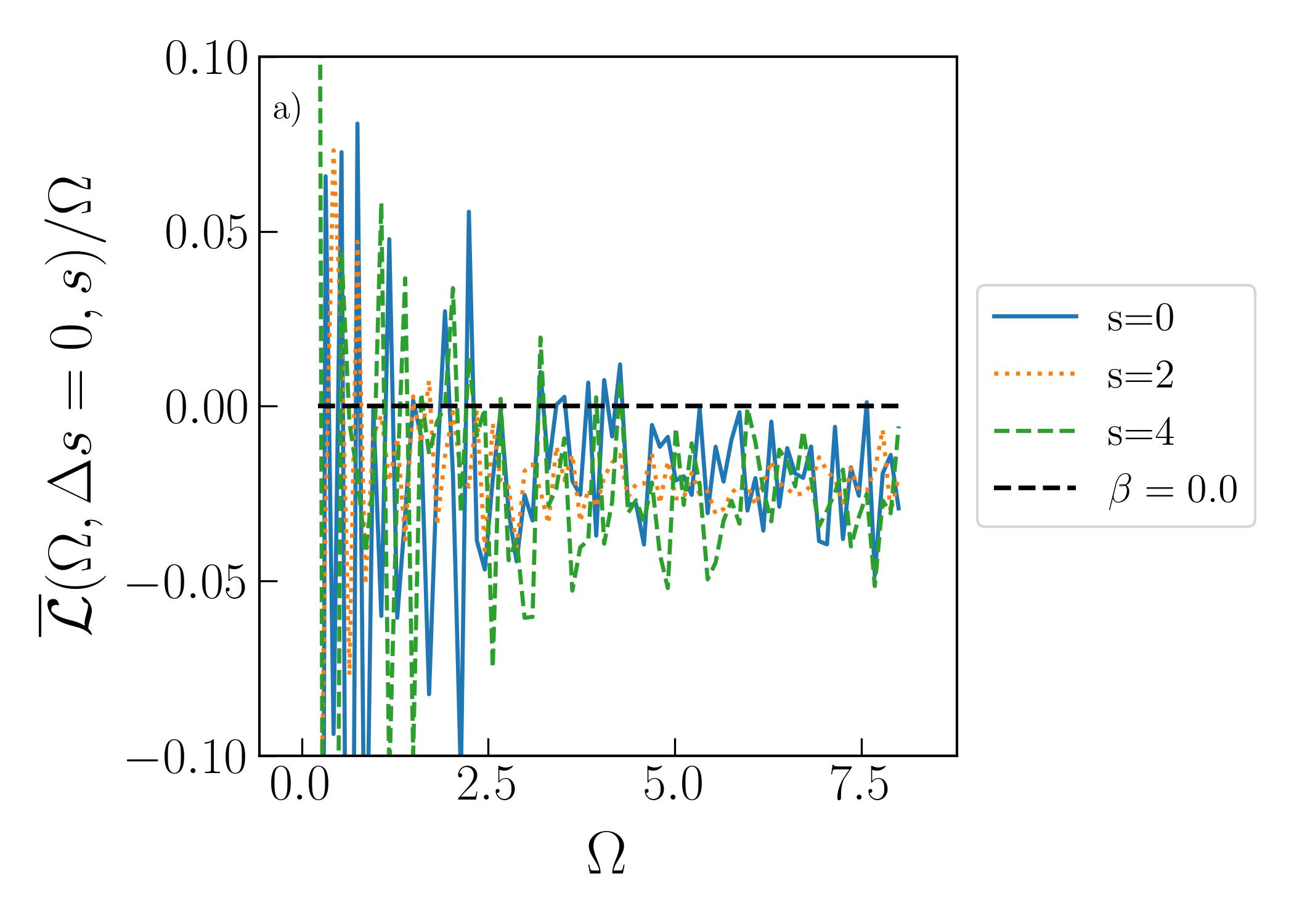}
    \end{subfigure}\hspace{1em}%
    \begin{subfigure}[t]{0.45\textwidth}
        \centering
        \includegraphics[width=0.9\textwidth]{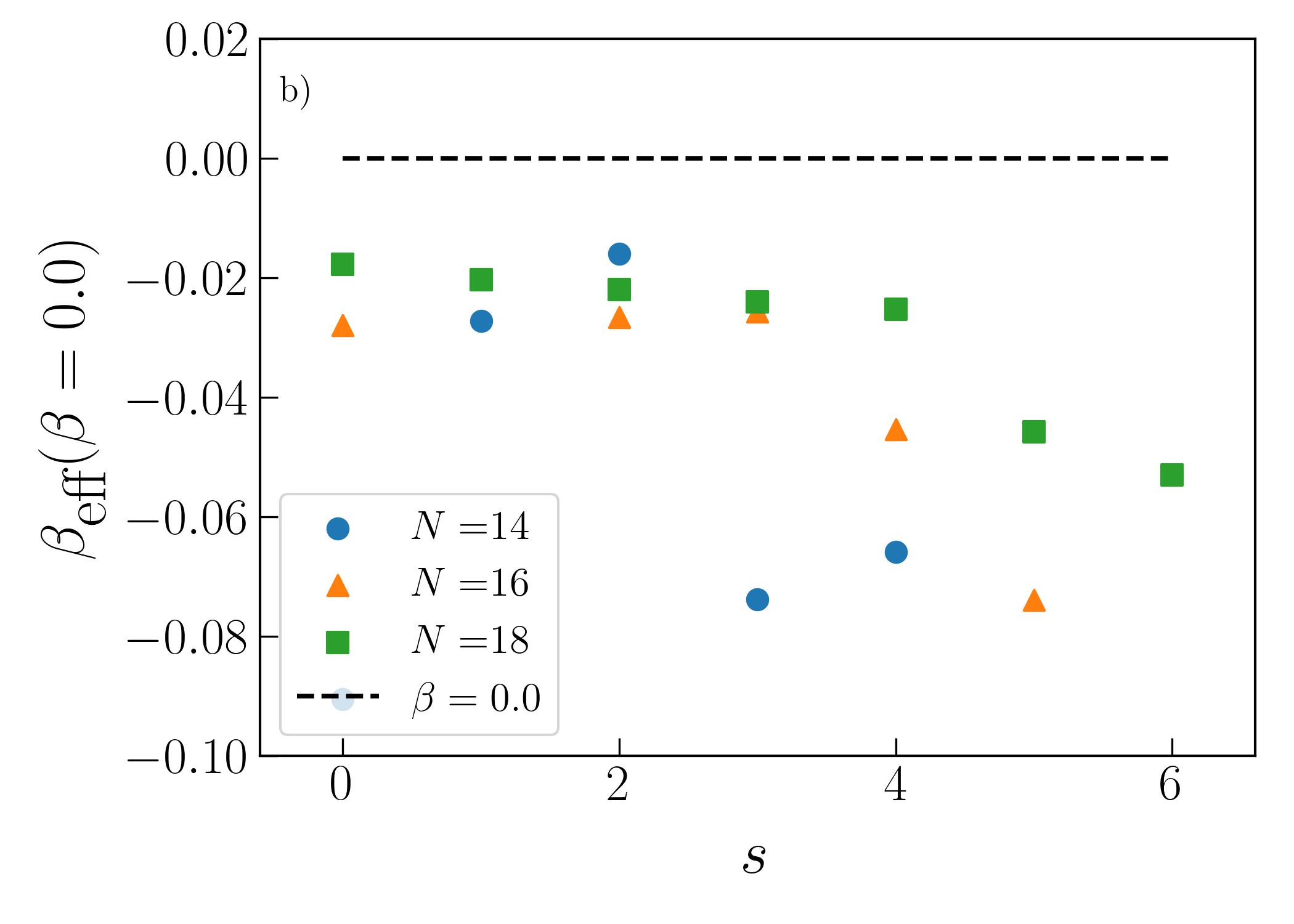}
    \end{subfigure}
    \caption{Test of the KMS relation within each $s$ eigenspace for $N=18$, at $\beta=0.0$.
    (a) Normalized average logarithmic ratio  $\overline{\mathcal{L}} (\Omega, \Delta s, s)  / \Omega$. Only even $s$ for  $s \leq 4$ are presented for clarity, but all $\overline{\mathcal{L}} (\Omega, \Delta s, s)  / \Omega$ show similar behavior.  (b) Effective inverse temperature $\beta_{\mathrm{eff}}$ at the true inverse temperature $\beta=0.0$ (dashed black line). Different marker colors (and shapes) correspond to different system sizes. Since $\overline{\mathcal{L}}  / \Omega$ lies close to $\beta$ (left-hand plot), $\beta_{\mathrm{eff}}$ agrees with $\beta$ well (right-hand plot).
    }
    \label{fig:AppE_fig3}
\end{figure}

\begin{figure}[H]
    \centering
    \begin{subfigure}[t]{0.45\textwidth}
        \centering
        \includegraphics[width=0.9\textwidth]{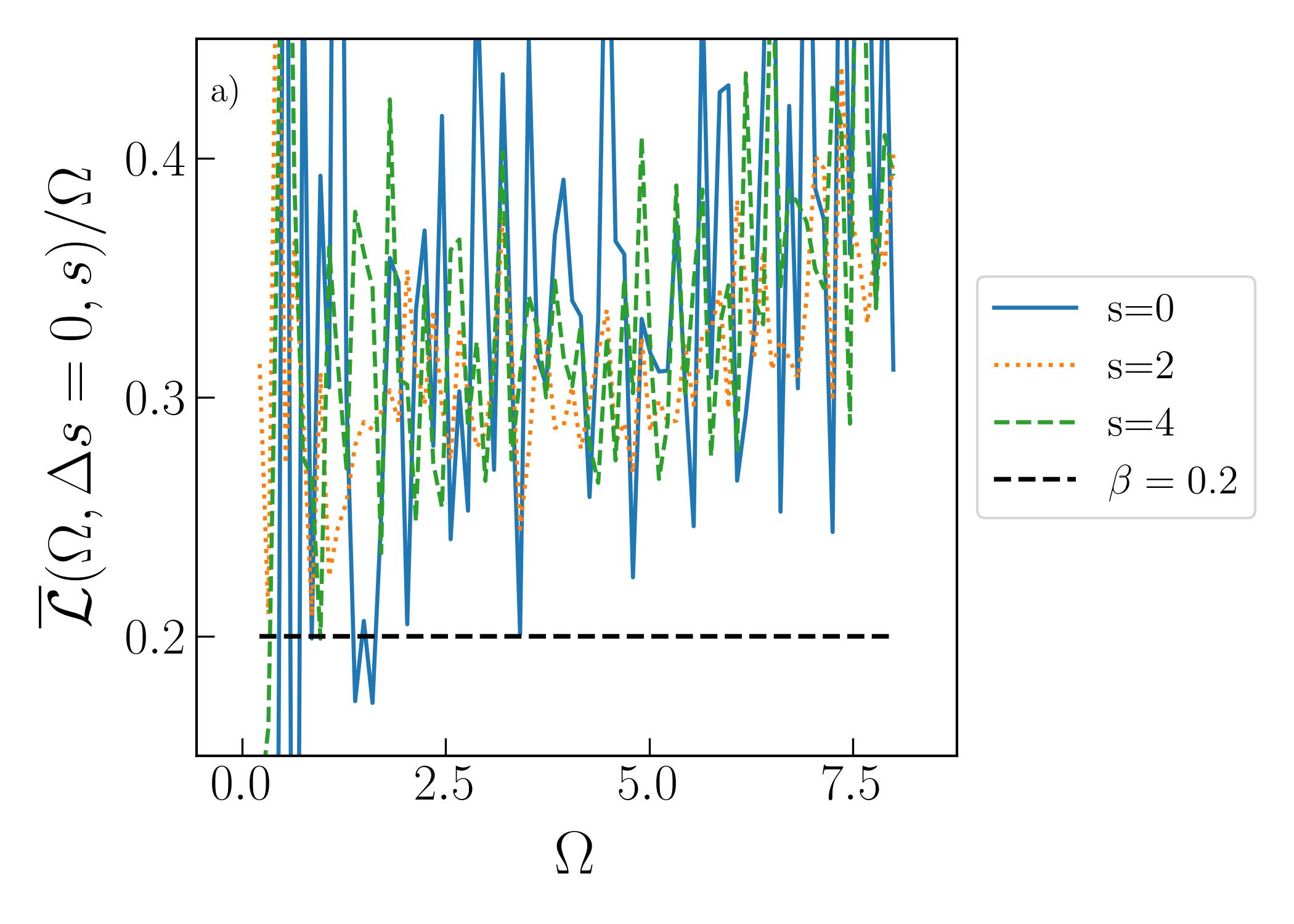}
    \end{subfigure}\hspace{1em}%
    \begin{subfigure}[t]{0.45\textwidth}
        \centering
        \includegraphics[width=0.9\textwidth]{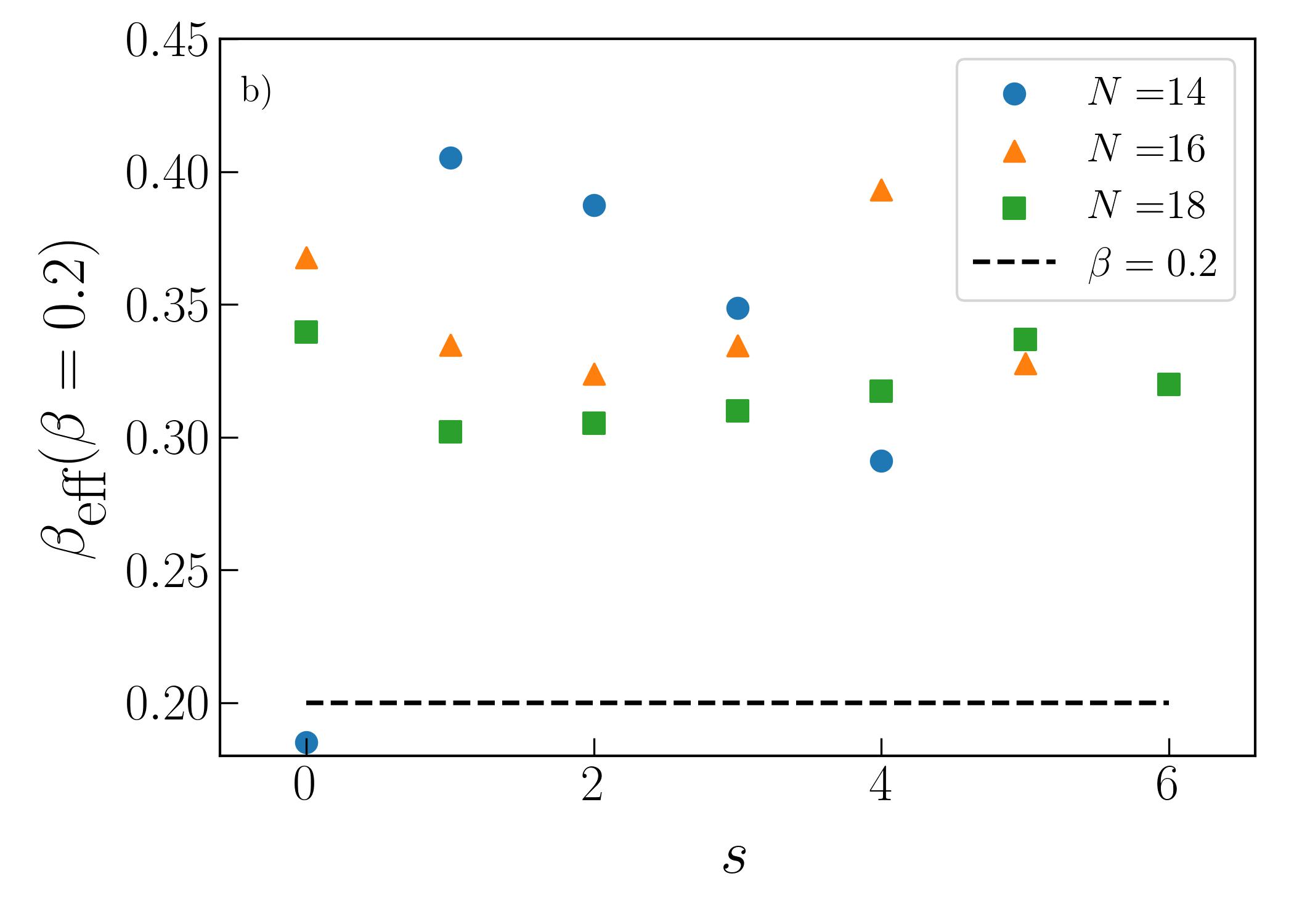}
    \end{subfigure}
    \caption{Test of the KMS relation within each $s$ eigenspace for $N=18$, at $\beta=0.2$.
    (a) Normalized average logarithmic ratio   $\overline{\mathcal{L}} (\Omega, \Delta s, s)  / \Omega$. Only even $s$ for $s \leq 4$ are presented for clarity, but all $\overline{\mathcal{L}} (\Omega, \Delta s, s)  / \Omega$ show similar behavior.
    (b) Effective inverse temperature  $\beta_{\mathrm{eff}}$ at true inverse temperature $\beta=0.2$. Different marker colors (and shapes) correspond to different system sizes. $\beta_{\mathrm{eff}}$ tends to overestimate $\beta$.
    }
    \label{fig:AppE_fig4}
\end{figure}

Now, we support the KMS relation (Eq.~\eqref{eq_KMS_eigen} ) within various $s$ subspaces. We calculated the normalized average logarithmic ratio 
$\overline{\mathcal{L}} / \Omega$ versus $\Omega$ at various $s$ values. Figure~\ref{fig:AppE_fig3}(a) shows the result at $\beta=0.0$; and Fig.~\ref{fig:AppE_fig4}(a), at $\beta=0.2$.
Different curves correspond to different spin quantum numbers $s \in \{0, 1,2,3,4\}$. We omit higher $s$ values for clarity and to eliminate low-quality data: if $s > 4$, the corresponding subspace is too small to yield enough data to produce high-quality results. From the aforementioned figures' data, we produced Figs.~\ref{fig:AppE_fig3}(b) and~\ref{fig:AppE_fig4}(b). The latter figures show the effective inverse temperature $\beta_\textrm{eff}$ at each $s$ value. (We have defined $\beta_{\rm eff}$ as the average of 
$\overline{\mathcal{L}}/\Omega$ across $\Omega \in [2, 5]$.) When $\beta=0.0$, $\beta_{\mathrm{eff}}$ agrees well with $\beta$. When $\beta=0.2$, $\beta_{\mathrm{eff}}$ overestimates $\beta$. $\overline{\mathcal{L}} / \Omega$ (from which $\beta_\textrm{eff}$ is derived) is considerably noisier at $\beta=0.2$ than at $\beta=0.0$, possibly because the corresponding subspace is smaller. The increased noise and overestimation of $\beta$ in Fig.~\ref{fig:AppE_fig4} is consistent with the $\beta{=}0.2$ data presented in Fig.~\ref{fig:AppE_fig1}.

\begin{figure}[H]
    \centering
    \begin{subfigure}[t]{0.45\textwidth}
        \centering
        \includegraphics[width=0.9\textwidth]{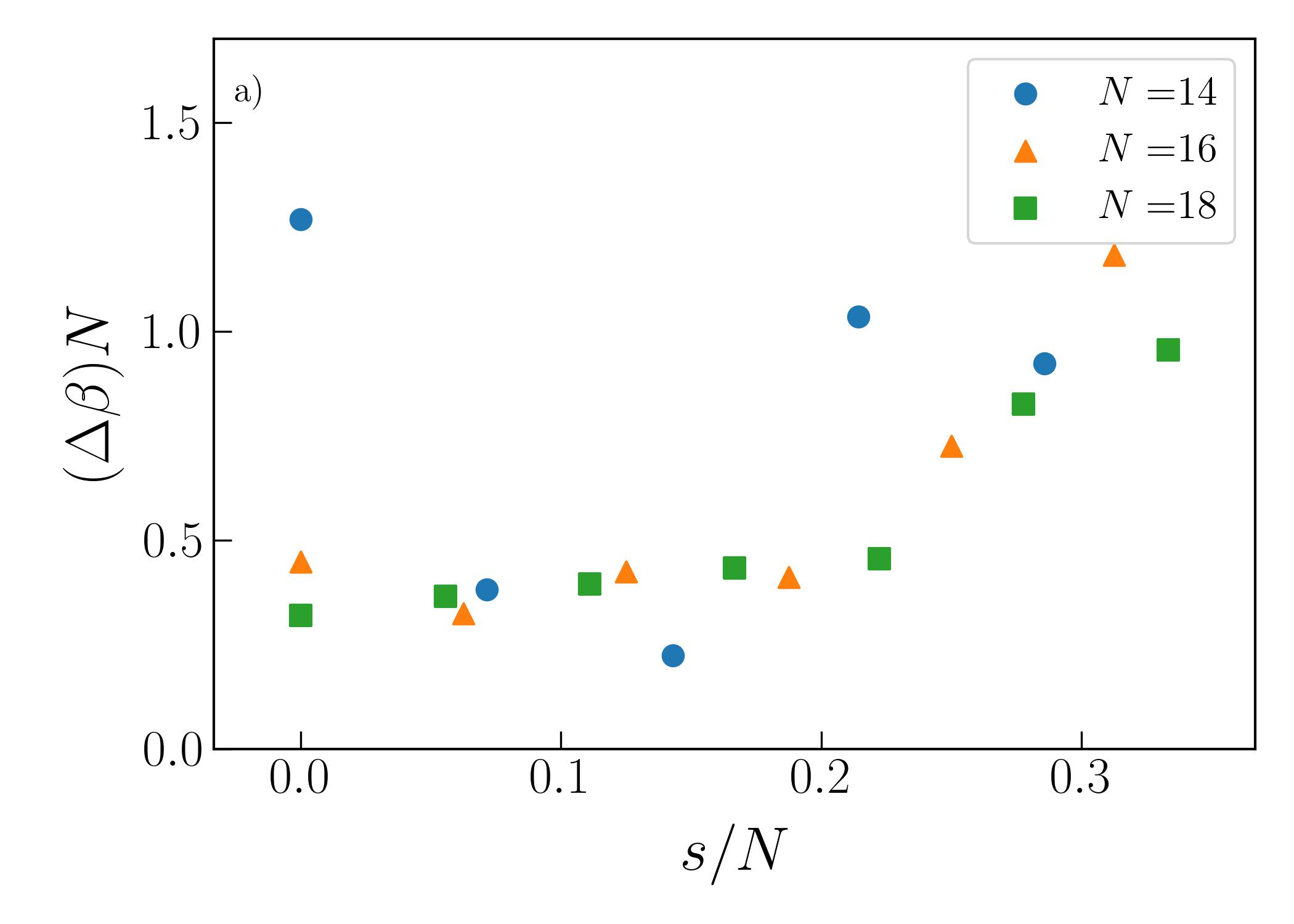}
        \label{fig:AppE_LvsOmega_varyS_beta02}
    \end{subfigure}\hspace{1em}%
    \begin{subfigure}[t]{0.45\textwidth}
        \centering
        \includegraphics[width=0.9\textwidth]{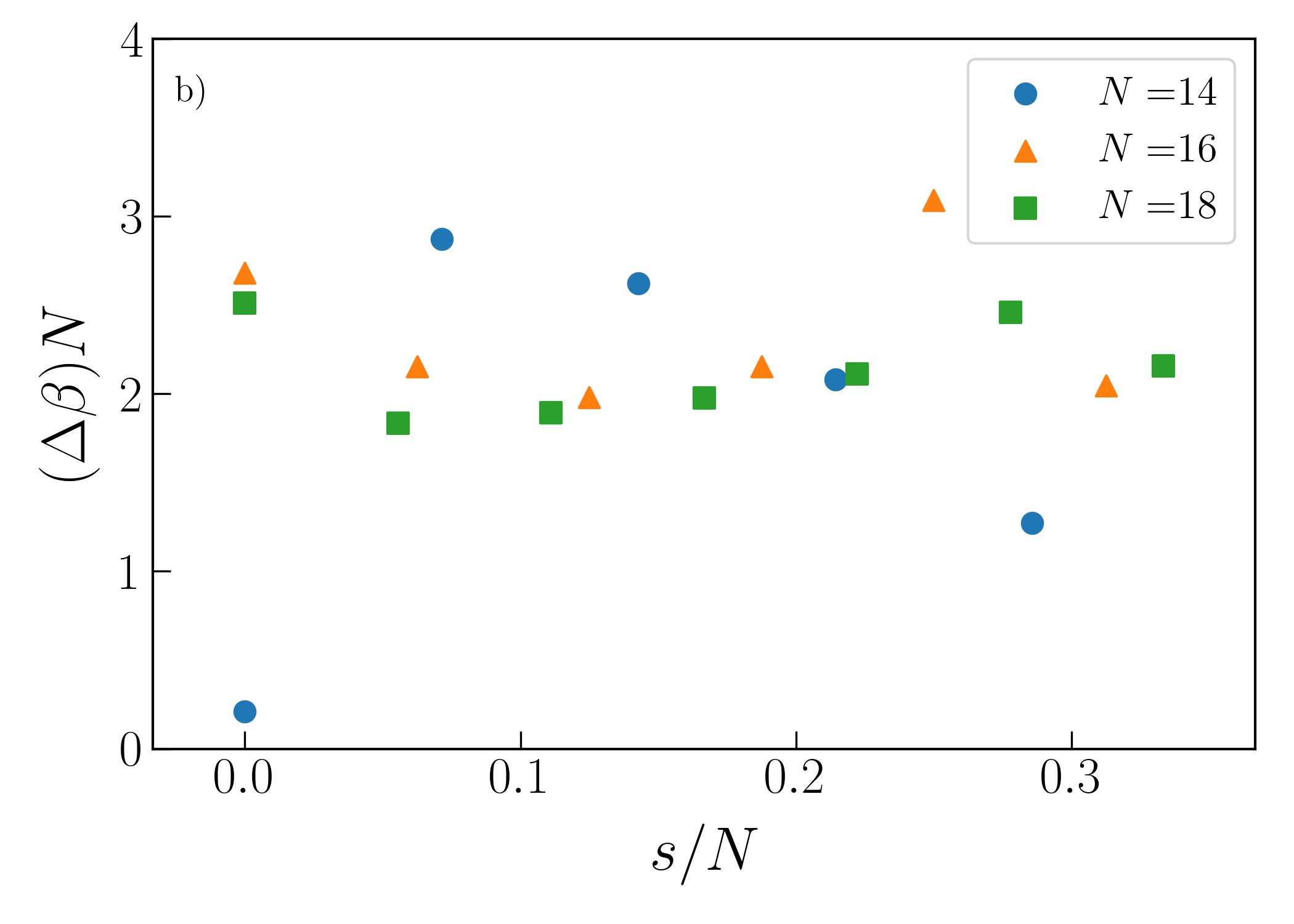}
        \label{fig:AppE_betaeffvsS_varyN_beta02}
    \end{subfigure}
    \caption{Effective inverse temperature's deviation $\Delta \beta = \beta_\textrm{eff}-\beta$, times the system size $N$, versus $s/N$. In (a), the inverse temperature $\beta=0.0$; and, in (b), $\beta=0.2$. Different marker colors (and shapes) correspond to different system sizes.}
    \label{fig:AppE_fig5}
\end{figure}

Figure~\ref{fig:AppE_fig5} shows the average finite-size correction $\Delta \beta$, times the system size $N$, as a function of $s/N$. The finite-size correction $\Delta \beta \coloneqq \beta_\textrm{eff} - \beta$.
Figure~\ref{fig:AppE_fig5}(a) shows the $\beta{=}0.0$ data; and Fig.~\ref{fig:AppE_fig5}(b), the $\beta{=}0.2$ data.
$\beta_\textrm{eff}$ agrees with the thermodynamic inverse temperature $\beta$ well when $\beta=0.0$. The KMS relation implies this result. The agreement is poorer when $\beta=0.0$. The reason is likely that the corresponding subspace is smaller when $\beta = 0.0$, such that the width-$(\Delta E)$ window contains fewer energy eigenstates to average over. This finding is consistent with Fig.~\ref{fig:AppE_fig4}, which shows that $\beta_{\rm eff}$ overestimates $\beta$ when $\beta=0.2$.

\end{appendices}

\bibliography{FDT_bib.bib}

\end{document}